\documentclass[a4paper,11pt]{article}

\usepackage{jheppub}
\usepackage{amsmath}
\usepackage{amssymb}
\usepackage{amsthm}
\usepackage{yfonts}
\usepackage{mathrsfs}
\usepackage{enumerate}
\usepackage{graphicx}
\usepackage{dsfont}
\usepackage{hyperref}
\usepackage{tikz}
\usepackage{bm}
\usepackage{subfig}
\usetikzlibrary{arrows, positioning, shapes.geometric, decorations.markings, patterns}
\newcommand{\bra}[1]{\langle #1 |}
\newcommand{\ket}[1]{| #1 \rangle}
\newcommand{\braket}[2]{\left\langle #1 \big\vert #2 \right\rangle}

\newcommand{\rank}{\operatorname{rank}}

\renewcommand{\tilde}{\widetilde}
\renewcommand{\bar}{\overline}
\newcommand\onenorm[1]{\lVert#1\rVert_1}
\newcommand{\tr}{\operatorname{Tr}}

\newtheorem{thm}{Theorem}

\title{Beyond toy models: \\distilling tensor networks in full AdS/CFT}

\author[a]{Ning Bao}
\author[b]{Geoffrey Penington}
\author[b]{Jonathan Sorce}
\author[c]{Aron C. Wall}

\affiliation[a]{Berkeley Center for Theoretical Physics, University of California, 366 Le Conte Hall, Berkeley, CA 94720-7300, U.S.A.}
\affiliation[b]{Stanford Institute for Theoretical Physics, Stanford University, 382 Via Pueblo Mall, Stanford, CA 94305-4060, U.S.A.}
\affiliation[c]{Centre for Mathematical Sciences, Cambridge University, Wilberforce Rd, Cambridge CB3 0WA, U.K.}

\emailAdd{ningbao75@gmail.com}
\emailAdd{geoffp@stanford.edu}
\emailAdd{jsorce@stanford.edu}
\emailAdd{aroncwall@gmail.com}

\abstract{We present a general procedure for constructing tensor networks that accurately reproduce holographic states in conformal field theories (CFTs). Given a state in a large-$N$ CFT with a static, semiclassical gravitational dual, we build a tensor network by an iterative series of approximations that eliminate redundant degrees of freedom and minimize the bond dimensions of the resulting network. We argue that the bond dimensions of the tensor network will match the areas of the corresponding bulk surfaces. For ``tree'' tensor networks (i.e., those that are constructed by discretizing spacetime with non-intersecting Ryu-Takayanagi surfaces), our arguments can be made rigorous using a version of one-shot entanglement distillation in the CFT. Using the known quantum error correcting properties of AdS/CFT, we show that bulk legs can be added to the tensor networks to create holographic quantum error correcting codes. These codes behave similarly to previous holographic tensor network toy models, but describe actual bulk excitations in continuum AdS/CFT.

By assuming some natural generalizations of the ``holographic entanglement of purification'' conjecture, we are able to construct tensor networks for more general bulk discretizations, leading to finer-grained networks that partition the information content of a Ryu-Takayanagi surface into tensor-factorized subregions. While the granularity of such a tensor network must be set larger than the string/Planck scales, we expect that it can be chosen to lie well below the AdS scale. However, we also prove a no-go theorem which shows that the bulk-to-boundary maps cannot all be isometries in a tensor network with intersecting Ryu-Takayanagi surfaces.}

\arxivnumber{1812.01171}

\begin{document}
\maketitle

\flushbottom

\section{Introduction}

Among the most striking predictions of the AdS/CFT correspondence \cite{Maldacena1999} is that the entanglement structure of a holographic CFT state is encoded in the geometry of its semiclassical gravitational dual. This conjectured correspondence is made precise for static spacetimes by the Ryu-Takayanagi (RT) formula \cite{RT2006-1, RT2006-2}, and for dynamical spacetimes by the Hubeny-Rangamani-Takayanagi (HRT) formula \cite{HRT2007}. These formulas relate the entanglement entropy of a CFT subregion to the area of an extremal surface in the bulk whose boundary coincides with that of the subregion, and were derived by path integral arguments in \cite{LM,DLR}.

This apparent holographic relationship between geometry and entanglement led to the proposal that tensor networks, which were originally developed as tools for the numerical analysis of condensed matter systems with restricted entanglement structure, might be a good toy model for the AdS/CFT correspondence \cite{Swingle2012-1, Swingle2012-2}. Tensor networks represent a quantum state on a $D$-dimensional lattice as a contraction of tensors lying on a $D+1$-dimensional graph with the lattice as its boundary, naturally disentangling the boundary state into a geometric ``bulk'' representation.

Since entanglement in the boundary state of a tensor network is related to the geometry of its bulk graph, tensor networks display holographic properties that are at least superficially similar to those of AdS/CFT. For example, all tensor networks follow a version of the Ryu-Takayanagi formula, in the sense that the entanglement entropy of a boundary subregion is bounded above by the ``minimal area'' bulk graph cut sharing a boundary with that subregion \cite{Swingle2012-1}. For a large class of tensor networks, this ``Swingle bound'' is exactly or approximately saturated \cite{HaPPY, HNQTWY2016}. Many tensor networks also display features of quantum error correction \cite{HaPPY, FP2014, KC2018}, which are expected to appear in the AdS/CFT correspondence \cite{ADH2015}. On the other hand, these models typically feature a flat (or almost flat) entanglement spectrum, which is at odds with known entanglement features of AdS/CFT (for more discussion on this point, see \cite{HNQTWY2016}).

In much of the existing literature, the idea that AdS/CFT can be explained in terms of tensor networks is taken ``seriously, but not literally.'' Holographic tensor networks are generally regarded as toy models for AdS/CFT that provide some intuition for how the geometric structure of a spacetime is encoded in the entanglement of its boundary dual. Even when a tensor network is interpreted literally as an approximate holographic description of a CFT state, it is often assumed that each tensor in the network must represent a volume of space that is at least of order $\ell_{AdS}^{d-1}$ (where $\ell_{AdS}$ is the characteristic scale of the semiclassical bulk spacetime, and $d$ is its spacetime dimension).\footnote{For a previous effort to construct tensor networks below the AdS scale, see \cite{YHQ2016}.}

This conclusion, however, is fundamentally at odds with the notion that tensor networks can be used to understand the Ryu-Takayanagi formula. Since the Ryu-Takayanagi formula is believed to hold exactly up to quantum and stringy corrections, it cannot be described adequately by a model that displays only AdS-scale locality. Any true tensor-network explanation of the Ryu-Takayanagi formula must therefore involve tensor network models that describe spacetime accurately at sub-AdS scales (so long as those scales remain above the string and Planck scales).

The Ryu-Takayanagi formula is not the only geometric formula for an information-theoretic quantity that is claimed to hold below AdS scales. Progress in understanding the holographic properties of tensor networks has been accompanied in recent years by a series of increasingly bold conjectures regarding the formal relationship between entanglement and geometry in AdS/CFT (e.g., the holographic entanglement of purification conjecture \cite{TU2017, NDHZS2017}, which will play a central role in this paper). While many of these conjectures are partially inspired by tensor network models, they too are supposed to hold up to quantum (and stringy) corrections --- well below the scale of existing tensor network models.

In this paper, we take literally the idea that there exists an approximate tensor network description for any holographic state, for essentially any discretization of the bulk, so long as the Planck and string scales are small compared to the discretization scale. As opposed to previous tensor network constructions, where a tensor network toy model is first defined and then shown to have properties similar to AdS/CFT, here we start with a holographic state in full AdS/CFT and construct a tensor network that describes it with high accuracy. Given a holographic CFT state with a static semiclassical dual, and assuming certain conjectures that naturally extend the holographic entanglement of purification conjecture, we provide an explicit procedure to construct a network that is ``geometrically appropriate'' for the holographic state in the following sense:
\begin{enumerate}[(i)]
	\item it approximately reproduces the original CFT state on its boundary, and 
	\item it has the same geometric features as the bulk dual to leading order in $N$. 
\end{enumerate}
Our networks include subleading fluctuations around a flat entanglement spectrum, which we interpret as corresponding to fluctuations of the areas of extremal surfaces in full AdS/CFT. As a result, our constructions do not suffer from the usual issue of a flat, non-physical entanglement spectrum.\footnote{Our tensor networks will not have the same R\'{e}nyi entropies as the AdS/CFT states from which they are constructed, nor do they need to in order for condition (i) to be satisfied. As mentioned in Section \ref{sec:bipartite_networks}, the R\'{e}nyi entropies are very sensitive to small changes in a state that do not alter its physical properties.}

If our assumptions hold up to quantum and stringy corrections (as is commonly assumed for the Ryu-Takayanagi formula and the holographic entanglement of purification conjecture), then essentially any discretization of a bulk geometry gives a corresponding tensor network description of the boundary state at sufficiently large $N$ (and strong coupling). The tensor network description of the AdS/CFT correspondence would therefore be valid even at sub-AdS scales, and could be interpreted not as a ``toy model'' but as a genuine description of the quantum gravitational theory.

When the chosen discretization of the bulk geometry is constructed using only non-intersecting minimal surfaces, we are able to prove the existence of a corresponding tensor network, whose graph will always form a tree, without resorting to the holographic entanglement of purification. The construction relies only on the Ryu-Takayanagi formula for the von Neumann entropy and its extension to more general formulas for holographic R\'{e}nyi entropies in \cite{Dong2016}; it is only when we extend our construction to finer-grained discretizations (and hence tensor networks with loops) that we require the use of entropies of purification. Regardless of the particular discretization in question, all of our procedures involve a process of disentangling and removing as many boundary degrees of freedom as possible to simplify the entanglement structure of a holographic state without altering its physical properties. In this sense, our constructions form a systematic approach for building tensor networks that describe their corresponding holographic states with maximal efficiency.

In Section \ref{sec:background}, we review the fundamental principles of tensor networks, introduce the abstract index notation that will be used in the remainder of the paper, and explain the basics of entanglement distillation from the tensor network perspective. We also sketch an important, but unpublished, result due to Hayden, Swingle, and Walter \cite{HS} on one-shot holographic entropies that allows us to apply the general entanglement distillation procedure in the holographic context. In Section \ref{sec:tree_networks}, we then identify a large class of tensor networks, which we call ``tree tensor networks,'' that can be constructed rigorously from holographic CFT states using one-shot entanglement distillation. In Section \ref{sec:qec}, we show how these constructions can be adapted to produce holographic quantum error correcting codes. In Section \ref{sec:MEP}, we review the holographic entanglement of purification conjecture, and show that it can be used to improve the granularity of our networks by localizing the information contained within a single Ryu-Takayanagi surface. In Section \ref{sec:iteration}, we explain how natural generalizations of the holographic entanglement of purification conjecture can be used to extend this procedure to produce even finer-grained tensor networks, where the discretization scale of the network may lie well below the AdS scale so long as it exceeds the string and Planck lengths. In Section \ref{sec:quantum_geo}, we discuss quantum fluctuations of the spacetime geometry and propose an interpretation in terms of quantum superpositions of tensor networks, which require significantly fewer degrees of freedom to describe than a full, ``fluctuating'' geometry. We identify an uncertainty relationship between the areas of intersecting Ryu-Takayanagi surfaces, and prove a related no-go theorem that limits the kinds of bulk-to-boundary isometries that can be obtained in a tensor network that accurately reproduces the geometry of AdS/CFT. Finally, in Section \ref{sec:discussion}, we summarize our essential results and present several potential avenues for future work.

\section{Tensor Networks, Entanglement Distillation, and One-Shot Quantum Information} \label{sec:background}

\subsection{Tensor Networks} \label{sec:TN_review}

A pure state $\ket{\psi}$ on a multipartite Hilbert space $\mathcal{H} = \mathcal{H}_{A_1} \otimes \dots \otimes \mathcal{H}_{A_n}$ may be thought of as a tensor with $n$ (abstract) up-indices, each one corresponding to a tensor factor of $\mathcal{H}$. In the tensor interpretation, we write such a state as
\begin{equation}
	\ket{\psi}
		\leftrightarrow \psi^{A_1 \dots A_n}.
\end{equation}
Such a tensor can generally be written as an outer product and contraction of other tensors, each of which acts only on some subset of the tensor factors $A_1$ through $A_n.$ For $n=2$, for example, a state might be written as
\begin{equation}
	\psi^{A_1 A_2}
		= P^{A_1 B}{}_{C} Q^{A_2 C}{}_{B}, \label{simple_network}
\end{equation}
where the indices $B$ and $C$ correspond to some auxiliary Hilbert spaces $\mathcal{H}_B$ and $\mathcal{H}_C$ (and their dual spaces $\mathcal{H}_B^*$ and $\mathcal{H}_C^*$), defined solely for the purpose of constructing tensors $P$ and $Q$. Note that up-indices always refer to vector spaces, while down-indices refer to their corresponding dual spaces.

While it is always possible to find \emph{some} outer product representation of any given tensor, there are significant computational advantages to finding one in which the contracted Hilbert spaces have small dimension compared to the physical Hilbert space factors. (These contracted Hilbert spaces are often referred to as ``bonds,'' with their dimensions referred to as the ``bond dimensions'' of the network.) Even in cases where the bond dimension is chosen to be of the same order as the physical Hilbert space dimension, many physically interesting states have outer product representations with some particular restricted structure that allows them to be simulated efficiently on a classical computer (see, e.g., \cite{Vidal2007} and \cite{VMC2008}).

\begin{figure}[h]
	\begin{center}
	\begin{tikzpicture}[->, >=latex, auto, tensor/.style={circle, draw, minimum width=3em}, label/.style={}]
		\node[tensor] (P) {$P$};
		\node[tensor] (Q) [right=4cm of P] {$Q$};
		\node[label] (A1) [left=1cm of P] {$A_1$};
		\node[label] (A2) [right=1cm of Q] {$A_2$};

		\draw (P) edge[thick, bend left] node {$B$} (Q);
		\draw	(Q) edge[thick, bend left] node {$C$} (P);
		\draw (P) edge[thick] node {} (A1);
		\draw (Q) edge[thick] node {} (A2);
	\end{tikzpicture}
	\caption{A simple network for the state given in equation (\ref{simple_network}). Outward-pointing arrows denote up-indices, inward-pointing arrow denote down-indices, and arrows
			connecting tensors denote contractions.}
	\label{fig:simple_network}
	\end{center}
\end{figure}
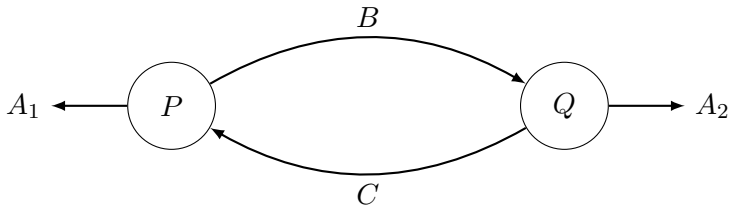

From the perspective of holography, one advantage of such an outer product representation of a quantum many-body state is that it has a natural geometric interpretation that shares features with holographic spacetimes. This geometric interpretation is called a \emph{tensor network}. A tensor network is constructed from an outer product representation of a quantum state by drawing a vertex for each tensor, with one edge for each of its indices. In our convention, edges corresponding to up-indices will be labeled with arrows that point away from the vertex, while edges corresponding to down-indices will be labeled with arrows that point toward the vertex. Contractions are denoted by connecting the corresponding edges. As a simple example, the tensor network corresponding to the state in equation (\ref{simple_network}) is given in Figure \ref{fig:simple_network}.

An important connection between tensor networks and quantum information theory arises from the fact that 
each bond in a tensor network can be thought of as a projection onto a maximally entangled state in the bond Hilbert space. For concreteness, consider the tensor network representation
\begin{equation}
	\psi^{AB}
		= P^{A \gamma} Q^{B}{}_{\gamma} \label{simple_contraction}
\end{equation}
of a state on the bipartite Hilbert space $\mathcal{H}_{A} \otimes \mathcal{H}_B$. The tensor network is constructed by contraction over some bond Hilbert space $\mathcal{H}_{\gamma}.$ The inner product on this bond Hilbert space selects a preferred maximally entangled state $\ket{\phi}$ on the product space $\mathcal{H}_{\gamma} \otimes \bar{\mathcal{H}}_{\gamma}$, where $\bar{\mathcal{H}}_{\gamma}$ is the complex conjugate vector space to $\mathcal{H}_{\gamma}.$\footnote{Formally, the inner product on $\mathcal{H}_{\gamma}$ is a bilinear map $L : \mathcal{H}_{\gamma} \times \bar{\mathcal{H}}_{\gamma} \rightarrow \mathbb{C},$ or, equivalently, a tensor in the space $\mathcal{H}_{\gamma}^{*} \otimes \bar{\mathcal{H}}_{\gamma}^{*}$. Since the inner product is nondegenerate by assumption, it has an inverse tensor $L^{-1}$ on $\mathcal{H}_{\gamma} \otimes \bar{\mathcal{H}}_{\gamma},$ which can be shown to be a maximally entangled state on the tensor product Hilbert space.} Since the state is maximally entangled, tracing out either of its tensor factors yields the identity operator on the remaining tensor factor (up to a normalization factor). In index notation, this statement is simply
\begin{eqnarray}
	\phi^{\gamma \bar{\gamma}} \phi^{*}_{\gamma' \bar{\gamma}}
		&=& \frac{1}{d}\delta^{\gamma}{}_{\gamma'}, \label{max_ent_metric_inverse} \\
	\phi^{\gamma \bar{\gamma}} \phi^{*}_{\gamma \bar{\gamma}'}
		&=& \frac{1}{d}\delta^{\bar{\gamma}}{}_{\bar{\gamma}'}, \label{max_ent_conjugate_metric_inverse}
\end{eqnarray}
where $\phi^*$ denotes the tensor corresponding to the dual state $\bra{\phi}$. Equations (\ref{max_ent_metric_inverse}) and (\ref{max_ent_conjugate_metric_inverse}) imply that $\phi$ and $\phi^*$ can be used to raise and lower indices between $\mathcal{H}_{\gamma}$ and $\bar{\mathcal{H}}_{\gamma}.$  Since $\phi^*$ can be obtained from $\phi$ by lowering its indices, we will generally drop the asterisk and refer to the tensors as $\phi^{\gamma \bar{\gamma}}$ and $\phi_{\gamma \bar{\gamma}},$ respectively.

Using $\phi$ to raise and lower indices, the state in equation (\ref{simple_contraction}) can be rewritten as
\begin{equation}
	\psi^{AB}
		= P^{A \gamma} Q^{B \bar{\gamma}} \phi_{\gamma \bar{\gamma}}, \label{simple_PEP}
\end{equation}
which is a slightly different tensor network representation of the same state. In the familiar bra-ket notation, this corresponds to projecting a state $\ket{P} \in \mathcal{H}_A \otimes \mathcal{H}_{\gamma}$ and a state $\ket{Q} \in \mathcal{H}_{B} \otimes \bar{\mathcal{H}}_{\gamma}$ onto a maximally entangled state $\ket{\phi} \in \mathcal{H}_{\gamma} \otimes \bar{\mathcal{H}}_{\gamma}$, i.e.,
\begin{equation}
	\ket{\psi}
		= \bra{\phi}(\ket{P} \otimes \ket{Q}). \label{PEP_bond}
\end{equation}

A tensor network in which every bond takes this form is sometimes referred to as a \emph{projected entangled-pair state} (PEPS) network (see, e.g., \cite{HNQTWY2016}), though the term PEPS is more commonly used to refer to a more highly-restricted class of tensor networks on a (usually square) lattice where each tensor has a single uncontracted physical index \cite{VC2004}. To avoid confusion, we refer to a tensor network with bonds of the form (\ref{PEP_bond}) as a \emph{projection of entangled pairs} (PEP). Note that any tensor network can be rewritten in this form by raising and lowering indices with the appropriate maximally entangled state on each bond.

\begin{figure}[h]
	\begin{center}
	\subfloat[\label{fig:simple_contraction_network}]{
		\begin{tikzpicture}[->, >=latex, auto, tensor/.style={circle, draw, minimum width=3em}, label/.style={}]
		\node[tensor] (P) {$P$};
		\node[tensor] (Q) [right=2cm of P] {$Q$};

		\node[label] (A) [above left=1cm of P]{$A$};
		\node[label] (B) [above right=1cm of Q] {$B$};

		\draw (P) edge[thick] node {$\gamma$} (Q);
		\draw (P) edge[thick] node {} (A);
		\draw (Q) edge[thick] node {} (B);
		\end{tikzpicture}
	}
	\hspace{1cm}
	\subfloat[\label{fig:simple_PEP_network}]{
		\begin{tikzpicture}[->, >=latex, auto, tensor/.style={circle, draw, minimum width=3em}, EPR/.style={regular polygon, regular polygon sides=4, minimum width=3.5em, draw},
					label/.style={}]
		\node[tensor] (P) {$P$};
		\node [EPR] (phi) [above right=1cm and 1cm of P] {$\phi$};
		\node[tensor] (Q) [below right=1cm and 1cm of phi] {$Q$};

		\node[label] (A) [above left=1cm of P]{$A$};
		\node[label] (B) [above right=1cm of Q] {$B$};

		\draw (P) edge[thick] node {$\gamma$} (phi);
		\draw (Q) edge[thick] node {$\bar{\gamma}$} (phi);
		\draw (P) edge[thick] node {} (A);
		\draw (Q) edge[thick] node {} (B);
		\end{tikzpicture}
	}
	\caption{Tensor networks for equations (\ref{simple_contraction}) and (\ref{simple_PEP}). (a) A tensor network on a bipartite system with a single contraction. (b) A PEP
			network for the same state created by replacing the contraction with a maximally entangled state.}
	\end{center}
\end{figure}
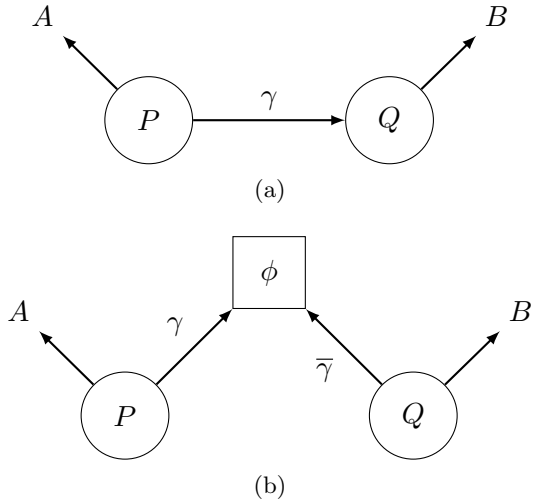

The original tensor network given by equation (\ref{simple_contraction}) is drawn in Figure \ref{fig:simple_contraction_network}, and its equivalent PEP network in Figure \ref{fig:simple_PEP_network}. Since PEP networks manifestly represent bonds as maximally entangled states, it is generally useful to consider them when drawing connections between tensor networks and AdS/CFT, where geometric features of a semiclassical spacetime correspond directly to entanglement features of its boundary dual.

Particular classes of tensor networks are known to reproduce various features of AdS/CFT \cite{HaPPY, HNQTWY2016}, such as a version of the Ryu-Takayanagi formula and holographic quantum error correction. In this paper, we work backwards, beginning with these known features of AdS/CFT and using them to \emph{produce} tensor networks with corresponding properties. Much of our protocol is reliant on the procedure of \emph{entanglement distillation}, which shows how the entanglement between subregions of some physical state can be distilled out of the state in the form of a large number of EPR pairs (which will, ultimately, become the maximally entangled bonds of a PEP-style network). This procedure is the subject of the following subsection.

\subsection{Entanglement Distillation} \label{sec:tree}

Consider a state $\ket{\psi}$ in a CFT with large central charge that is known to have a static, semiclassical gravitational dual.\footnote{In fact, it suffices to consider states whose semiclassical gravitational duals contain a moment of time reflection symmetry around the Cauchy slice being considered. This assumption is crucial to our construction, however, in Section \ref{sec:discussion}, we discuss the possibility of lifting this restriction in future work.}  If the domain of the CFT is partitioned into connected regions $A$ and $A^c$, then the Ryu-Takayanagi formula states that the entanglement entropy of $\ket{\psi}$ between $A$ and $A^c$ is given to leading order in the gravitational constant $G_N$ by the area of the minimal codimension-2 bulk surface anchored on $\partial A$ and homologous to $A$, i.e.,
\begin{equation}
	S(\psi_{(A)}) = S(\psi_{(A^c)}) = \min\limits_{\gamma, \partial \gamma = \partial A} \frac{\operatorname{area}(\gamma)}{4 G_N} + O\left(1 \right). \label{RT}
\end{equation}
Implicit in this statement is a simultaneous regularization procedure where an ultraviolet cutoff is chosen in the CFT alongside a matching radial cutoff in the bulk spacetime \cite{RT2006-1}.\footnote{That the subleading corrections to the Ryu-Takayanagi formula are $O(1)$ in $G_N$ was established in \cite{FLM2013} and \cite{EW2014}.}

The Ryu-Takayanagi formula encourages us to think of the information encoded in the entanglement spectrum of $A$ and $A^c$ as lying physically on the extremal surface in the bulk that partitions $A$ and $A^c$. For example, the Ryu-Takayanagi formula is sometimes interpreted as counting the number of ``bit-threads'' of entanglement, each of which occupy a Planckian area $1/4G_N$ of the extremal surface \cite{FH2017}. One way to more concretely justify this intuition comes from \emph{entanglement distillation}, which makes precise the statement that entanglement entropy measures the number of qubits (or, more precisely, ``ebits'') of entanglement shared between a region and its complement.

Entanglement distillation is the procedure by which a large number $m$ of copies of some bipartite quantum state $\ket{\psi} \in \mathcal{H}_{A} \otimes \mathcal{H}_{A^c}$ can be converted into a large number $n$ of Bell pairs with some fixed asymptotic ratio $n/m \approx S(A) / \ln{(2)}$.\footnote{In most of the original literature, ``entanglement distillation'' is used to refer to the general case of a mixed state $\rho \in S(\mathcal{H}_{A} \otimes \mathcal{H}_{A^c})$, while ``entanglement concentration'' is used in the special case where $\rho$ is pure. Here, we use the terms interchangeably.} For our purposes, the most useful formulation of this principle is that for large $m$, the state $\ket{\psi}^{\otimes m}$ can be expressed with high fidelity as
\begin{equation}
	\ket{\psi}^{\otimes m}
		\approx (V \otimes W) \left[ \frac{1}{\sqrt{D}} \sum_{i=0}^{e^{S(A)m-O(\sqrt{m})}} \ket{i \bar{i}} \otimes \sum_{j=0}^{e^{O(\sqrt{m})}} \sqrt{p_j} \ket{j \bar{j}} \right], \label{asymptotic_distillation}
\end{equation}
where $V$ and $W$ are isometries\footnote{In the literature, the word ``isometry'' is commonly used to refer to a Hilbert space map $V$ satisfying $V^{\dagger} V = \mathds{1}.$ These maps are not generally isomorphisms in the mathematical sense, i.e., they are not invertible, unless the domain and target spaces have the same dimension.} that embed Hilbert spaces of size $e^{S(A) m - O(\sqrt{m})}$ and $e^{O(\sqrt{m})}$, along with their complex conjugate Hilbert spaces, back into the physical space $\mathcal{H}_{A} \otimes \mathcal{H}_{A^c}.$\footnote{This expression follows from a procedure of smoothing and binning the entanglement spectrum of $\ket{\psi}^{\otimes m}$ in a way that is analogous to the procedure for a holographic state detailed in Section \ref{sec:tree_networks}.}  (See, e.g., \cite{HW2003} for further details.) 

The first factor of the tensor product in equation (\ref{asymptotic_distillation}) is just a maximally entangled Bell state on $O(n)$ qubits, while the second term lives in a Hilbert space of subleading dimension in the asymptotic entanglement entropy $n \ln{(2)} = m S(A).$ The fact that asymptotically many copies of $\ket{\psi}$ can be approximately represented as a number of Bell pairs determined by the entanglement entropy gives partial justification for thinking of $S(A)$ as a measure of the number of degrees of freedom entangled between $A$ and $A^c$, which in turn encourages us to think of all the information in the entanglement spectrum of a subregion of a holographic state as living physically on its Ryu-Takayanagi surface.

Using the notation of Section \ref{sec:TN_review}, equation (\ref{asymptotic_distillation}) is a tensor network of the form
\begin{equation}
	\psi^{A_{(1)} A^c_{(1)} \dots A_{(m)} A^c_{(m)}}
		\approx V^{A_{(1)} \dots A_{(m)}}{}_{\gamma f} W^{A^c_{(1)} \dots A^c_{(m)}}{}_{\bar{\gamma} \bar{f}} \phi^{\gamma \bar{\gamma}} \sigma^{f \bar{f}}, 
		\label{asymptotic_TN}
\end{equation}
where $\phi^{\gamma \bar{\gamma}}$ is the maximally entangled state on $O(n)$ qubits and $\sigma^{f \bar{f}}$ is the leftover state on a Hilbert space of subleading dimension.

We see from equation (\ref{asymptotic_TN}) that entanglement distillation can be used to construct a simple tensor network that reproduces $\ket{\psi}^{\otimes m}$ with high fidelity for large $m$ and any given quantum state $\ket{\psi}$. We could easily apply this procedure to a holographic CFT state (or, indeed, a non-holographic CFT state) and obtain a tensor network for the product state $\ket{\psi^{\text{CFT}}}^{\otimes m}$. In the holographic case, this tensor network will respect the geometry of AdS/CFT in the sense that its internal bond dimensions are inherited from the areas of the Ryu-Takayanagi surfaces.

Of course, in quantum gravity, people do not generally consider a large number of copies of a single holographic state. Any hope of understanding the entanglement structure of a \emph{single} holographic state using the tools of quantum information theory is inhibited by the fact that almost all operational interpretations of a state's von Neumann entropy  involve an asymptotic number of copies of the state in question. For our purposes, therefore, it is instead useful to consider ``one-shot''
or ``smooth'' entropies, which determine how well procedures like entanglement distillation can be carried out using only a single copy of a state. Luckily, the one-shot entropies of holographic states are highly constrained by the large central charge of the CFT, which is the focus of the following section.

\subsection{Holographic One-Shot Entropies} \label{sec:smoothminmax}
If the von Neumann entropy only has a physically meaningful interpretation in the limit of asymptotically many copies of a state, then it might seem peculiar that the holographic dictionary relates the von Neumann entropy of a single reduced state to the (physically significant) area of a corresponding Ryu-Takayanagi surface. The resolution to this puzzle is quite simple: as we shall see, the semiclassical limit of large $N$ (or, equivalently, $G_N \rightarrow 0$) in a holographic state fulfills the same information-theoretic purpose as the asymptotic limit of a large number of identical states in non-holographic quantum information theory.

Two important entropy measures for a quantum state $\rho$ are the max-entropy
\begin{equation}
	S_{\text{max}}
		= \log \left(\rank(\rho)\right)
\end{equation}
and the min-entropy
\begin{equation}
	S_{\text{min}}
		= \log (\lambda^{-1}_{\text{max}}(\rho)),
\end{equation}
where $\lambda_{\text{max}}(\rho)$ is the largest eigenvalue of $\rho$. These quantities agree with the von Neumann entropy for a state with a flat probability spectrum, but generically differ for arbitrary density matrices while satisfying $S_{\text{max}} \geq S \geq S_{\text{min}}$. The max- and min-entropies can be interpreted as R\'{e}nyi entropies
\begin{equation}
	S_{\alpha}
		=\frac{1}{1 - \alpha} \log \tr (\rho^\alpha)
\end{equation}
in the limits $\alpha \to 0$ and $\alpha \to \infty$ respectively. (Note that the von Neumann entropy $S$ is given by the R\'{e}nyi entropy in the limit $\alpha \to 1$.)

Since we are only interested in constructing tensor network states to within some small tolerance $\varepsilon$, it is more natural for our purposes (and in many similar situations) to consider the smooth max-entropy
\begin{equation}
	S^\varepsilon_{\text{max}}
		= \min_{\onenorm{\rho - \sigma} < \varepsilon} \log \left(\rank(\sigma)\right) \label{eq:smooth_max_def}
\end{equation}
and the smooth min-entropy
\begin{equation}
	S^\varepsilon_{\text{min}}
		= \max_{\onenorm{\rho - \sigma} < \varepsilon}\log (\lambda^{-1}_{\text{max}}(\sigma)). \label{eq:smooth_min_def}
\end{equation}
In other words, we consider the minimum max-entropy (respectively the maximum min-entropy) of any state $\sigma$ lying within a ball of radius $\varepsilon$ around the state $\rho$, where we have used the trace norm $\lVert \rho -\sigma \rVert_1 = \tr \sqrt{(\rho - \sigma)^{\dagger} (\rho - \sigma)}$ as a metric on the space of density matrices.

For holographic theories, the \emph{smooth} min- and max-entropies are expected to agree with the von Neumann entropy to leading order in $1/G_N.$ This was shown in \cite{CHLS2015} for single intervals in ground and thermal states of $1+1$-dimensional holographic CFTs (where the density of states of the modular Hamiltonian may be computed explicitly), and in \cite{HS} for arbitrary regions in holographic theories of arbitrary dimension.\footnote{For discussion of smooth max-entropies in general quantum field theories, see \cite{WE2018}.} Somewhat counterintuitively, this generic behavior of the smooth min- and max-entropies follows from the fact that in holographic theories, the R\'{e}nyi entropies are given to leading order by
\begin{align}
	S_\alpha = \frac{s_\alpha}{G_N}, \label{eq:holographic_renyi}
\end{align}
where $s_\alpha$ is independent of $G_N$ but depends non-trivially on $\alpha$ and is related to the areas of surfaces in particular backreacted geometries \cite{Headrick2010, Dong2016}. Since \cite{HS} remains unpublished, we include here a simplified version of the derivation of the smooth min- and max-entropies for holographic states, which follows on very general grounds from \eqref{eq:holographic_renyi}.\footnote{While this article was in preparation, a pair of articles \cite{AR,DHM} appeared with a similar analysis of the R\'{e}nyi entropy spectrum; we will discuss their proposal that tensor networks correspond to area-eigenstates in more detail in section \ref{sec:cobwebs}.}

Let $K = - \log (\rho)$ be the modular Hamiltonian corresponding to $\rho$. Let $\{E_i\}$ be the eigenvalues of $K$, with density of states defined by
\begin{equation}
	D(E) \equiv \sum_i \delta(E - E_i). \label{eq:density_of_states}
\end{equation}
Then the partition function 
\begin{equation}
	Z(\alpha)
		= \int_0^\infty \, \text{d}E \,D(E)\, e^{-\alpha E} \label{eq:renyi_partition_function}
\end{equation}
is related to the R\'{e}nyi entropies by
\begin{equation}
	e^{(1-\alpha) S_\alpha}
		= \tr{(\rho^\alpha)} = Z(\alpha).
\end{equation}
Hence $e^{(1-\alpha) S_\alpha}$ is the Laplace transform of $D(E)$ and thus the density of states is given by
\begin{equation} \label{eq:laplace}
	D(E)
		= IL(e^{(1-\alpha) S_\alpha}) (E) = \int_C \, \text{d}\alpha\, e^{\alpha E} e^{(1-\alpha) S_\alpha},
\end{equation}
where $IL(e^{(1-\alpha) S_\alpha})$ is the inverse Laplace transform of $e^{(1-\alpha) S_\alpha}$ and $C$ is a contour parallel to the imaginary axis with sufficiently large positive real part.\footnote{In this case, it suffices to take real part greater than or equal to one, and probably greater than or equal to zero. (Assuming the modular Hamiltonian has no maximum temperature state.)} If $S_\alpha$ has the form given in \eqref{eq:holographic_renyi}, then $D(E)$ can be evaluated by a saddle point approximation for sufficiently small $G_N$. To leading order, we must therefore find
\begin{equation}
	D(E)
		= e^{f(G_N E)/G_N + o(1/G_N)}
\end{equation}
for some function $f(G_N E)$ that can be found by evaluating the exponent in \eqref{eq:laplace} at the saddle point. If we substitute this expression for the density of states back into the expression given in equation \eqref{eq:renyi_partition_function}, and substitute $E' = E G_N$, we find that the trace of $\rho$ can be written as
\begin{equation} \label{eq:one-norm}
	Z(1)
		=  \int\, \text{d}E\, D(E) e^{-E} = \int\, \text{d}E'\,  e^{(f(E')-E')/G_N) + o(1/G_N)}.
\end{equation}
Note that the coefficient of $G_N$ that would appear from substituting $E'$ for $E$ in the measure of the integral has been absorbed into the subleading corrections of order $e^{o(1/G_N)}$.

By the same argument given above for the density of states, we find that the integral for $Z(1)$ will be dominated at small $G_N$ by the leading saddle point $E'_0$. We can therefore approximate the integral to within any arbitrarily small precision $\varepsilon$ by integrating over a restricted range of eigenvalues of the modular Hamiltonian, constraining $E$ to lie in the range
\begin{equation}
	E_0'/ G_N - O(\sqrt{\log{(1/\varepsilon)} /G_N }) < E < E_0'/ G_N + O(\sqrt{\log{(1/\varepsilon)} / G_N}). \label{eq:mod_eig_range}
\end{equation}
Since $E'_0$ is the solution to the saddle point equation $f'(E') = 1$, it is independent of $G_N$.

The error in approximating this integral controls the error induced by shaving off the largest and smallest eigenvalues of $\rho$. More precisely, if we define a ``smoothed state'' $\sigma = P \rho P / \tr(P \rho P)$, where $P$ is the projector onto the eigenspaces of $K = - \log (\rho)$ with eigenvalues in the range given in \eqref{eq:mod_eig_range}, then it is clear that (i) $\sigma$ lies within an $O(\varepsilon)$-ball of $\rho$, and (ii) it has maximal and minimal eigenvalues given by
\begin{eqnarray}
	\lambda_{\text{max}}(\sigma)
		& = & e^{-E_0'/G_N +  O(1/\sqrt{G_N})}, \label{eq:lmax_smooth} \\
	\lambda_{\text{min}}(\sigma)
		& = & e^{-E_0'/G_N -  O(1/\sqrt{G_N})}. \label{eq:lmin_smooth}
\end{eqnarray}
It follows immediately from the definitions given in equations \eqref{eq:smooth_max_def} and \eqref{eq:smooth_min_def} that the smooth min- and max-entropies of $\rho$ agree with one another to leading order in $G_N$, since we have
\begin{equation}
	\rank(\sigma) \lambda_{\text{min}}(\sigma)
		\leq \tr(\sigma)
		\leq \rank(\sigma) \lambda_{\text{max}}(\sigma).
\end{equation}
More precisely, since $\sigma$ is normalized with $\tr{(\sigma)}=1,$ equations \eqref{eq:lmax_smooth} and \eqref{eq:lmin_smooth} imply
\begin{equation}
	S_{\text{max}}(\sigma) = \log(\rank(\sigma)) \leq \log\left(\frac{1}{\lambda_{\text{min}}(\sigma)}\right) =  \frac{E'_0}{G_N} + O\left(\frac{1}{\sqrt{G_N}}\right)
	\label{eq:sigma_max}
\end{equation}
and
\begin{equation}
	S_{\text{min}}(\sigma) = \log(\lambda^{-1}_{\text{max}}(\sigma)) = \frac{E'_0}{G_N} - O\left(\frac{1}{\sqrt{G_N}}\right). \label{eq:sigma_min}
\end{equation}
Since $S_{\text{max}}(\sigma)$ must be greater than $S_{\text{min}}(\sigma)$, equations \eqref{eq:sigma_max} and \eqref{eq:sigma_min} together imply that the min and max entropies of $\sigma$ agree with one another to leading order in $G_N.$ Since $\sigma$ is within an $O(\varepsilon)$-ball of $\rho$, the same statement holds true for the \emph{smooth} min and max entropies of $\rho$, as we previously claimed.

The only remaining question is to find the saddle point value $E_0'$. The von Neumann entropy of $\rho$ is given by
\begin{equation}
	S(\rho)
		= \int\, \text{d}E\, D(E)\, E\,e^{-E} = \frac{E_0'}{G_N} + O(1),
\end{equation}
where we have used the form of $D(E)$ given in \eqref{eq:density_of_states}. It follows from the Ryu-Takayanagi formula that the saddle point value $E_0'$ is given by $A/4$, where $A$ is the area of the corresponding RT surface. The smooth min- and max-entropies are therefore equal to the von Neumann entropy up to $O(1/\sqrt{G_N})$ corrections for any fixed nonzero $\varepsilon$. Since the von Neumann entropy $S$ is of order $O(1/G_N)$, the corrections grow as the square root of the entropy (for fixed values of the UV cutoff). It follows that for holographic states, the smooth min and max entropies satisfy
\begin{eqnarray}
	S_{\text{min}}
		& = & S - O(\sqrt{S}), \label{eq:smoothmin} \\
	S_{\text{max}}
		& = & S + O(\sqrt{S}). \label{eq:smoothmax}
\end{eqnarray}
This is exactly the same scaling that is seen when we take the asymptotic limit of a large number of copies of a state.\footnote{Until now, we have assumed that $\varepsilon$ is some fixed small number that is independent of $G_N$. However, the range of integration required to approximate \eqref{eq:one-norm} depends only weakly on the allowed error $\varepsilon$ as $\sqrt{\log(1/\varepsilon)}$. Hence we can make the error $\varepsilon$ non-perturbatively small with respect to $G_N$, at the small cost of allowing the smooth min- and max-entropies be separated by $O(\sqrt{S f(S)})$ for some super-logarithmic function $f(S)$. (We can then obtain $\varepsilon = e^{-f(S)}$, which is non-perturbatively small.) A natural choice might be $f(S) = S^\delta$ for some small $\delta$, or maybe $f(S) = (\log(S))^2$. Regardless of the specific function chosen, it is easy to ensure that $O(\sqrt{S f(S)})$ is subleading compared to the entropy $S$.}
 The semiclassical holographic limit of large central charge is therefore replicating, at least partially, the effects of the asymptotic i.i.d.~limit of a large number of identical copies of a single, non-holographic state. When we construct holographic tensor networks in the following section, we will see that these subleading corrections to the smooth min- and max-entropies can be related to known subleading contributions to holographic entanglement in AdS/CFT.
\section{Tree Tensor Networks from Holographic Entanglement Distillation} \label{sec:tree_networks}

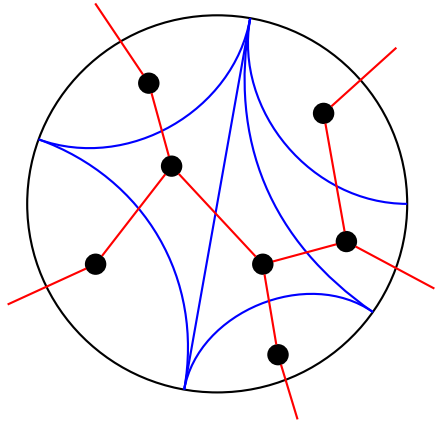
\begin{figure}[h]
\begin{center}
	\begin{tikzpicture}[thick, vertex/.style={draw, shape=circle, fill=black, scale=0.7}]
		\draw (0, 0) circle (2.5cm);

		\draw [blue] (2.5, 0) to[out=180, in=260] (0.434, 2.462);
		\draw [blue] (0.434, 2.462) to[out=260, in=340] (-2.349, 0.855);
		\draw [blue] (-2.349, 0.855) to[out=340, in=80] (-0.434, -2.462);
		\draw [blue] (-0.434, -2.462) to[out=80, in=145] (2.048, -1.434);
		\draw [blue] (0.434, 2.462) to (-0.434, -2.462);
		\draw [blue] (0.434, 2.462) to[out=260, in=145] (2.048, -1.434);

		\node [vertex] at (-0.6, 0.5) (1) {};
		\node [vertex] at (-0.9, 1.6) (2) {};
		\node [vertex] at (1.4, 1.2) (3) {};
		\node [vertex] at (1.7, -0.5) (4) {};
		\node [vertex] at (0.6, -0.8) (5) {};
		\node [vertex] at (-1.6, -0.8) (6) {};
		\node [vertex] at (0.8, -2) (7) {};

		\draw [red] (1) to (2);
		\draw [red] (1) to (6);
		\draw [red] (1) to (5);
		\draw [red] (3) to (4);
		\draw [red] (4) to (5);
		\draw[red] (5) to (7);

		\node at (-1.7, 2.8) (A) {};
		\node at (2.5, 2.2) (B) {};
		\node at (3, -1.2) (C) {};
		\node at (1.1, -3) (D) {};
		\node at (-2.9, -1.4) (E) {};

		\draw [red] (2) to (A);
		\draw [red] (3) to (B);
		\draw [red] (4) to (C);
		\draw [red] (7) to (D);
		\draw [red] (6) to (E);
	\end{tikzpicture}
	\caption{A bulk discretization of vacuum $AdS_3$ by non-intersecting RT surfaces. Blue curves represent extremal surfaces in the bulk, red lines are edges in the dual graph, and
			black dots are vertices in the dual graph. In the resulting tensor network, the dangling edges passing through the boundary of the spacetime will correspond to uncontracted
			(physical) Hilbert space indices.}
	\label{fig:tree_discretization}
\end{center}
\end{figure}

In order to construct a meaningful tensor network for a state in the AdS/CFT correspondence, it is necessary to produce a network that (i) reproduces the correct boundary state with high fidelity, and (ii) has a bulk geometry that matches the bulk spacetime. Since tensor networks have discrete geometries, property (ii) must be interpreted in terms of some discretization of the bulk spacetime. In this section, we consider \emph{tree networks} --- those constructed by discretizing the bulk with non-intersecting Ryu-Takayanagi surfaces. Given such a discretization, the underlying graph of the corresponding tensor network is taken to be the dual graph of the set of Ryu-Takayanagi surfaces and their corresponding boundary regions. A sample discretization of vacuum $AdS_3$, along with the corresponding dual graph, is shown in Figure \ref{fig:tree_discretization}.

Once a network is constructed on this graph, it is straightforward to quantify how well the resulting state satisfies property (i) by looking at the inner product between the state constructed by the tensor network and the target state. We will say that the network satisfies property (ii) if the bond dimension of each edge in the network matches the area of the Ryu-Takayanagi surface through which it passes. More precisely, we will require that each bond $\gamma$ in the network satisfies $\dim(\mathcal{H}_{\gamma}) = e^{A_{\gamma}/ 4G_N + o(1/G_N)}$, where $A_{\gamma}$ is the area of the corresponding RT surface.\footnote{Both sides of this equality are infinite. As usual, equations involving the entanglement entropy or the area of extremal surfaces should be interpreted in the context of a regularization scheme in which the CFT state is regulated on a lattice with finite spacing and the bulk spacetime is regulated with a radial cutoff.} Since the dimension of any bond Hilbert space could be made arbitrarily large without altering the state by adding zero probability states, we also require that each contraction is \emph{full rank in the bond space} in the sense that it contains no trivial contractions.\footnote{Formally, for a bond of the form $P^{A \gamma} Q^{B}{}_{\gamma}$, we require that there is no nonzero vector $v^{\gamma}$ in $\mathcal{H}_{\gamma}$ or dual vector $\omega_{\gamma}$ in $\mathcal{H}_{\gamma}^*$ such that $P^{A \gamma} \omega_{\gamma}$ or $Q^{B}{}_{\gamma} v^{\gamma}$ identically vanishes.} In fact, this condition is not quite strong enough, as the dimension of $\mathcal{H}_{\gamma}$ could still be made arbitrarily large by the addition of states with arbitrarily small probability --- in our construction, this possibility is avoided by ensuring that the eigenvalues of each bond are bounded below by a nontrivial function of the bond entropy (cf. the smoothing process of Section \ref{sec:smoothminmax}, where each ``smoothed state'' has bounded minimal eigenvalue).

\subsection{Bipartite Tensor Networks} \label{sec:bipartite_networks}

To explain the procedure of constructing tree networks via one-shot entanglement distillation, we restrict temporarily to the case where the bulk is discretized by a single Ryu-Takayanagi surface. For such a discretization, the boundary is partitioned into two connected regions $A$ and $A^c.$ The generic procedure for arbitrary non-intersecting bulk partitions is detailed in Section \ref{sec:larger_tree_networks}.

As discussed in Section \ref{sec:smoothminmax}, the smooth min- and max-entropies of a holographic state agree with the von Neumann entropy to leading order in $G_N$. It follows that for the reduced CFT density matrix $\psi_{(A)}$ of a holographic CFT state $\ket{\psi}$, there exists a normalized state $\psi_{(A)}^{\varepsilon}$ within $\varepsilon$ trace distance of $\psi_{(A)}$ satisfying
\begin{eqnarray}
	\rank\left(\psi_{(A)}^{\varepsilon}\right)
		& = & e^{S(A) + O(\sqrt{S})} \label{bipartite_smooth_max}, \\ 
	\lambda_{\text{max}}\left(\psi_{(A)}^{\varepsilon}\right)
		& = & e^{- S(A) + O(\sqrt{S})} \label{bipartite_smooth_min},
\end{eqnarray}
where $\lambda_{\text{max}}\left(\psi_{(A)}^{\varepsilon}\right)$ is the largest eigenvalue of $\psi_{(A)}^{\varepsilon}$.

Given the full, pure CFT state $\ket{\psi} \in \mathcal{H}_{A} \otimes \mathcal{H}_{A^c}$,\footnote{In reality, the Hilbert space of the actual CFT will not factorize in this way, due to ultraviolet issues. However, since we have already regularized the theory, there is no problem splitting the Hilbert space into tensor factors.} one can write the Schmidt decomposition
\begin{equation}
	\ket{\psi}
		= \sum_{n} \sqrt{\lambda_n} \ket{n}_{A} \ket{n}_{A^c}, \label{unsmoothed_TN_state}
\end{equation}
where $\{\lambda_n\}$ are the eigenvalues of the reduced states $\psi_{(A)}$ and $\psi_{(A^c)}.$ If $\{\tilde{\lambda}_n\}$ are the eigenvalues of the smoothed state $\psi^{\varepsilon}_{(A)}$, then it is easy to verify that the state
\begin{equation}
	\ket{\psi^{\varepsilon}}
		= \sum_{n} \sqrt{\tilde{\lambda}_n} \ket{n}_{A} \ket{n}_{A^c} \label{smoothed_TN_state}
\end{equation}
approximates the original state $\ket{\psi}$ with very high fidelity. In particular, we have
\begin{equation}
\left|\braket{\psi^\varepsilon}{\psi} \right|^2 = F(\psi_{(A)}, \psi_{(A)}^{\varepsilon}) \geq \left(1 - \frac{1}{2} \onenorm{\psi_{(A)} - \psi_{(A)}^\varepsilon} \right)^2 \geq 1 - \varepsilon, \label{smoothing_error}
\end{equation}
where $F(\rho, \sigma) = \left[ \tr\sqrt{\sqrt{\rho} \sigma \sqrt{\rho}}\right]^2$ is the \emph{fidelity} of two quantum states. The first equality in (\ref{smoothing_error}) follows from the definition of fidelity and the form of the states (\ref{unsmoothed_TN_state}) and (\ref{smoothed_TN_state}), while the subsequent inequality is one of the Fuchs-van de Graaf inequalities \cite{FvdG1997}.

If we re-order the probability spectrum $\tilde{\lambda}_n$ such that it is monotonically decreasing, i.e. $\tilde{\lambda}_{n+1} \leq \tilde{\lambda}_{n}$, and break the resulting sum into blocks of size $\Delta$, then we may rewrite (\ref{smoothed_TN_state}) as
\begin{equation}
	\ket{\psi^{\varepsilon}}
		= \sum_{n=0}^{\rank[\psi^{\varepsilon}_{(A)}]/\Delta - 1}
			\sum_{m=0}^{\Delta - 1} \sqrt{\tilde{\lambda}_{n \Delta + m}} \ket{n \Delta + m}_{A} \ket{n \Delta + m}_{A^c}
\end{equation}
Now, suppose we discard the $m$-dependence of the eigenvalues $\tilde{\lambda}_{n \Delta + m}$ and replace all of the eigenvalues in each block with the average value of that block, $\tilde{\lambda}_{n \Delta}^{\text{avg}}$. The resulting state (which is still correctly normalized) is
\begin{equation}
	 \ket{\Psi^{\varepsilon}}
		= \sum_{n=0}^{\rank[\psi^{\varepsilon}_{(A)}]/\Delta - 1}
			\sum_{m=0}^{\Delta - 1} \sqrt{\tilde{\lambda}^{\text{avg}}_{n \Delta}} \ket{n \Delta + m}_{A} \ket{n \Delta + m}_{A^c},
\end{equation}
and satisfies
\begin{equation}
	\onenorm{\Psi^{\varepsilon} - \psi^{\varepsilon}_{(A)}}
		\leq \lambda_{\text{max}}[\psi^{\varepsilon}_{(A)}] \cdot \Delta \equiv \delta. \label{eigenvalue_averaging_error}
\end{equation}
By the same arguments as in \eqref{smoothing_error}, the overlap between $\ket{\Psi^{\varepsilon}}$ and the original CFT state $\ket{\psi}$ is bounded below by
\begin{align} \label{eq:totalerror}
\left|\braket{\psi}{ \Psi^\varepsilon}\right|^2 \geq \left|\braket{\psi}{\psi^\varepsilon} \right|^2\left|\braket{\psi^{\varepsilon}}{ \Psi^\varepsilon}\right|^2 \geq 1 - \varepsilon - \delta.
\end{align}
If we choose $\Delta = e^{S-O(\sqrt{S})},$ with the $O(\sqrt{S})$ dependence chosen to approximately cancel the $O(\sqrt{S})$ dependence of $\lambda_{\text{max}}[\psi^{\varepsilon}_{(A)}]$ given in equation (\ref{bipartite_smooth_min}), then we can ensure that $\delta$ remains small, while the state becomes
\begin{equation}
	\ket{\Psi^{\varepsilon}}
		= \sum_{n=0}^{e^{O(\sqrt{S})}}
			\sum_{m=0}^{e^{S - O(\sqrt{S})}}
				\sqrt{\tilde{\lambda}^{\text{avg}}_{n\Delta}} \ket{n \Delta + m}_{A} \ket{n \Delta + m}_{A^c}, \label{bipartite_TN_state}.
\end{equation}

To properly distill the EPR pairs out of this state, we define auxiliary Hilbert spaces $\mathcal{H}_{\gamma}$ and $\mathcal{H}_{f}$ with dimensions given by
\begin{eqnarray}
	\dim\mathcal{H}_{f} &=& e^{O(\sqrt{S})}, \\
	\dim\mathcal{H}_{\gamma} &=& e^{S - O(\sqrt{S})},
\end{eqnarray}
where the precise values are chosen to match the range of the sums in equation (\ref{bipartite_TN_state}). We define the isometries $\mathcal{H}_{f} \otimes \mathcal{H}_{\gamma} \hookrightarrow \mathcal{H}_{A}$ and $\bar{\mathcal{H}}_f \otimes \bar{\mathcal{H}}_{\gamma} \hookrightarrow \mathcal{H}_{A^c}$ by
\begin{eqnarray}
	V\ket{n}_{f} \ket{m}_{\gamma}
		& = \ket{n \Delta + m}_{A}, \\
	W\ket{\bar{n}}_{\bar{f}} \ket{\bar{m}}_{\bar{\gamma}}
		& = \ket{n \Delta + m}_{A^c}
\end{eqnarray}
for some arbitrarily chosen bases of the auxiliary Hilbert spaces and the corresponding bases in their complex conjugate Hilbert spaces. We may then rewrite the state $\ket{\Psi^{\varepsilon}}$ as
\begin{equation}
	\ket{\Psi^{\varepsilon}}
		= (V \otimes W)
			\left[ \sum_{n=0}^{e^{O(\sqrt{S})}} \sqrt{\tilde{\lambda}^{\text{avg}}_{n \Delta}} \ket{n \bar{n}}_{f \bar{f}} \right]
			\otimes
			\left[ \sum_{m=0}^{e^{S - O(\sqrt{S})}} \ket{m \bar{m}}_{\gamma \bar{\gamma}} \right]. \label{bipartite_distilled_state}
\end{equation}
This expression is essentially identical to (\ref{asymptotic_distillation}), except that it approximates a \emph{single} copy of the original CFT state $\ket{\psi}.$

Approximate states such as those given in equation \eqref{bipartite_distilled_state} are considerably more common in the quantum information literature than in discussions of quantum gravity, so we pause briefly to discuss their physical relevance. The first and most important point to note is that expectation value of any bounded operator $\hat O$ on the Hilbert space of $\ket{\psi}$ is well approximated by its expectation value in the approximate state. In particular, since the overlap between the original CFT state $\ket{\psi}$ and the new state $\ket{\Psi^{\varepsilon}}$ takes the form given in \eqref{eq:totalerror}, we can guarantee that the expectation values of bounded operators between the two states differ at most by
\begin{align}
\left| \braket{\psi}{\hat O | \psi} - \braket{\Psi^\varepsilon}{\hat O |\Psi^ \varepsilon} \right| \leq 2 \sqrt{\varepsilon + \delta} \,\lVert \hat O \rVert.
\end{align}
If both $\varepsilon$ and $\delta$ can be made sufficiently small, then this bound is quite narrow. Since the time evolution operator $e^{iHt}$ is bounded above by $\lVert e^{iHt}\rVert \leq 1$ for any $t$ that is either real or in the upper half-plane, a similar bound holds for correlation functions at arbitrary times. In particular, correlation functions of bounded operators satisfy 
\begin{align}
\left| \braket{\psi}{\hat O_1(t_1) \dots \hat O_n(t_n) | \psi} - \braket{\Psi^\varepsilon}{\hat O_1(t_1) \dots \hat O_n(t_n) |\Psi^ \varepsilon} \right| \leq 2 \sqrt{\varepsilon + \delta} \,\lVert \hat O_1 \rVert \dots \lVert \hat O_n \rVert.
\end{align}
The states $\ket{\psi}$ and $\ket{\Psi^{\varepsilon}}$ therefore generally produce approximately the same values for any Euclidean or Lorentzian correlation function with arbitrarily large time gaps, including out-of-time order Lorentzian correlation functions.

There are a few scenarios in which $\ket{\psi}$ and $\ket{\Psi^{\varepsilon}}$ can display qualitatively different behavior; we argue that none of these scenarios are actually physically important. If $\braket{\psi}{\hat O | \psi}$ is itself very small compared to $\lVert\hat O \rVert$, then there may $O(1)$ differences in the \emph{relative} size of $\braket{\psi}{\hat O | \psi}$ and $\braket{\Psi^\varepsilon}{\hat O | \Psi^\varepsilon}$. However, in this case, since $\varepsilon$ can be arbitrarily small, both the expectation values would have to be zero at leading order; they would still agree up to $O(\varepsilon)$ corrections.

Secondly, non-observable quantities such as R\'{e}nyi entropies for $\alpha \neq 1$ may (and will) look very different for $\ket{\psi}$ and $\ket{\Psi^\varepsilon}$.\footnote{Here, we mean that the R\'{e}nyi entropies are non-observable when provided with only a single copy of the state --- they can be computed from matrix moments of a state, which are observable given multiple replicas of the state in question.} This is unsurprising, as the R\'{e}nyi entropies are very sensitive to small and (physically) insignificant perturbations, and can vary drastically without significantly altering the expectation values of bounded operators. 

Finally, we are often interested in unbounded operators. In this case we do not have any bound on the error in expectation values for the approximate state. However, in such circumstances we can generally replace an unbounded self-adjoint operator $\hat O$ by some bounded function $f(\hat O)$ of the operator without affecting the important physics. We will then obtain a tight bound on the error of the expectation value $\langle f(\hat O) \rangle$ when approximating it in the distilled state $\ket{\Psi^{\varepsilon}}$.

\begin{figure}[h]
\begin{center}
	\begin{tikzpicture}[thick, >=latex, auto, vertex/.style={draw, shape=circle, fill=black, scale=0.7}, tensor/.style={circle, draw, minimum width=2.5em},
				EPR/.style={regular polygon, regular polygon sides=4, fill=white, draw, inner sep=2pt},
				cobwebs/.style={regular polygon, fill=white, regular polygon sides=3, draw, inner sep=2pt}]
		\draw (0, 0) circle (2.5cm);

		\draw [blue] (0, 2.5) to (0, -2.5);

		\node[tensor] (V) at (-1.5, 0) {$V$};
		\node[tensor] (W) at (1.5, 0) {$W$};
		\node[EPR] (phi) at (0, 1.5) {$\phi$};
		\node[cobwebs] (sigma) at (0, -1.5) {$\sigma$};

		\node[label] (A) at (-4, 0) {$A$};
		\node[label] (Ac) at (4, 0) {$A^c$};

		\draw (phi) edge[->] node {$\gamma$} (V);
		\draw (phi) edge[->] node {$\bar{\gamma}$} (W);
		\draw (sigma) edge[->] node {$f$} (V);
		\draw (sigma) edge[->] node {$\bar{f}$} (W);
		\draw (V) edge[->] node {} (A);
		\draw (W) edge[->] node {} (Ac);
	\end{tikzpicture}
	\caption{A tensor network for a bipartite discretization of $AdS_3$ by a single Ryu-Takayanagi surface. $\phi^{\gamma \bar{\gamma}}$ is a maximally entangled state on a Hilbert
			space of dimension $e^{O(S(A))}$, while $\sigma^{f \bar{f}}$ is a (generally not maximally entangled) state on a Hilbert space of dimension $e^{O\left(\sqrt{S(A)}\right)}$.
			The tensors $V$ and $W$ embed these states isometrically into the boundary.}
	\label{fig:bipartite_network}
\end{center}
\end{figure}
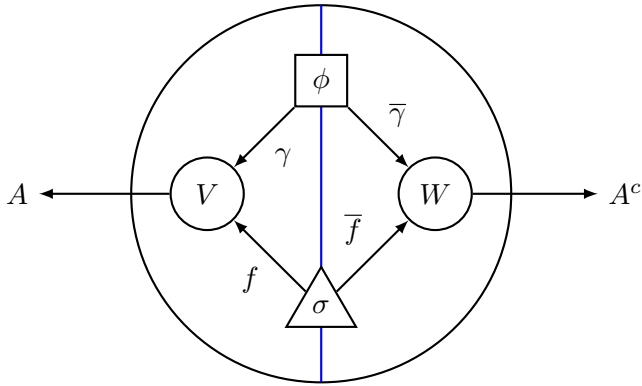

So long as we originally chose $\varepsilon$ small and chose $\Delta$ correctly to ensure that $\delta$ is small, then \eqref{eq:totalerror} ensures that the distilled state $\ket{\Psi^{\varepsilon}}$ is a good approximation of the original CFT state $\ket{\psi}$. As discussed above, this implies that the physics of the two states should be the same up to non-perturbatively small corrections. Moreover, the expression given in (\ref{bipartite_distilled_state}) is a tensor network with a geometry matching the semiclassical dual of $\ket{\psi}$. In the abstract index notation of section \ref{sec:TN_review}, the state in (\ref{bipartite_distilled_state}) is written as
\begin{equation}
	\Psi^{A A^c}
		= V^{A}{}_{f \gamma} W^{A^c}{}_{\bar{f} \bar{\gamma}} \phi^{\gamma \bar{\gamma}} \sigma^{f \bar{f}}, \label{bipartite_network_rep}
\end{equation}
where
\begin{eqnarray}
	\ket{\phi}
		&=&  \sum\limits_{m=0}^{e^{S - O(\sqrt{S})}} \ket{m \bar{m}}_{\gamma \bar{\gamma}}, \\
	\ket{\sigma}
		&=& \sum\limits_{n=0}^{e^{O(\sqrt{S})}} \sqrt{\tilde{\lambda}^{\text{avg}}_{n \Delta}} \ket{n \bar{n}}_{f \bar{f}}.
\end{eqnarray}
The network corresponding to (\ref{bipartite_network_rep}) is sketched in Figure \ref{fig:bipartite_network}, superposed over the corresponding discretization of vacuum $AdS_3$. The tensors $V$ and $W$ correspond to the entanglement wedges of regions $A$ and $A^c$, respectively, as isometries that embed the states $\phi$ and $\sigma$ into the boundary. Since $\phi$ is a maximally entangled state on a Hilbert space of dimension $e^{O(S(A))},$ it has the right entanglement to reproduce the Ryu-Takayanagi surface that separates $A$ from $A^c.$ Because the bond dimension of the legs of the state $\ket{\sigma}$ is very small compared to the large bond dimensions of the $\phi$-legs, we can think of the $\sigma$-legs as thin ``cobwebs'', attached to the thick ``girders'' of the $\phi$-legs.

It is important to be clear about our motivation in including a second approximation step, where we average $\tilde \lambda_i$ within each block and hence extract the dependence of $\tilde \lambda_i$ on $i$ into the state $\ket{\sigma}$ in the (relatively) small auxiliary Hilbert space $\mathcal{H}_f \otimes \bar{\mathcal{H}}_f$. After all, we could already have constructed a tensor network with the correct discretized geometry by using the smoothed, but unflattened state
\begin{align}
\sum_i \sqrt{\tilde \lambda_i} \ket{i}_\gamma \ket{i}_{\bar \gamma}.
\end{align}
Our purpose in flattening the $\tilde \lambda_i$ spectrum is not to claim that the state produced is in any sense ``better'' or ``more holographic'' than the smooth state that was constructed prior to this flattening procedure. In fact, the degrees of freedom extracted into $\ket{\sigma}$ are highly non-unique, as they depend a great deal on the exact block size $\Delta$ chosen while flattening the spectrum. However, by showing that the original state $\ket{\psi}$ can be approximated in this way, we are showing explicitly that the entanglement spectrum is so flat (up to smoothing) that the \emph{effective} number of degrees of freedom in $\ket{\psi}$ that describe the gradient of the entanglement spectrum is subleading compared to the effective number of degrees of freedom that are simply maximally entangled between the two sides.

Note that to obtain a tensor network description with the correct leading order bond dimensions, we only needed the smooth max-entropy to be sufficiently small. The additional requirement that the smooth min-entropy agree with smooth max-entropy to leading order was imposed to ensure that the auxiliary Hilbert space $\mathcal{H}_f$ on which the state is not maximally entangled has subleading dimension $e^{O(\sqrt{S})}$.

Before moving on to more general tree tensor networks, we pause to consider the role of the ``cobweb'' state $\ket{\sigma}$. This state does not have an immediate interpretation in the usual Ryu-Takayanagi picture of bulk entanglement, at least at leading order in $G_N.$ Since the state $\ket{\sigma}$ is of subleading size in the entanglement entropy and hence in the central charge, the most natural interpretation is that it arises from quantum fluctuations in the bulk geometry (e.g., graviton fluctuations) that alter the areas of Ryu-Takayanagi surfaces at subleading order in $G_N$.\footnote{An alternative approach \cite{livingedge, RTQEC} describes the fluctuations of the areas of extremal surfaces as ``edge modes,'' i.e. superselection sectors that commute with the algebra of observables on both sides of the surface. In our approach, however, these fluctuations are described explicitly by the states of the $\ket{\sigma}$ tensor that lies on the Ryu-Takayanagi surface.} Generally, such fluctuations are expected to be suppressed by a factor of $O(\sqrt{G_N})$, meaning that the resulting fluctuations in the entropy are also of order $O(1/\sqrt{G_N})$.\footnote{To see that this is the correct scaling of metric fluctuations, note that a one graviton state with order-unity frequency has an $O(1/G_N)$ energy, but the energy is the square of the amplitude of the metric strain.  Note however that in contexts where we are only interested in the \emph{average} area, the $O(1/\sqrt{G_N})$ term does not appear because, for linearized gravitons, positive fluctuations are just as likely as negative fluctuations.  That is why the quantum corrections in the holographic von Neumann entropy \eqref{RT}, which traces over the whole probability distribution, are merely $O(1)$.  In our tensor network contexts, however, we need to keep track of the Hilbert space dimensions, which do not average out.} Since this matches the rank of $\ket{\sigma},$ it seems natural to associate the subleading state with these geometric fluctuations. We discuss this proposal in more detail in Section \ref{sec:cobwebs}.

It is worth commenting that we are also at liberty to absorb the state $\ket{\sigma}$ into one of (or a combination of) the isometries $V$ and $W$. This simplifies the picture of the tensor network, but comes at the cost of at least one of the operators $V$ and $W$ no longer being an isometry. In fact, they will not even be approximate isometries, although they will remain isometries if interpreted as operators $V:\mathcal{H}_\gamma \hookrightarrow \bar{\mathcal{H}}_f \otimes \mathcal{H}_A$ and $W: \bar{\mathcal{H}}_\gamma \hookrightarrow \mathcal{H}_f \otimes \mathcal{H}_{A^c}$. The question of whether tensors in the network are (at least approximate) isometries is important for various reasons, both in ensuring that the boundary state of the network correctly approximates the original CFT state and in understanding the error correcting properties of the network. As such we shall always keep the state $\ket{\sigma}$, and its generalizations in more complicated networks, explicit.

\subsection{General Tree Networks} \label{sec:larger_tree_networks}

The argument given above can be extended to construct a tree tensor network for an arbitrary discretization of the bulk by (non-intersecting) Ryu-Takayanagi surfaces. This generalization works roughly as one would expect: one simply localizes degrees of freedom to each RT surface in turn, each time creating an additional link and tensor in the network. However, some important subtleties arise during this process. It is easy to construct a superficially-reasonable procedure that will not actually approximate the original state with high fidelity. We will therefore describe the procedure for constructing generic tree networks in some detail.\footnote{For similar work in other contexts, see \cite{SDV2006} and \cite{DF2013}.}

Our argument is inductive: we assume that we have successfully constructed a tree tensor network for a simpler discretization with one fewer RT surface, and then show that we can always add an additional RT surface while (approximately) preserving the bulk bond dimensions and the boundary CFT state. After arbitrarily many inductive steps, the final network will still approximate the original ``target'' CFT state on its boundary. However, to obtain rigorous bounds on our final error in approximating the original CFT state, we must be somewhat careful in the order in which we choose to add RT surfaces. Specifically, there must exist some choice of boundary node, which we shall label the ``root'' node, such that each ``parent'' tensor was added after all of its ``children.'' (In practice, it seems likely that one will obtain a correct approximation of the original state even when the RT surfaces are added in an arbitrary order. Without adding them according to a particularly nice orientation, however, it is hard to guarantee that one couldn't obtain a large boundary error by sheer bad luck.)

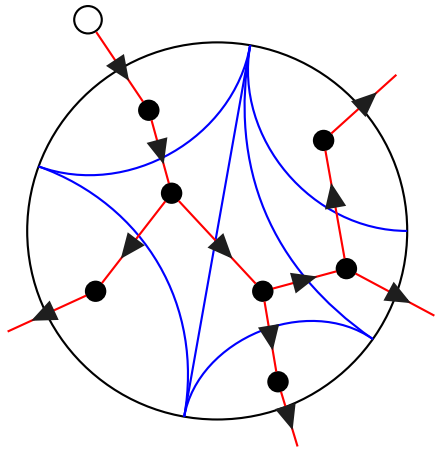
\begin{figure}[h]
\begin{center}
	\begin{tikzpicture}[thick, vertex/.style={draw, shape=circle, fill=black, scale=0.7},
				decoration={markings, mark=at position 0.7 with {\arrow[color={rgb,255:red,30;green,30;blue,30}, very thick]{triangle 45}}}]
		\draw (0, 0) circle (2.5cm);

		\draw [blue] (2.5, 0) to[out=180, in=260] (0.434, 2.462);
		\draw [blue] (0.434, 2.462) to[out=260, in=340] (-2.349, 0.855);
		\draw [blue] (-2.349, 0.855) to[out=340, in=80] (-0.434, -2.462);
		\draw [blue] (-0.434, -2.462) to[out=80, in=145] (2.048, -1.434);
		\draw [blue] (0.434, 2.462) to (-0.434, -2.462);
		\draw [blue] (0.434, 2.462) to[out=260, in=145] (2.048, -1.434);

		\node [vertex] at (-0.6, 0.5) (1) {};
		\node [vertex] at (-0.9, 1.6) (2) {};
		\node [vertex] at (1.4, 1.2) (3) {};
		\node [vertex] at (1.7, -0.5) (4) {};
		\node [vertex] at (0.6, -0.8) (5) {};
		\node [vertex] at (-1.6, -0.8) (6) {};
		\node [vertex] at (0.8, -2) (7) {};

		\draw [red, postaction={decorate}] (2) to (1);
		\draw [red, postaction={decorate}] (1) to (6);
		\draw [red, postaction={decorate}] (1) to (5);
		\draw [red, postaction={decorate}] (4) to (3);
		\draw [red, postaction={decorate}] (5) to (4);
		\draw [red, postaction={decorate}] (5) to (7);

		\node [shape=circle, draw] at (-1.7, 2.8) (A) {};
		\node at (2.5, 2.2) (B) {};
		\node at (3, -1.2) (C) {};
		\node at (1.1, -3) (D) {};
		\node at (-2.9, -1.4) (E) {};

		\draw [red, postaction={decorate}] (A) to (2);
		\draw [red, postaction={decorate}] (3) to (B);
		\draw [red, postaction={decorate}] (4) to (C);
		\draw [red, postaction={decorate}] (7) to (D);
		\draw [red, postaction={decorate}] (6) to (E);
	\end{tikzpicture}
	\caption{The bulk discretization of vacuum $AdS_3$ that was originally sketched in Figure \ref{fig:tree_discretization} has here been given a root-leaf orientation on its dual graph
			by choosing an arbitrary boundary edge as the ``root'' (represented here by a white circle).}
	\label{fig:tree_flow}
\end{center}
\end{figure}

We begin by choosing a ``target'' discretization of the bulk by non-intersecting RT surfaces, such as the one sketched above in Figure \ref{fig:tree_discretization}. This will be the graph of our final tree tensor network. Designating one of its nodes as a root  picks out a preferred orientation for the tree by flowing away from the root, as sketched in Figure \ref{fig:tree_flow}. In general, we will choose the root node to lie on the boundary, although our construction works even if the root is chosen to lie in the bulk.  All boundary nodes that are not the root are now leaves of this oriented graph. Note that this orientation is defined only for the purpose of ordering the RT surfaces that make up the discretization, and is independent of the orientation imposed on the tensor network to denote up- and down-indices (cf. Section \ref{sec:TN_review}).

To construct a tree tensor network for this graph, edges will be added to the network inductively from leaves up to the root. To preserve the isometry properties of the tensor network, no RT surface can be added to the network before all of its children have been added (according to the orientation induced by choosing a boundary root). Different choices of root node on the boundary, and even different orderings of RT surfaces that are consistent with a single root-leaf orientation, will in general produce different tree tensor networks. However, all such networks are geometrically appropriate for the AdS/CFT correspondence in the sense that they have bond dimensions that match the holographic geometry and boundary states that approximately reproduce the original CFT state.

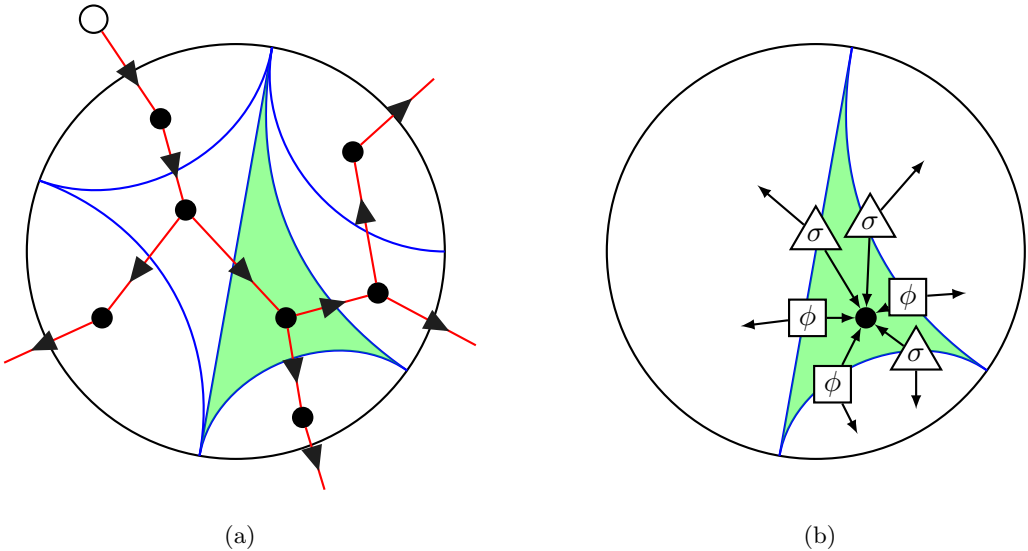
\begin{figure}[h]
\begin{center}
	\subfloat[\label{fig:tree_state_shading}]{
		\begin{tikzpicture}[scale=1.1, thick, vertex/.style={draw, shape=circle, fill=black, scale=0.7},
				decoration={markings, mark=at position 0.7 with {\arrow[color={rgb,255:red,30;green,30;blue,30}, very thick]{triangle 45}}}]
		\draw (0, 0) circle (2.5cm);

		\draw [blue] (2.5, 0) to[out=180, in=260] (0.434, 2.462);
		\draw [blue] (0.434, 2.462) to[out=260, in=340] (-2.349, 0.855);
		\draw [blue] (-2.349, 0.855) to[out=340, in=80] (-0.434, -2.462);
		\draw [blue] (-0.434, -2.462) to[out=80, in=145] (2.048, -1.434);
		\draw [blue] (0.434, 2.462) to (-0.434, -2.462);
		\draw [blue] (0.434, 2.462) to[out=260, in=145] (2.048, -1.434);

		\begin{scope}
			  \clip (0.434, 2.462) to[out=260, in=145] (2.048, -1.434) to[out=145, in=80] (-0.434, -2.462)
					to (0.434,2.462);
			\draw [draw=green, fill=green, opacity=0.4] (-2.5,-2.5) rectangle (2.5, 2.5);
		\end{scope}

		\node [vertex] at (-0.6, 0.5) (1) {};
		\node [vertex] at (-0.9, 1.6) (2) {};
		\node [vertex] at (1.4, 1.2) (3) {};
		\node [vertex] at (1.7, -0.5) (4) {};
		\node [vertex] at (0.6, -0.8) (5) {};
		\node [vertex] at (-1.6, -0.8) (6) {};
		\node [vertex] at (0.8, -2) (7) {};

		\draw [red, postaction={decorate}] (2) to (1);
		\draw [red, postaction={decorate}] (1) to (6);
		\draw [red, postaction={decorate}] (1) to (5);
		\draw [red, postaction={decorate}] (4) to (3);
		\draw [red, postaction={decorate}] (5) to (4);
		\draw [red, postaction={decorate}] (5) to (7);

		\node [shape=circle, draw] at (-1.7, 2.8) (A) {};
		\node at (2.5, 2.2) (B) {};
		\node at (3, -1.2) (C) {};
		\node at (1.1, -3) (D) {};
		\node at (-2.9, -1.4) (E) {};

		\draw [red, postaction={decorate}] (A) to (2);
		\draw [red, postaction={decorate}] (3) to (B);
		\draw [red, postaction={decorate}] (4) to (C);
		\draw [red, postaction={decorate}] (7) to (D);
		\draw [red, postaction={decorate}] (6) to (E);
		\end{tikzpicture}
	}
	\hspace{0.5cm}
	\subfloat[\label{fig:tree_state_network}]{
		\begin{tikzpicture}[scale=1.1, thick, >=latex, vertex/.style={draw, shape=circle, fill=black, scale=0.7},
				decoration={markings, mark=at position 0.7 with {\arrow[color={rgb,255:red,30;green,30;blue,30}, very thick]{triangle 45}}},
				cobwebs/.style={regular polygon, fill=white, regular polygon sides=3, draw, inner sep=0.5pt},
				EPR/.style={regular polygon, regular polygon sides=4, fill=white, draw, inner sep=0pt}]
		\draw (0, 0) circle (2.5cm);

		\draw [blue] (-0.434, -2.462) to[out=80, in=145] (2.048, -1.434);
		\draw [blue] (0.434, 2.462) to (-0.434, -2.462);
		\draw [blue] (0.434, 2.462) to[out=260, in=145] (2.048, -1.434);
		\begin{scope}
			  \clip (0.434, 2.462) to[out=260, in=145] (2.048, -1.434) to[out=145, in=80] (-0.434, -2.462)
					to (0.434,2.462);
			\draw [draw=green, fill=green, opacity=0.4] (-2.5,-2.5) rectangle (2.5, 2.5);
		\end{scope}

		\node [vertex] at (0.6, -0.8) (5) {};

		\node [EPR] at (1.1, -0.55) (phi1) {$\phi$};
		\node [cobwebs] at (0.65, 0.35) (sigma1) {$\sigma$};
		\node [EPR] at (0.2, -1.6) (phi2) {$\phi$};
		\node [cobwebs] at (1.2, -1.25) (sigma2) {$\sigma$};
		\node [EPR] at (-0.1, -0.8) (phi3) {$\phi$};
		\node [cobwebs] at (0, 0.2) (sigma3) {$\sigma$};

		\node at (-1.7, 2.8) (A) {};
		\node at (2.5, 2.2) (B) {};
		\node at (3, -1.2) (C) {};
		\node at (1.1, -3) (D) {};
		\node at (-2.9, -1.4) (E) {};

		\draw (phi1) edge[->] (5);
		\draw (sigma1) edge[->] (5);
		\draw (phi2) edge[->] (5);
		\draw (sigma2) edge[->] (5);
		\draw (phi3) edge[->] (5);
		\draw (sigma3) edge[->] (5);

		\draw (phi1) edge[->] (1.8, -0.5);
		\draw (sigma1) edge[->] (1.3, 1.1);
		\draw (phi2) edge [->] (0.5, -2.2);
		\draw (sigma2) edge[->] (1.2, -1.9);
		\draw (phi3) edge[->] (-0.9, -0.9);
		\draw (sigma3) edge[->] (-0.7, 0.8);
		\end{tikzpicture}
	}
	\caption{(a) The bulk discretization of vacuum $AdS_3$ shown in Figure \ref{fig:tree_flow} can be divided into bulk states by selecting regions of the bulk that are bounded by
			Ryu-Takayanagi surfaces and boundary subregions. Such a region is shaded here. (b) The bulk state obtained from truncating a tree tensor network to the shaded region.
				Each edge of the tree tensor network is composed of a maximally entangled state $\ket{\phi}$ and a subleading state $
				\ket{\sigma}$, just as in the bipartite construction of Section \ref{sec:bipartite_networks}. The bulk state on the	shaded region is defined by removing all tensors
				outside of the shaded region while keeping the edge states $\ket{\phi}$ and $\ket{\sigma}$ that define the edges at the boundary of the shaded region. The result is
				a truncated state on the tensor product Hilbert space of the edges that cross the boundary of the shaded region.}
	\label{fig:tree_bulk_states}
\end{center}
\end{figure}

To define the isometry properties of a tree tensor network precisely, we first define the state associated to any bulk region bounded by a mixture of Ryu-Takayanagi surfaces and subregions of the boundary to be the state produced by a truncation of the tree tensor network to that region. This is sketched in Figure \ref{fig:tree_bulk_states} for a subregion of the tree tensor network that was introduced in Figures \ref{fig:tree_discretization} and \ref{fig:tree_flow}. Importantly, the edge states $\ket{\phi}$ and $\ket{\sigma}$ associated to each RT surface are included in the state assigned to the region. There is a natural map from the state associated to a bulk region to the state associated to any larger region which contains the smaller region. This is essentially the ``inclusion'' map of the tensor network, which consists of all tensors that are included in the state of the larger region but not included in the smaller one; we call this the \emph{extension map}. The extension map does not include the edge states associated to its ``input'' RT surfaces, as those are already included in the state of the smaller subregion; however, it will include the edge states associated to any ``output'' RT surfaces that bound the larger bulk subregion. This convention is chosen so that an extension map ``beginning'' on a given RT surface can always be composed with an extension map that ``ends'' on the same surface.

In the bipartite construction of Section \ref{sec:bipartite_networks}, the bulk tensors $V$ and $W$ each served as the extension map from the state in the complementary bulk region out to the global boundary. In this construction, the maps $V$ and $W$ were exact isometries. For a tree network constructed from a boundary root orientation (as sketched in Figure \ref{fig:tree_flow}), we will show inductively that extension maps flowing entirely along the direction of the orientation are always exact isometries.  An extension map which flows partially against the orientation of the graph will not in general be an exact isometry; however, it will be an approximate isometry with respect to a particular state-dependent metric. (For example, in Figure \ref{fig:tree_state_shading}, the extension map from the shaded region through the network out to the ``right half'' of the boundary is an exact isometry, as it flows along the orientation of the graph. The extension map from the shaded region out to the global boundary, however, is only an approximate isometry, as it must flow against the orientation of the graph to reach the root node at the boundary.)

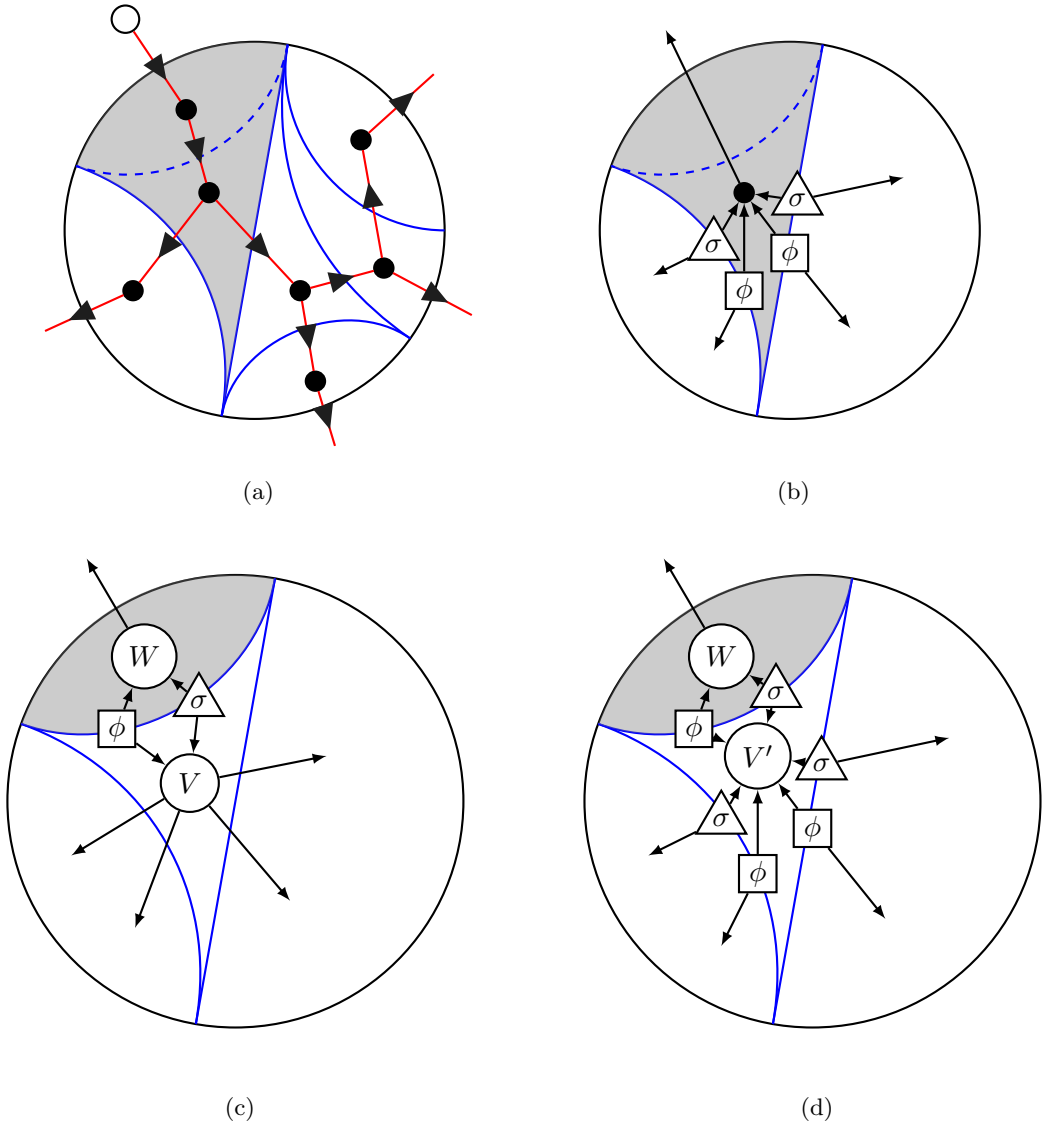
\begin{figure}[p!]
\begin{center}
	\subfloat[\label{fig:final_tree_shading}]{
		\begin{tikzpicture}[scale=1, thick, vertex/.style={draw, shape=circle, fill=black, scale=0.7},
				decoration={markings, mark=at position 0.7 with {\arrow[color={rgb,255:red,30;green,30;blue,30}, very thick]{triangle 45}}}]
		\draw (0, 0) circle (2.5cm);

		\draw [blue] (2.5, 0) to[out=180, in=260] (0.434, 2.462);
		\draw [blue] (-2.349, 0.855) to[out=340, in=80] (-0.434, -2.462);
		\draw [blue] (-0.434, -2.462) to[out=80, in=145] (2.048, -1.434);
		\draw [blue] (0.434, 2.462) to (-0.434, -2.462);
		\draw [blue] (0.434, 2.462) to[out=260, in=145] (2.048, -1.434);

		\begin{scope}
			  \clip (0.434, 2.462) to[out=260, in=80] (-0.434, -2.462) to[out=80, in=340] (-2.349, 0.855)
						to[out=70, in=170] (0.434, 2.462);
			\draw [draw=gray, fill=gray, opacity=0.4] (-2.5,-2.5) rectangle (2.5, 2.5);
		\end{scope}

		\node [vertex] at (-0.6, 0.5) (1) {};
		\node [vertex] at (-0.9, 1.6) (2) {};
		\node [vertex] at (1.4, 1.2) (3) {};
		\node [vertex] at (1.7, -0.5) (4) {};
		\node [vertex] at (0.6, -0.8) (5) {};
		\node [vertex] at (-1.6, -0.8) (6) {};
		\node [vertex] at (0.8, -2) (7) {};

		\draw [blue, dashed] (0.434, 2.462) to[out=260, in=340] (-2.349, 0.855);

		\draw [red, postaction={decorate}] (2) to (1);
		\draw [red, postaction={decorate}] (1) to (6);
		\draw [red, postaction={decorate}] (1) to (5);
		\draw [red, postaction={decorate}] (4) to (3);
		\draw [red, postaction={decorate}] (5) to (4);
		\draw [red, postaction={decorate}] (5) to (7);

		\node [shape=circle, draw] at (-1.7, 2.8) (A) {};
		\node at (2.5, 2.2) (B) {};
		\node at (3, -1.2) (C) {};
		\node at (1.1, -3) (D) {};
		\node at (-2.9, -1.4) (E) {};

		\draw [red, postaction={decorate}] (A) to (2);
		\draw [red, postaction={decorate}] (3) to (B);
		\draw [red, postaction={decorate}] (4) to (C);
		\draw [red, postaction={decorate}] (7) to (D);
		\draw [red, postaction={decorate}] (6) to (E);

		\end{tikzpicture}
	}
	\hspace{0.5cm}
	\subfloat[\label{fig:final_tree_substate}]{
		\begin{tikzpicture}[scale=1, thick, >=latex, vertex/.style={draw, shape=circle, fill=black, scale=0.7},
				decoration={markings, mark=at position 0.7 with {\arrow[color={rgb,255:red,30;green,30;blue,30}, very thick]{triangle 45}}},
				cobwebs/.style={regular polygon, inner sep=0.5pt, fill=white, regular polygon sides=3, draw},
				EPR/.style={regular polygon, regular polygon sides=4, inner sep=0pt, fill=white, draw}]
		\draw (0, 0) circle (2.5cm);

		\draw [blue] (-2.349, 0.855) to[out=340, in=80] (-0.434, -2.462);
		\draw [blue] (0.434, 2.462) to (-0.434, -2.462);

		\begin{scope}
			  \clip (0.434, 2.462) to[out=260, in=80] (-0.434, -2.462) to[out=80, in=340] (-2.349, 0.855)
						to[out=70, in=170] (0.434, 2.462);
			\draw [draw=gray, fill=gray, opacity=0.4] (-2.5,-2.5) rectangle (2.5, 2.5);
		\end{scope}

		\node [vertex] at (-0.6, 0.5) (1) {};

		\draw [blue, dashed] (0.434, 2.462) to[out=260, in=340] (-2.349, 0.855);

		\node [EPR] at (-0.6, -0.8) (phi2) {$\phi$};
		\node [cobwebs] at (-1, -0.2) (sigma2) {$\sigma$};
		\node [EPR] at (0, -0.3) (phi3) {$\phi$};
		\node [cobwebs] at (0.1, 0.4) (sigma3) {$\sigma$};

		\node at (-1.7, 2.8) (A) {};
		\node at (2.5, 2.2) (B) {};
		\node at (3, -1.2) (C) {};
		\node at (1.1, -3) (D) {};
		\node at (-2.9, -1.4) (E) {};

		\draw (phi2) edge[->] (1);
		\draw (sigma2) edge[->] (1);
		\draw (phi3) edge[->] (1);
		\draw (sigma3) edge[->] (1);

		\draw (phi2) edge [->] (-1, -1.6);
		\draw (sigma2) edge[->] (-1.8, -0.6);
		\draw (phi3) edge[->] (0.8, -1.3);
		\draw (sigma3) edge[->] (1.5, 0.7);
		\draw (1) edge[->] (A);
		\end{tikzpicture}
	}
	\hspace{0.5cm}
	\subfloat[\label{fig:final_tree_bipartite}]{
		\begin{tikzpicture}[scale=1.2, thick, >=latex, vertex/.style={draw, shape=circle, fill=black, scale=0.7},
				decoration={markings, mark=at position 0.7 with {\arrow[color={rgb,255:red,30;green,30;blue,30}, very thick]{triangle 45}}},
				cobwebs/.style={regular polygon, inner sep=0.5pt, fill=white, regular polygon sides=3, draw},
				EPR/.style={regular polygon, regular polygon sides=4, inner sep=0pt, fill=white, draw},
				tensor/.style={circle, draw, fill=white, minimum width=1.5em}]
		\draw (0, 0) circle (2.5cm);

		\draw [blue] (-2.349, 0.855) to[out=340, in=80] (-0.434, -2.462);
		\draw [blue] (0.434, 2.462) to (-0.434, -2.462);
		\draw [blue] (0.434, 2.462) to[out=260, in=340] (-2.349, 0.855);

		\begin{scope}
			  \clip (0.434, 2.462) to[out=260, in=340] (-2.349, 0.855)
						to[out=70, in=170] (0.434, 2.462);
			\draw [draw=gray, fill=gray, opacity=0.4] (-2.5,-2.5) rectangle (2.5, 2.5);
		\end{scope}

		\node [tensor] at (-0.5, 0.2) (1) {$V$};
		\node [tensor] at (-1, 1.6) (2) {$W$};

		\node [EPR] at (-1.3, 0.8) (phi1) {$\phi$};
		\node [cobwebs] at (-0.4, 1.1) (sigma1) {$\sigma$};

		\node at (-1.7, 2.8) (A) {};
		\node at (2.5, 2.2) (B) {};
		\node at (3, -1.2) (C) {};
		\node at (1.1, -3) (D) {};
		\node at (-2.9, -1.4) (E) {};

		\draw (phi1) edge[->] (1);
		\draw (sigma1) edge[->] (1);
		\draw (phi1) edge[->] (2);
		\draw (sigma1) edge[->] (2);

		\draw (1) edge [->] (-1.1, -1.4);
		\draw (1) edge[->] (-1.8, -0.6);
		\draw (1) edge[->] (0.6, -1.1);
		\draw (1) edge[->] (1, 0.5);
		\draw (2) edge[->] (A);
		\end{tikzpicture}
	}
	\subfloat[\label{fig:final_tree_network}]{
		\begin{tikzpicture}[scale=1.2, thick, >=latex, vertex/.style={draw, shape=circle, fill=black, scale=0.7},
				decoration={markings, mark=at position 0.7 with {\arrow[color={rgb,255:red,30;green,30;blue,30}, very thick]{triangle 45}}},
				cobwebs/.style={regular polygon, inner sep=0.5pt, fill=white, regular polygon sides=3, draw},
				EPR/.style={regular polygon, regular polygon sides=4, inner sep=0.1pt, fill=white, draw},
				tensor/.style={circle, draw, fill=white, minimum width=1.5em}]
		\draw (0, 0) circle (2.5cm);

		\draw [blue] (-2.349, 0.855) to[out=340, in=80] (-0.434, -2.462);
		\draw [blue] (0.434, 2.462) to (-0.434, -2.462);
		\draw [blue] (0.434, 2.462) to[out=260, in=340] (-2.349, 0.855);

		\begin{scope}
			  \clip (0.434, 2.462) to[out=260, in=340] (-2.349, 0.855)
						to[out=70, in=170] (0.434, 2.462);
			\draw [draw=gray, fill=gray, opacity=0.4] (-2.5,-2.5) rectangle (2.5, 2.5);
		\end{scope}

		\node [tensor] at (-0.6, 0.5) (1) {$V'$};
		\node [tensor] at (-1, 1.6) (2) {$W$};

		\node [EPR, inner sep=0pt] at (-1.3, 0.8) (phi1) {$\phi$};
		\node [cobwebs] at (-0.4, 1.2) (sigma1) {$\sigma$};
		\node [EPR] at (-0.6, -0.8) (phi2) {$\phi$};
		\node [cobwebs] at (-1, -0.2) (sigma2) {$\sigma$};
		\node [EPR] at (0, -0.3) (phi3) {$\phi$};
		\node [cobwebs] at (0.1, 0.4) (sigma3) {$\sigma$};

		\node at (-1.7, 2.8) (A) {};
		\node at (2.5, 2.2) (B) {};
		\node at (3, -1.2) (C) {};
		\node at (1.1, -3) (D) {};
		\node at (-2.9, -1.4) (E) {};

		\draw (phi1) edge[->] (1);
		\draw (sigma1) edge[->] (1);
		\draw (phi1) edge[->] (2);
		\draw (sigma1) edge[->] (2);
		\draw (phi2) edge[->] (1);
		\draw (sigma2) edge[->] (1);
		\draw (phi3) edge[->] (1);
		\draw (sigma3) edge[->] (1);

		\draw (phi2) edge [->] (-1, -1.6);
		\draw (sigma2) edge[->] (-1.8, -0.6);
		\draw (phi3) edge[->] (0.8, -1.3);
		\draw (sigma3) edge[->] (1.5, 0.7);
		\draw (2) edge[->] (A);
		\end{tikzpicture}
	}
	\caption{The final step of a tree tensor network construction for the bulk discretization of vacuum $AdS_3$ with orientation given in Figure \ref{fig:tree_flow}.
			(a) It is assumed that a tree network has been constructed for the discretization that consists of all but the final (dashed) RT surface. The bulk subregion $a$, shaded in
			gray, is the smallest bulk subregion containing this surface in its interior.
			(b) The state of the network on $a$ is sketched explicitly. Since the final RT surface has not yet been added to the network, the bulk subregion $a$ contains only one tensor.
				As all extension maps away from $a$ follow the root-leaf orientation of the network, they are all exact isometries.
			(c) The subregion state on $a$ is distilled across the new RT surface into a bipartite tensor network such as the one given in \eqref{bipartite_distilled_state}.
			(d) To preserve the structure of the states $\ket{\phi_i}$ and $\ket{\sigma_i}$ on the already-constructed RT surfaces, the bipartite isometry $V$ is replaced by the map 
				$V'$ given in equation \eqref{eq:adjusted_isometry}. Note that all extension maps that begin on the new smallest bulk subregion (shaded here) are still
				exact isometries.}
	\label{fig:tree_final_step}
\end{center}
\end{figure}

We now present the inductive argument for constructing a tree tensor network for an arbitrary (non-intersecting) bulk discretization. For clarity, this whole procedure is sketched in Figure \ref{fig:tree_final_step} for the final step of the oriented discretization sketched in Figure \ref{fig:tree_flow}. After a boundary edge has been designated as the root and a root-leaf orientation has been imposed on the dual graph, we pick one of the ``uppermost'' Ryu-Takayanagi surfaces (i.e., one of the surfaces that neighbors the root node) to be the last surface added to the network, and assume that we have already constructed a tree tensor network for the discretization that includes all but this final surface.\footnote{The choice of ``uppermost'' RT surface is in general non-unique, and adding surfaces to the network in different orders will produce different tensor networks that all approximate the original CFT state with high fidelity.} To make an inductive argument, we assume that the tensor network for the ``all-but-one'' discretization has been constructed so that it:
\begin{enumerate}[(a)]
\item approximately reproduces the original ``target'' CFT state on the boundary, 
\item has internal bond dimensions that match the areas of the discretization surfaces, and 
\item has the isometry properties detailed above (i.e., extension maps that follow the flow of the root-leaf orientation are exact isometries).
\end{enumerate}

We now consider the smallest bulk subregion $a$ containing the new RT surface in its interior (cf. Figure \ref{fig:final_tree_shading}, where the bulk subregion $a$ is shaded). In the tensor network for the reduced discretization, the state associated to this region is formed by a single tensor together with $\ket{\phi}$ and $\ket{\sigma}$ states on each already-constructed RT surface on the boundary of $a$. Because we chose the RT surface to neighbor the root node, the extension map from this state to entire boundary state flows entirely along the orientation of the tree and so is an exact isometry by assumption. Furthermore, we have assumed that the entire boundary state is approximately equal to the target CFT state. 

Since the smooth min- and max-entropies of the target state depend only weakly on the error $\varepsilon$, and since the extension map from $a$ to the global boundary is an isometry, the smooth min- and max-entropies of the subregion state on $a$ will agree with those of the target state to leading order. We can therefore apply the exact same bipartite distillation procedure to the subregion state on $a$ that we used to construct the global bipartite tensor network in Section \ref{sec:bipartite_networks}. This will produce a bipartite tensor network $(V \otimes W) \ket{\phi} \ket{\sigma}$ of the form given in \eqref{bipartite_distilled_state} that approximates the state of the subregion $a$. We label our isometries so that $W$ is the ``upwards'' isometry that includes the root node in its image, while $V$ is the ``downwards'' isometry that maps away from the root node of the discretization  (cf. Figure \ref{fig:final_tree_bipartite}).

This newly-distilled state approximately reproduces the subregion state on $a$, and could be substituted directly into the network as in Figure \ref{fig:final_tree_bipartite}. Doing so, however, would require erasing all of the $\ket{\phi}$ and $\ket{\sigma}$ states associated to each of the neighboring, previously-added RT surfaces, replacing them with the outward-pointing legs of $V$. Instead, we wish only to replace the central tensor associated to $a$. Fortunately, as discussed in Section \ref{sec:TN_review}, there is a canonical isomorphism between the states $\ket{\phi} \in \mathcal{H}_\gamma \otimes \bar{\mathcal{H}}_\gamma$ and $\ket{\sigma} \in \mathcal{H}_f \otimes \bar{\mathcal{H}}_f$ and operators $\phi : \mathcal{H}_\gamma \to \mathcal{H}_\gamma$ and $\sigma : \mathcal{H}_f \to \mathcal{H}_f$. Since $\ket{\phi}$ and $\ket{\sigma}$ are full-rank, these operators are invertible. We can therefore simply replace the central tensor in the subregion by $V' W \ket{\phi} \ket{\sigma}$, where
\begin{equation}
	V' = \left(\prod_i \sigma_i^{-1} \phi_i^{-1} \right) V \label{eq:adjusted_isometry}
\end{equation}
and the product is taken over all RT surfaces on the boundary of $a$. In the case where there is a unique ``uppermost'' RT surface for our discretization, the isometry $W$ will map directly to the boundary and will not require modification. In the case of a discretization where two or more RT surfaces both neighbor the root node, the isometry $W$ will need to be modified on any of its outgoing legs that pass through already-constructed RT surfaces, as in equation \eqref{eq:adjusted_isometry}.

By construction, the new state associated to the bulk subregion will approximate the old state, and since by assumption the rest of the tensor network forms an isometry from $V' W \ket{\phi} \ket{\sigma}$ (or $V' W' \ket{\phi} \ket{\sigma}$ in the case of multiple ``uppermost'' RT surfaces) to the global boundary, the new state constructed by the entire network will continue to approximate the target CFT state. The map $V$, which maps the new bulk subregion across the new RT surface into the rest of the network, is an exact isometry even though $V'$ is not. This validates our assumption that extension maps that follow the root-leaf orientation of the network should always be isometries. By induction, we can therefore construct a tensor network state whose geometry agrees with an arbitrary tree discretization and which approximately reproduces the original CFT state.

The only claim that we have made but are yet to prove is that extension maps which flow partially against the root-leaf orientation of the underlying graph are still approximate, if not exact, isometries. This claim was not required as part of our inductive procedure for constructing tree networks, but will be useful for constructing more general networks in following sections. Consider an arbitrary RT surface in some tree network discretization. We have two extension maps, one in each direction, that map the edge state $\ket{\phi} \ket{\sigma}$ associated with this surface to the entire boundary. One map $V: \bar{\mathcal{H}}_{\gamma} \otimes \bar{\mathcal{H}}_{f} \hookrightarrow \mathcal{H}_{A^c}$ flows entirely with the orientation of the graph and so is an exact isometry. When the RT surface was first added to the network, the other extension map $W: \mathcal{H}_{\gamma} \otimes \mathcal{H}_f \hookrightarrow \mathcal{H}_A$ was also an exact isometry for exactly the same reasons. However, because this extension map flows partially against the root-leaf orientation of the graph, it will continue to change as additional surfaces are added to the network. This is in contrast to the ``downwards'' isometry $V$, which remains unaltered because of the order in which we chose to add the RT surfaces. In the final network, once all RT surfaces have been added, we call the extension map that flows partially against the root-leaf orientation $X: \mathcal{H}_{\gamma} \otimes \mathcal{H}_f \to \mathcal{H}_A$. This map will general not be an exact isometry; we will show that it is still an approximately isometry in an appropriate sense.

Because our construction is designed so that the tensor network approximately reproduces the boundary state at every stage in its construction, we find
\begin{align} \label{eq:approx_isom}
V \otimes W \ket{\phi} \ket{\sigma} \approx V \otimes X \ket{\phi} \ket{\sigma}.
\end{align}
The left-hand side of \eqref{eq:approx_isom} describes the state produced by the entire tensor network when the RT surface is first added to the network, while the right-hand side is the state of the final tensor network. Since $\ket{\phi} \ket{\sigma}$ is fully entangled (i.e., its reduced density matrices on either side of the RT surface are full-rank), exactness of \eqref{eq:approx_isom} would imply that $X = W$ and so the extension map that flows partially against the network orientation would have to remain an exact isometry. As such, we can interpret \eqref{eq:approx_isom} as showing that $X$ is an approximate isometry with respect to a particular metric that is adapted to the state $\ket{\phi} \ket{\sigma}$. It can be equivalently written as
\begin{align} \label{eq:HSapproxisom}
\lVert (X - W) \phi \otimes \sigma \rVert_2 = \lVert (X - W) \rho_\phi^{1/2} \rho_\sigma^{1/2} \rVert_2 \leq \varepsilon,
\end{align}
where $\lVert A \rVert_2 \equiv \sqrt{\tr(A^{\dagger} A)}$ is the Hilbert-Schmidt norm, $\rho_\phi \rho_\sigma = \phi^2 \sigma^2$ is the reduced density matrix of $\ket{\phi} \ket{\sigma}$, and $\varepsilon > 0$ is small. This removes the isometry $V$ from \eqref{eq:approx_isom}, and so gives a distance that depends only on $X$, $W$ and the reduced density matrix $\rho_\phi \rho_\sigma$. 

Finally, we can look at the partial trace of \eqref{eq:approx_isom} over $\mathcal{H}_f \otimes \mathcal{H}_\gamma$.  Using the fact that the trace norm $\lVert \rho \rVert_1 = \tr{\sqrt{\rho^{\dagger} \rho}}$ is monotonically decreasing under the partial trace, it follows from \eqref{eq:approx_isom} that
\begin{align} \label{eq:2ndapproxisom}
\left\lVert \tr_{f \gamma} \left(X \ket{\phi} \ket{\sigma} \bra{\phi} \bra{\sigma} X^\dagger\right)  - \rho_\phi \rho_{\sigma} \right\rVert_1 = \left\lVert \rho_\phi^{1/2} \rho_\sigma^{1/2} (X^\dagger X - \mathds{1}) \rho_\phi^{1/2} \rho_{\sigma}^{1/2} \right\rVert_1 \leq \varepsilon
\end{align}
This is a strictly weaker condition than \eqref{eq:approx_isom} and \eqref{eq:HSapproxisom}: inefficiencies in the Fuchs-van de Graaf inequalities \cite{FvdG1997} mean that we cannot recover \eqref{eq:approx_isom} and \eqref{eq:HSapproxisom} without some loss of precision. However, it is perhaps the easiest condition to interpret.  

Let $\{x_i\}$ be the eigenvalues of $X^\dagger X$. If $\ket{\sigma}$ were maximally mixed, we could rewrite \eqref{eq:2ndapproxisom} as
\begin{align} \label{eq:averageerror}
\frac{1}{d} \sum_i |x_i - 1| \leq \varepsilon.
\end{align}
In other words, we would understand \eqref{eq:2ndapproxisom} as saying that \emph{on average} the eigenvalues of $X^\dagger X$ is close to one. Hence $X^\dagger X \simeq \mathds{1}$ and $X$ is an approximate isometry. Since $\ket{\sigma}$ is full rank but is not maximally mixed, we should instead think of \eqref{eq:2ndapproxisom} as a \emph{weighted} average of $|x_i - 1|$. Of course, since in general $X^\dagger X$ and $\rho_\sigma$ will not commute, this interpretation is not quite literal. It does, however, provide the correct intuition.

Thus far, we have only considered the approximate isometry condition for extension maps that go from an RT surface all the way to the boundary. What about an extension map $X'$ that flows only partially through the network, whose image lies on RT surfaces in addition to, or instead of, the boundary? In this case, the ``output'' RT surfaces may not have been added to the network when the ``input'' RT surface was added, so it may not necessarily be possible to compare the final extension map $X'$ to some intermediary extension map $W'$ as we did in equation \eqref{eq:approx_isom}. However, it will always be possible to \emph{compose} $X'$ with some exact isometry $W'$ so that the resulting operator $W' X'$ is an extension map from the ``input'' RT surfaces of $X'$ out to the global boundary.\footnote{In the final network, in general, the extension map $W'$ from the output surfaces of $X'$ to the global boundary will not be an exact isometry. However, it will always be possible to find some \emph{intermediary} network, constructed as part of the inductive procedure, where $W'$ is exact. Since both final and intermediary networks reproduce the boundary state to within tolerance $\varepsilon$, either can be used to prove equation \eqref{eq:approx_ext_map}.}

Since $W'$ is an exact isometry, applying \eqref{eq:2ndapproxisom} to the map $W' X'$ yields the following inequality:
\begin{align}\label{eq:approx_ext_map}
\left\lVert \rho_\phi^{1/2} \rho_\sigma^{1/2} (X'^\dagger X' - \mathds{1}) \rho_\phi^{1/2} \rho_{\sigma}^{1/2} \right\rVert_1 = \left\lVert \rho_\phi^{1/2} \rho_\sigma^{1/2} (X'^\dagger W'^\dagger W' X' - \mathds{1}) \rho_\phi^{1/2} \rho_{\sigma}^{1/2} \right\rVert_1 \leq \varepsilon
\end{align}
We conclude that extension maps that do not flow all the way to the boundary, such as $X'$, satisfy the same approximate isometry condition given in \eqref{eq:2ndapproxisom} for extension maps that \emph{do} flow all the way to the boundary.

Unfortunately, the approximate isometry conditions \eqref{eq:approx_isom}, \eqref{eq:HSapproxisom}, \eqref{eq:2ndapproxisom}, and \eqref{eq:approx_ext_map} are not quite as powerful as one might want. They tell us that $X$ and $X'$ are close to isometries ``on average'', but, because of the large dimensions of the Hilbert spaces involved, they do not say much about the ``worst-case'' error. For example, a single eigenvalue $x_i$ of $X^\dagger X$ can be very far from one without making a large contribution to the averaged error \eqref{eq:averageerror}.

More formally, to bound the ``worst-case'' error, we would want to bound the operator norm
\begin{align} \label{eq:op_norm}
\lVert X - W \rVert_\infty = \sup_{\ket{\psi}} \frac{\lVert(X-W) \ket{\psi}\rVert}{\lVert \ket{\psi} \rVert}.
\end{align}
However, the tightest bound that we can place on this norm is 
\begin{align} \label{eq:opnorm}
\lVert X - W \rVert_\infty \leq \lVert (X-W \phi \otimes \sigma)\rVert_\infty \lVert \phi^{-1}\otimes \sigma^{-1}\rVert_\infty \leq \varepsilon \lVert \phi^{-1}\otimes \sigma^{-1}\rVert_\infty,
\end{align}
where the first inequality follows from the submultiplicativity of the operator norm and the second follows from the monotonicity of the Schatten norms and \eqref{eq:HSapproxisom}. Since the operator norm of $\phi^{-1} \otimes \sigma^{-1}$ is quite large, na\"{i}vely satisfying
\begin{equation}
	\lVert \phi^{-1}\otimes \sigma^{-1}\rVert_\infty = e^{O(S)},
\end{equation}
we cannot make $\varepsilon$ small enough to make this a tight bound. 

If the RT surfaces of a tree tensor network are added in an arbitrary order, rather than adding child surfaces before parent surfaces, then the potentially large error in \eqref{eq:opnorm} prevents us from guaranteeing that the final network correctly approximates the original boundary state. In this section, we avoided this possibility by imposing a root-leaf orientation on our construction; however, when constructing more complicated sub-AdS scale tensor networks in Section \ref{sec:iteration}, the issue of errors in our approximate isometries will arise once again without the guarantee of a root-leaf orientation to stop them from blowing up.

However, in practice, \eqref{eq:approx_isom} and \eqref{eq:2ndapproxisom} should ensure that ``generic'' small perturbations in the subregion states of our network that occur during an unoriented induction should only lead to a small error in the final state. Large errors would only occur if these perturbations were somehow finely tuned to blow up when mapped out to the boundary via extension maps. We therefore think it very likely that our distillation procedure will produce tensor networks that approximate the original holographic state correctly \emph{regardless} of the order in which Ryu-Takayanagi surfaces are added to the network, and more generally will produce accurate sub-AdS scale tensor networks even when we no longer have the luxury of a root-leaf orientation to constrain accumulated errors precisely.

\section{Bulk Legs and Holographic Quantum Error Correction} \label{sec:qec}

Thus far, we have focused on constructing a tensor network for a single, arbitrary holographic CFT state. In the literature, however, it is common to consider tensor networks that describe not only a single holographic state, but an entire code subspace of holographic states. These tensor networks have some dangling bulk legs that turn the entire tensor network into a bulk-to-boundary map that encodes bulk excitations into a subspace of the boundary (see, e.g., \cite{HaPPY} and \cite{HNQTWY2016}). Ideally, the map from the bulk to the boundary will have some appropriate error correcting properties that allow bulk operator reconstruction to be interpreted in the language of quantum error correction.

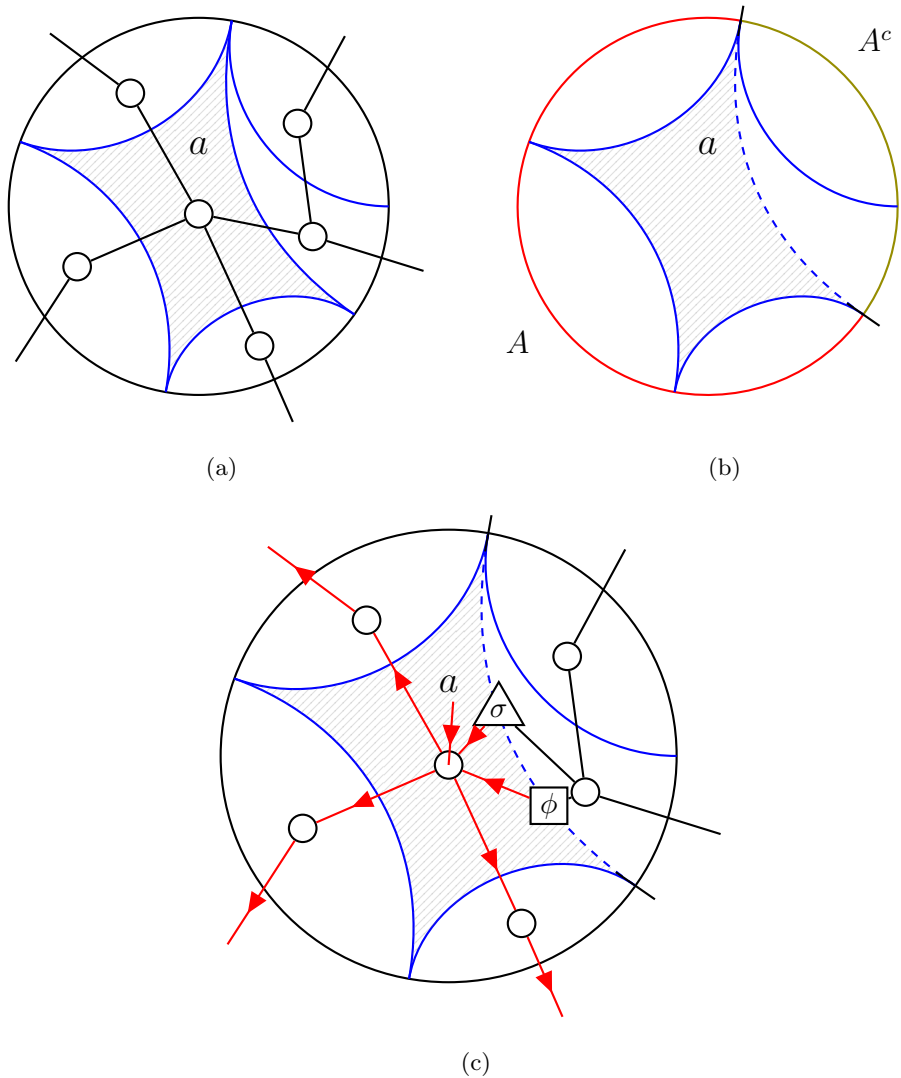
\begin{figure}[p!]
\begin{center}
	\subfloat[\label{fig:qec_shading}]{
		\begin{tikzpicture}[scale=1, thick,
					EPR/.style={regular polygon, inner sep=-1pt, fill=white, regular polygon sides=3, draw},
					cobwebs/.style={rectangle, fill=white, draw},
					tensor/.style={circle, draw, fill=white, minimum width=0.2em}]
		\draw (0, 0) circle (2.5cm);

		\draw [blue] (2.5, 0) to[out=180, in=260] (0.434, 2.462);
		\draw [blue] (-2.349, 0.855) to[out=340, in=80] (-0.434, -2.462);
		\draw [blue] (-0.434, -2.462) to[out=80, in=145] (2.048, -1.434);
		\draw [blue] (0.434, 2.462) to[out=260, in=145] (2.048, -1.434);
		\draw [blue] (0.434, 2.462) to[out=260, in=340] (-2.349, 0.855);

		\begin{scope}
			  \clip (0.434, 2.462) to[out=260, in=145] (2.048, -1.434) to[out=145, in=80] (-0.434, -2.462)
						to[out=80, in=340] (-2.349, 0.855) to[out=340, in=260] (0.434, 2.462);
			\draw [draw=gray, pattern=north east lines, pattern color=gray, opacity=0.5] (-2.5,-2.5) rectangle (2.5, 2.5);

		\end{scope}

		\node [tensor] at (-0.9, 1.5) (2) {};
		\node [tensor] at (1.3, 1.1) (3) {};
		\node [tensor] at (1.5, -0.4) (4) {};
		\node [tensor] at (0, -0.1) (5) {};
		\node [tensor] at (-1.6, -0.8) (6) {};
		\node [tensor] at (0.8, -1.85) (7) {};

		\node at (0, 0.8) {\Large$a$};

		\node at (-2.1, 2.4) (A) {};
		\node at (2, 2.4) (B) {};
		\node at (3.1, -0.9) (C) {};
		\node at (1.3, -3) (D) {};
		\node at (-2.5, -2.2) (E) {};

		\draw (2) to (5) to (6);
		\draw (7) to (5) to (4) to (3);

		\draw (2) to (A);
		\draw (3) to (B);
		\draw (4) to (C);
		\draw (7) to (D);
		\draw (6) to (E);

		\end{tikzpicture}
	}
	\hspace{0.3cm}
	\subfloat[\label{fig:qec_RT_choice}]{
		\begin{tikzpicture}[scale=1, thick,
					EPR/.style={regular polygon, inner sep=-1pt, fill=white, regular polygon sides=3, draw},
					cobwebs/.style={rectangle, fill=white, draw},
					tensor/.style={circle, draw, fill=white, minimum width=0.2em}]
		\draw [red] (0.434, 2.462) arc (80:360-35:2.5);
		\draw [olive] (2.048, -1.434) arc (-35:80:2.5);

		\draw [blue] (2.5, 0) to[out=180, in=260] (0.434, 2.462);
		\draw [blue] (-2.349, 0.855) to[out=340, in=80] (-0.434, -2.462);
		\draw [blue] (-0.434, -2.462) to[out=80, in=145] (2.048, -1.434);
		\draw [blue, dashed] (0.434, 2.462) to[out=260, in=145] (2.048, -1.434);
		\draw [blue] (0.434, 2.462) to[out=260, in=340] (-2.349, 0.855);

		\begin{scope}
			  \clip (0.434, 2.462) to[out=260, in=145] (2.048, -1.434) to[out=145, in=80] (-0.434, -2.462)
						to[out=80, in=340] (-2.349, 0.855) to[out=340, in=260] (0.434, 2.462);
			\draw [draw=gray, pattern=north east lines, pattern color=gray, opacity=0.5] (-2.5,-2.5) rectangle (2.5, 2.5);

		\end{scope}

		\draw (0.4, 2.26) to (0.47, 2.66);
		\draw (1.833, -1.284) to (2.262, -1.584);

		\node at (-2.5, -1.8) (Aboundary) {\large$A$};
		\node at (2.2, 2.2) (Ac) {\large$A^c$};

		\node at (0, 0.8) {\Large$a$};

		\node at (-2.1, 2.4) (A) {};
		\node at (2, 2.4) (B) {};
		\node at (3.1, -0.9) (C) {};
		\node at (1.3, -3) (D) {};
		\node at (-2.5, -2.2) (E) {};

		\end{tikzpicture}
	}
	\hspace{0.3cm}
	\subfloat[\label{fig:qec_pushed}]{
		\begin{tikzpicture}[scale=1.2, thick,
					cobwebs/.style={regular polygon, inner sep=0.5pt, fill=white, regular polygon sides=3, draw},
					EPR/.style={regular polygon, regular polygon sides=4, inner sep=0pt, fill=white, draw},
					tensor/.style={circle, draw, fill=white, minimum width=0.2em},
					decoration={markings, mark=at position 0.7 with {\arrow[color=red, thick]{triangle 45}}}]
		\draw (0, 0) circle (2.5cm);

		\draw [blue] (2.5, 0) to[out=180, in=260] (0.434, 2.462);
		\draw [blue] (-2.349, 0.855) to[out=340, in=80] (-0.434, -2.462);
		\draw [blue] (-0.434, -2.462) to[out=80, in=145] (2.048, -1.434);
		\draw [blue, dashed] (0.434, 2.462) to[out=260, in=145] (2.048, -1.434);
		\draw [blue] (0.434, 2.462) to[out=260, in=340] (-2.349, 0.855);

		\begin{scope}
			  \clip (0.434, 2.462) to[out=260, in=145] (2.048, -1.434) to[out=145, in=80] (-0.434, -2.462)
						to[out=80, in=340] (-2.349, 0.855) to[out=340, in=260] (0.434, 2.462);
			\draw [draw=gray, pattern=north east lines, pattern color=gray, opacity=0.5] (-2.5,-2.5) rectangle (2.5, 2.5);

		\end{scope}

		\draw (0.4, 2.26) to (0.47, 2.66);
		\draw (1.833, -1.284) to (2.262, -1.584);

		\node [tensor] at (-0.9, 1.5) (2) {};
		\node [tensor] at (1.3, 1.1) (3) {};
		\node [tensor] at (1.5, -0.4) (4) {};
		\node [tensor] at (0, -0.1) (5) {};
		\node [tensor] at (-1.6, -0.8) (6) {};
		\node [tensor] at (0.8, -1.85) (7) {};

		\node at (0, 0.8) {\Large$a$};

		\node [EPR] at (1.1, -0.55) (phi) {$\phi$};
		\node [cobwebs] at (0.55, 0.5) (sigma) {$\sigma$};

		\node at (-2.1, 2.4) (A) {};
		\node at (2, 2.4) (B) {};
		\node at (3.1, -0.9) (C) {};
		\node at (1.3, -3) (D) {};
		\node at (-2.5, -2.2) (E) {};

		\draw [red, postaction=decorate] (5) to (2);
		\draw [red, postaction=decorate] (5) to (6);
		\draw [red, postaction=decorate] (5) to (7);
		\draw [red, postaction=decorate] (phi) to (5);
		\draw [red, postaction=decorate] (sigma) to (5);
		\draw (sigma) to (4);
		\draw (phi) to (4);
		\draw (4) to (3);

		\draw [red, postaction=decorate] (0.05, 0.6) to (0, -0.1);

		\draw [red, postaction=decorate] (2) to (A);
		\draw (3) to (B);
		\draw (4) to (C);
		\draw [red, postaction=decorate] (7) to (D);
		\draw [red, postaction=decorate] (6) to (E);

		\end{tikzpicture}
	}
	\caption{(a) A tree tensor network for a particular discretization of vacuum $AdS_3$. A bulk subregion $a$ of the discretization (shaded here) has been chosen to create a code subspace
			of excitations around the vacuum generated by operators whose support lies entirely within $a$. Note that the $\ket{\phi}$ and $\ket{\sigma}$ edge states have been
			suppressed in this sketch for the sake of visual clarity.
			(b) Choosing a particular RT surface that bounds $a$, denoted here with a dashed line, partitions the global boundary
			into two regions: $A$, which contains $a$ in its entanglement wedge, and its complement $A^c$.
			(c) As explained in the text, a code subspace of excitations localized in $a$ can be represented by adding a single bulk leg to the tensor in region $a$. For a particular
				choice of $RT$ surface bounding $a$, the ``controlled extension map'', which maps the bulk leg in $a$ plus the edge states on the chosen RT surface into the
				boundary region $A$, is sketched with red arrows. In order to clarify where the extension map begins, the edge states $\ket{\phi}$ and $\ket{\sigma}$ are shown
				explicitly on the chosen RT surface, and suppressed on all other RT surfaces in the diagram.}
	\label{fig:qec_figs}
\end{center}
\end{figure}

We discuss here, briefly, how our construction can be extended to create such code spaces. We shall focus on a code space that consists of a single bulk qubit, localized within a single bulk region $a$ of the tree network discretization shown in Figure \ref{fig:qec_shading}. The generalization to more complicated bulk code spaces in tree tensor networks is straightforward, and we would optimistically expect similar results to hold for the more general constructions of Sections \ref{sec:MEP} and \ref{sec:iteration}.

In much of the existing literature on holographic quantum error correcting codes (see, e.g., \cite{HaPPY} and \cite{HNQTWY2016}), one begins with a particular tensor network construction and then shows that it has quantum error correcting properties analogous to those of AdS/CFT. As in the rest of this paper, we reverse this script. We begin with the known quantum error correcting properties of AdS/CFT \cite{ADH2015}, and especially entanglement wedge reconstruction \cite{DHW2016,CHPSSW2017, Hayden:2018khn, FL2017}, and show that they imply the existence of tensor network constructions for code spaces of \emph{actual} holographic states, which behave in exactly the same way as the toy models that have previously appeared in the literature. We emphasize that this should not be taken as an independent proof of entanglement wedge reconstruction. Instead it demonstrates that the error-correction properties of AdS/CFT itself and of holographic tensor networks toy models are not merely analogous to one another, but are in fact different examples of exactly the same phenomenon.

Consider, for example, a 2-dimensional ``code subspace'' of states in AdS/CFT that can be built from a single starting state by applying low-energy unitary bulk operators in only a single bulk region $a$ of some tree network discretization.\footnote{If the energy of the bulk matter excitations is $O(1)$ in a $G_N$ expansion, then the backreaction on the geometry is $O(G_N)$, resulting in an $O(1)$ change in the Bekenstein-Hawking entropy of the surfaces bounding the region $a$. This is smaller than the $O(G_N^{-1/2})$ fluctuations already present in our network, and can therefore be neglected without changing our error estimates on the bond dimensions of the network.} Any Ryu-Takayanagi surface neighboring the bulk region $a$ splits the boundary into two regions: region $A$, the entanglement wedge of which contains the bulk region $a$, and its complement $A^c$. This is sketched for a particular choice of bulk region and Ryu-Takayanagi surface in Figure \ref{fig:qec_RT_choice}. Because of the usual error correcting properties of AdS/CFT \cite{ADH2015, DHW2016, CHPSSW2017, RTQEC}, the reduced density matrix on $A^c$ is approximately the same for every state in the code space. Moreover, any two orthogonal states in the code space will have reduced density matrices on $A$ that are approximately orthogonal; the trace distance between them will be close to maximal. This is in fact equivalent to the fact that the reduced density matrix on $A^c$ is the same for every state, including superpositions, in the code space (see, e.g., the weak decoupling duality described in \cite{HW2012,hayden2017approximate}).

Let us suppose we have successfully constructed a tree tensor network for a single state $\ket{\psi}$ in the code space using the techniques of Section \ref{sec:tree_networks}. Because the reduced density matrices on $A^c$ are the same for every RT surface bounding $a$ and every state in the code space, any state in the code subspace can be represented by a tensor network that is identical to the one constructed for $\ket{\psi}$ \emph{except} that it differs in the tensor associated to the bulk subregion $a$. It follows by linearity that we can describe the entire two-dimensional code space adding a two-dimensional bulk leg to the tensor in region $a$ (cf. Figure \ref{fig:qec_pushed}). The resulting tensor network can be interpreted as a map $T : \mathcal{H}_{\text{bulk}} \hookrightarrow \mathcal{H}_{\text{CFT}}$ from the bulk to the boundary. Furthermore, by using our freedom to choose an inner product on $\mathcal{H}_{\text{bulk}}$, we can ensure that $T$ is an exact isometry.

Showing that our tensor networks are quantum error correcting in the sense of \cite{HaPPY} requires showing that for any choice of RT surface bounding $a$ and corresponding boundary subregion $A$, any bulk operator on $\mathcal{H}_{\text{bulk}}$ has an equivalent boundary operator whose support lies only on $A$. By ``equivalent,'' we mean that acting with one of these boundary operators on any state in the code subspace produces approximately the same state that one would obtain by acting with the original, bulk operator on $\mathcal{H}_{\text{bulk}}$. Such boundary representations of bulk operators are obtained in \cite{HaPPY} by using the tensor network to ``push the bulk operator through the network'' and into the boundary.

To be precise, the map that will be used to push bulk operators to the boundary region $A$ is the ``controlled extension map'' $X: \mathcal{H}_f \otimes \mathcal{H}_{\gamma} \otimes \mathcal{H}_{\text{bulk}} \rightarrow A,$ sketched in Figure \ref{fig:qec_pushed}. We call this a \emph{controlled extension map} because once a state $\ket{\psi}$ is fixed on the bulk leg $\mathcal{H}_{\text{bulk}}$, the resulting map $X\ket{\psi}$ is an extension map in the corresponding tensor network in the sense of Section \ref{sec:larger_tree_networks}. We will show first that $X$ is an approximate isometry in the sense of equation \eqref{eq:2ndapproxisom}, and second that this condition allows us to use $X$ to push bulk operators to equivalent operators on the boundary.

Let $\ket{0}, \ket{1}$ form a basis for $\mathcal{H}_{\text{bulk}}$. The controlled extension map $X$ can then be represented in this basis as
\begin{equation}
	X
		= X_0 \bra{0} + X_1 \bra{1}, 
\end{equation}
where $X_0$ and $X_1$ are the extension maps from the RT surface to $A$ for the tree tensor networks defined by specifying bulk leg states $\ket{0}$ and $\ket{1}$ respectively. We showed in Section \ref{sec:larger_tree_networks} that extension maps in tree tensor networks are approximate isometries, and so $X_0$ and $X_1$ satisfy 
\begin{equation}
	X_0^\dagger X_0
		\simeq X_1^\dagger X_1 \simeq \mathds{1}
\end{equation}
in the sense of equation \eqref{eq:2ndapproxisom}. Because orthogonal bulk states are almost orthogonal in region $A$, these extension maps satisfy
\begin{equation}
	X_1^\dagger X_0
		\simeq X_0^\dagger X_1 \simeq 0
\end{equation}
in the same sense. It follows immediately that $X^\dagger X \simeq \mathds{1}$ and so $X$ is an approximate isometry in the sense of \eqref{eq:2ndapproxisom}. By the same arguments as in Section \ref{sec:larger_tree_networks}, we can also show that controlled extension maps that end on RT surfaces are also approximate isometries.

Given an operator $\hat{O}_b$ acting on $\mathcal{H}_{\text{bulk}}$, we wish to use the fact that $X$ is an approximate isometry to produce a boundary operator $\hat{O}_{A}$ supported on $A$ whose action on the code subspace is approximately the same as $\hat{O}_b.$ More precisely, for a state $\ket{\psi} \in \mathcal{H}_{\text{bulk}}$, we wish to show
\begin{equation} \label{eq:bulk_boundary_equiv}
	\hat{O}_{A} T \ket{\psi} \approx T \hat{O}_{b} \ket{\psi}.
\end{equation}
This is exactly the same sense of bulk reconstruction through quantum error correction that was developed for exact tensor network toy models in \cite{HaPPY}. In terms of the controlled extension map $X$ and the bulk tensor network state $\ket{T_{A^c}}$ associated to the entanglement wedge of $A^c$, the tensor network map $T$ can be decomposed as
\begin{equation}
	T \ket{\psi} = X \ket{T_{A^c}} \ket{\psi}.
\end{equation}
Equation \eqref{eq:bulk_boundary_equiv} therefore becomes
\begin{equation} \label{eq:operator_equiv}
	\hat{O}_{A} X \ket{T_{A^c}} \ket{\psi} \approx X \hat{O}_{b} \ket{T_{A^c}} \ket{\psi}.
\end{equation}

As in \cite{HaPPY}, we can find a boundary representation of a bulk operator by simply conjugating with $X$, i.e.,
\begin{equation} \label{eq:pushed1}
	\hat{O}_{A} \equiv X \hat{O}_{b} X^{\dagger}.
\end{equation}
Using the approximate isometry condition that $X^{\dagger} X$ acts approximately as the identity on $\ket{T_{A^c}} \ket{\psi}$ for any bulk state $\ket{\psi},$ we see that $\hat{O}_{A}$ satisfies equation \eqref{eq:bulk_boundary_equiv}. If $X$ were an exact isometry, as in \cite{HaPPY}, the map from bulk operators to boundary operators given in \eqref{eq:pushed1} would be unital on the image of $X$ (i.e., it maps the identity on $\mathcal{H}_{\text{bulk}}$ to the boundary projector onto the image of $X$). This condition is desirable from the perspective of quantum information theory, as it ensures that the map given in \eqref{eq:pushed1} is a quantum channel in the Heisenberg picture.

In order to ensure that this condition holds for our approximate isometries, thus making the bulk-to-boundary operator map into a genuine quantum channel, we instead define the boundary representation of a bulk operator as
\begin{equation} \label{eq:pushed2}
	\hat{O}_{A} \equiv X (X^{\dagger} X)^{-1/2} \hat{O}_{b} (X^{\dagger} X)^{-1/2} X^{\dagger}.
\end{equation}
From the isometry condition  \eqref{eq:2ndapproxisom}, we see that $\hat{O}_{A}$ still satisfies equation \eqref{eq:bulk_boundary_equiv}, and thus this an approximate boundary representation of the bulk operator $\hat{O}_{b}.$ We can also see plainly that the bulk-to-boundary operator map given by equation \eqref{eq:pushed2} maps the identity on $\mathcal{H}_{\text{bulk}}$ to the identity on the image of $X$, and is thus a quantum channel on operators in the code subspace (i.e., a completely positive, unital map on operators). We conclude that every bulk operator $\hat{O}_{b}$ has an equivalent boundary operator representation $\hat{O}_{A}$ whose support lies only on $A$ and whose action on the code subspace is approximately the same as the action of $\hat{O}_{b}$, as desired.

The boundary operator obtained via the quantum channel given in equation \eqref{eq:pushed2} may be very different from the boundary operator on the same region obtained through the extrapolate dictionary and the HKLL reconstruction procedure \cite{HKLL}; however, these operators will agree on the boundary code subspace obtained by taking the image of $\mathcal{H}_{\text{bulk}}$  under the map defined by the tensor network. Our tensor network description of a state in the AdS/CFT correspondence therefore displays exactly the same bulk-to-boundary operator mapping properties of the celebrated HaPPY class of holographic codes \cite{HaPPY}. We conclude that tensor networks models of bulk operator reconstruction may be taken quite literally. They are not just toy models!

\section{Loop Tensor Networks from Holographic Entanglement of Purification} \label{sec:MEP}

The tree networks constructed in Section \ref{sec:tree_networks} already constitute a large class of tensor networks for the AdS/CFT correspondence, but we might still hope for a more general construction. A single bond in a tree tensor network corresponds to a complete Ryu-Takayanagi surface in the bulk spacetime from which the network was constructed, so the information contained in that bond is \emph{a priori} distributed nonlocally across the entire corresponding surface. In order to localize degrees of freedom at sub-AdS scales within the bulk geometry, we must therefore find a way to divide each bond geometrically along its corresponding Ryu-Takayanagi surface. In the bulk discretization picture of Section \ref{sec:tree_networks}, this corresponds to constructing a holographic tensor network with \emph{loops}.

To construct such a tensor network, we will need to understand the \emph{holographic entanglement of purification} \cite{TU2017, NDHZS2017}, a geometric quantity in the bulk spacetime that is conjectured to correspond to an information-theoretic quantity involving a tensor factorization of the information on an RT surface. In this section, we will assume the holographic entanglement of purification conjecture, and then use it to construct a geometrically accurate tensor network with \emph{a single loop} for an arbitrary holographic CFT state. With some further assumptions introduced in Section \ref{sec:iteration}, we will then be able to extend this procedure to construct tensor networks with arbitrarily many loops, and therefore to localize information in the bulk arbitrarily well within the regime of validity of the Ryu-Takayanagi formula (i.e., above the string/Planck length scales).

\subsection{Holographic Entanglement of Purification}
For a quantum state $\rho_{(AB)}$ on a bipartite Hilbert space $\mathcal{H}_{A} \otimes \mathcal{H}_{B}$, the entanglement of purification \cite{THLD2002} between subsystems $A$ and $B$ is defined as
\begin{equation}
	E_P(A:B)
		= \inf_{\ket{\Psi}_{A A' B B'}} S(AA'), \label{eop}
\end{equation}
where $\ket{\Psi}_{AA'BB'} \in \mathcal{H}_{A} \otimes \mathcal{H}_{A'} \otimes \mathcal{H}_{B} \otimes \mathcal{H}_{B'}$ is a purification of $\rho_{(AB)}$. The infimum in equation (\ref{eop}) is taken over all possible purifications of $\rho_{(AB)}$ in all possible auxiliary Hilbert spaces $\mathcal{H}_{A'}$ and $\mathcal{H}_{B'}.$

If $\rho_{(AB)}$ is a mixed state, then the von Neumann entropy $S(A)$ no longer measures the entanglement (or even the correlation) between $A$ and $B$ in any meaningful sense, since some portion of $S(A)$ is inherited from the nonzero von Neumann entropy of $\rho_{(AB)}$. The von Neumann entropy $S(A)$ may be non-zero for a product state $\rho_{(A)} \otimes \rho_{(B)}$, which has no correlation between the two subsystems. The entanglement of purification $E_P$ is a somewhat better measure of the degree to which $A$ and $B$ are entangled (or at least correlated), as it measures the minimal entanglement between $AA'$ and $BB'$ for any purification. As a result, the entanglement of purification is zero for product states, and is non-increasing under local operations. It is not a true entanglement monotone in the sense of \cite{V2000}, however, since it may be non-zero even for separable states (which only have classical correlation), and it may be increased by classical communication.

\begin{figure}[h]
\begin{center}
	\subfloat[\label{fig:EP_EW_connected}]{
		\begin{tikzpicture}[thick, vertex/.style={draw, shape=circle, fill=black, scale=0.7}]
		\draw (0, 0) circle (2.5cm);

		\draw [blue] (2.45, 0.5) to[out=191.5, in=348.5] (-2.45, 0.5);

		\draw [purple] (0, 2.5) to (0, .2);

		\node at (-2.2, 2) (A1) {\large $A_1$};
		\node at (2.2, 2) (A2) {\large $A_2$};
		\node at (0, -2.8) (Ac) {\large $A^c$};

		\node at (0.7, 1.3) (EW) {$\Sigma_{A_1 : A_2}$};

		\node at (-1.3, -0.1) (A1p) {$A_1'$};
		\node at (1.3, -0.1) (A2p) {$A_2'$};
		\end{tikzpicture}
	}
	\hspace{1cm}
	\subfloat[\label{fig:EP_EW_disconnected}]{
		\begin{tikzpicture}[thick, vertex/.style={draw, shape=circle, fill=black, scale=0.7}]
		\draw (0, 0) circle (2.5cm);

		\draw [blue] (-1.5, 2) to[out=306.8, in=233.1] (1.5, 2);
		\draw [blue] (-1.5, -2) to[out=53.1, in=126.8] (1.5, -2);

		\draw [purple] (0, 1.3) to (0, -1.3);

		\node at (-2.9, 0) (A1) {\large $A_1$};
		\node at (2.9, 0) (A2) {\large $A_2$};
		\node at (0, 2.8) (Ac1) {\large $A^c$};
		\node at (0, -2.8) (Ac2) {\large $A^c$};

		\node at (0.7, 0) (EW) {$\Sigma_{A_1 : A_2}$};

		\node at (-1.1, 1.2) (A1p1) {$A_1'$};
		\node at (-1.1, -1.2) (A1p2) {$A_1'$};
		\node at (1.0, 1.2) (A2p1) {$A_2'$};
		\node at (1.0, -1.2) (A2p2) {$A_2'$};
		\end{tikzpicture}
	}
	\caption{The entanglement wedge cross-section for two different configurations of boundary subregions $A_1$ and $A_2$ in vacuum $AdS_3.$ In both (a) and (b), the entanglement
			wedge cross-section divides the entanglement wedge (the bulk region bounded by $A_1, A_2$, and RT surface of $A_1 \cup A_2$) into two disjoint regions, each of which
			contains only one of either $A_1$ or $A_2$ in its boundary. The subregions of each RT surface inherited from this partition have been instructively labeled $A_1'$ and $A_2'$
			for reasons explained in Section \ref{sec:TN_MEP}.	}
	\label{fig:EP_EW}
\end{center}
\end{figure}
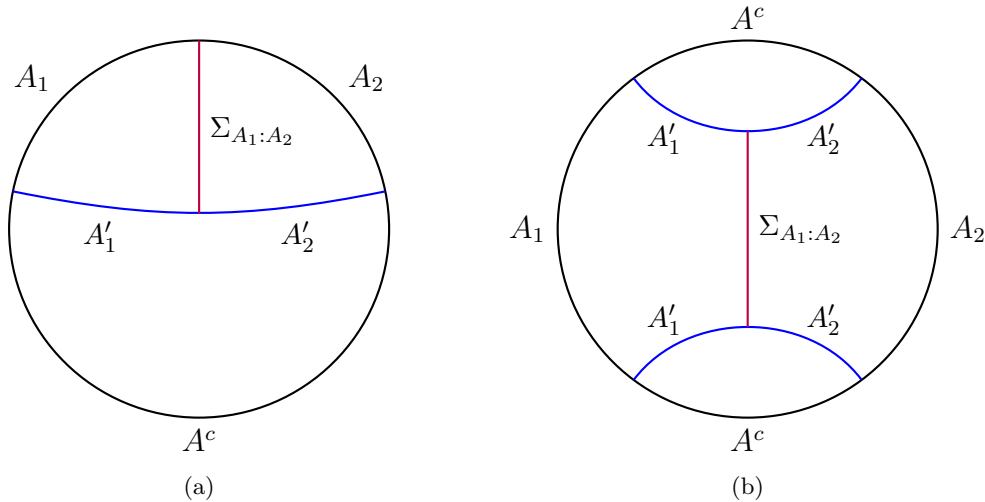

As with other information-theoretic quantities, one might hope that the entanglement of purification has a geometric dual in the context of AdS/CFT. For two subregions $A_1$ and $A_2$ of a holographic CFT state, it has recently been conjectured in \cite{TU2017, NDHZS2017} that $E_P(A_1:A_2)$ is given to leading order in $G_N$ by the area of the \emph{entanglement wedge cross-section}, $E_W(A_1 : A_2).$ Formally, $E_W(A_1 : A_2)$ is defined as
\begin{equation}
	E_W(A_1 : A_2)
		= \frac{\operatorname{area}(\Sigma_{A_1 : A_2})}{4 G_N},
\end{equation}
where $\Sigma_{A_1 : A_2}$ is the minimal surface anchored to the Ryu-Takayanagi surface of $A_1 \cup A_2$ that partitions the entanglement wedge into a portion whose boundary contains all of $A_1$ and a disjoint portion whose boundary contains all of $A_2.$ This surface is sketched in Figure \ref{fig:EP_EW} in vacuum $AdS_3$ for two different typical configurations of boundary regions $A_1$ and $A_2.$

If $A_1$ and $A_2$ are connected subregions of some larger connected region $A = A_1 \cup A_2$ (i.e., if $A_1$ and $A_2$ form a connected partition of $A$ as in Figure \ref{fig:EP_EW_connected}, then we call the state $\ket{\Psi}_{A_1 A_1' A_2 A_2'}$ that saturates the infimum in equation (\ref{eop}) the \emph{minimally entangled purification} (MEP) of the partition $A_1 : A_2.$\footnote{Of course, since \eqref{eop} contains an infimum rather than a minimum, it is not necessarily saturated by any state $\ket{\Psi}_{A_1 A_1' A_2 A_2'}$. In this case, we take the MEP to be a fixed state that saturates the infimum to within tolerance $\varepsilon$. Since one might expect that the infimum in \eqref{eop} could be saturated arbitrarily well by taking the purifying spaces $A_1'$ and $A_2'$ to be arbitrarily large, allowing this finite tolerance in the MEP prevents the dimensions of $A_1'$ and $A_2'$ from blowing up.}\textsuperscript{,}\footnote{A similar object, defined by a procedure in which one is also permitted to minimize over all possible partitions $A_1 : A_2$, has previously been studied in \cite{B2018}.}

The holographic entanglement of purification conjecture has been generalized to multipartite and conditional entropies (and their geometric duals) in \cite{bao2018holographic, bao2018conditional, umemoto2018entanglement, BCDR2018}. For the moment, however, we will concern ourselves only with the minimally entangled purification of a connected partition $A_1 : A_2$, and the associated entanglement of purification $E_P(A_1 : A_2)$.

\subsection{Tensor Networks from Minimally Entangled Purifications} \label{sec:TN_MEP}

Consider a holographic CFT state $\ket{\psi}$ with some subregion $A$ that is further divided into a connected partition $A_1 : A_2$, as sketched in Figure \ref{fig:EP_EW_connected}. The reduced state $\rho_{(A_1 A_2)}$ has a minimally entangled purification $\ket{\Psi}_{A_1 A_1' A_2 A_2'}$ chosen so that $S(A_1 A_1') = E_P(A_1 : A_2).$ According to the holographic entanglement of purification conjecture, this implies
\begin{equation}
	S(A_1 A_1') = E_W(A_1 : A_2) = \frac{\operatorname{area}(\Sigma_{A_1 : A_2})}{4 G_N}. \label{surface_state_RT}
\end{equation}

Examining Figure \ref{fig:EP_EW_connected}, it is easy to see that $\Sigma_{A_1 : A_2}$ is the minimal bulk surface anchored to the boundary region $A_1$ and a subregion of the Ryu-Takayanagi surface of $A_1 \cup A_2$, which we have instructively labeled $A_1'.$ Equation (\ref{surface_state_RT}) looks just like the Ryu-Takayanagi formula (\ref{RT}) if one supposes that $\ket{\Psi}_{A_1 A_1' A_2 A_2'}$ is a holographic state in some boundary theory with a domain corresponding to the codimension-2 bulk surface made up of $A_1$ and $A_2$ along with the Ryu-Takayanagi surface of their union.

Indeed, as explained in \cite{TU2017, NDHZS2017}, it makes sense to interpret the MEP $\ket{\Psi}_{A_1 A_1' A_2 A_2'}$ as a geometric state on some subregion of the bulk, where the auxiliary Hilbert spaces $\mathcal{H}_{A_1'}$ and $\mathcal{H}_{A_2'}$ are identified with the bulk surfaces $A_1'$ and $A_2'$, respectively. The conjectured surface-state correspondence \cite{MT2015} of holographic CFTs suggests that any codimension-2 surface in a holographic spacetime corresponds to some state in a boundary theory that encodes the physics in the entanglement wedge of that surface.\footnote{The ``entanglement wedge'' of a codimension-2 bulk surface $\Sigma$ is defined in analogy with the entanglement wedge of a boundary interval as the subregion of the bulk bounded by $\Sigma$ and by the minimal surface homologous to $\Sigma$ that shares its boundary.} If this is to believed, then it is natural to assume that the MEP of the partition $A_1 : A_2$ corresponds to a state on the codimension-2 bulk surface $\Sigma_{A_1 A_1' A_2 A_2'} = A_1 \cup A_2 \cup A_1' \cup A_2'.$

To make this proposal precise, we assume that the entanglement entropies of subsystems of the MEP are given at leading order by the Ryu-Takayanagi formula applied to the codimension-2 surface $\Sigma_{A_1 A_1' A_2 A_2'}$. In particular, since $A_1'$ and $A_2'$ are subregions of a minimal surface and hence minimal themselves, this assumption tells us that $A_1'$ and $A_2'$ are themselves the Ryu-Takayanagi surfaces for the Hilbert Space factors $\mathcal{H}_{A_1'}$ and $\mathcal{H}_{A_2'}$. In other words, their von Neumann entropies satisfy:
\begin{eqnarray}
	S(A_1')
		&=& \frac{\operatorname{area}(A_1')}{4 G_N} + o\left(\frac{1}{G_N}\right), \label{purified_entropy_1} \\
	S(A_2')
		&=& \frac{\operatorname{area}(A_2')}{4 G_N} + o\left(\frac{1}{G_N}\right). \label{purified_entropy_2}
\end{eqnarray}
We require one more assumption to build a tensor network using the holographic entanglement of purification, which is that the MEP has the same smooth min- and max-entropy properties \eqref{eq:smoothmin} and \eqref{eq:smoothmax} as holographic CFT states. In Section \ref{sec:smoothminmax}, we argued that this will be true for any state with von Neumann entropies given by equations of the form \eqref{purified_entropy_1} and \eqref{purified_entropy_2}, and with extensive R\'{e}nyi entropies that take the geometrical form given in equation \eqref{eq:holographic_renyi}. If the MEP is indeed a holographic state for the portion of the bulk bounded by $\Sigma_{A_1 A_1' A_2 A_2'}$, then it should satisfy both of these properties.

If the MEP has entropies given to leading order by the Ryu-Takayanagi formula on subregions of $\Sigma_{A_1 A_1' A_2 A_2'},$ and if the smooth min- and max-entropies of those subregions are also given by \eqref{eq:smoothmin} and \eqref{eq:smoothmax}, then the MEP satisfies all the conditions required to build a tree tensor network (cf. Section \ref{sec:tree_networks}).\footnote{As in Section \ref{sec:larger_tree_networks}, one must choose a root-leaf orientation for the MEP network in order to determine the order in which its subregions should be distilled. The choice of ordering will not matter for our purposes, though as usual it will determine which isometries are exact and which are only approximate.} This means that we can build a tree network for the MEP that matches the bulk geometry contained within $\Sigma_{A_1 A_1' A_2 A_2'}.$ Specifically, there exists some state $\ket{\Psi_{(\varepsilon)}}_{A_1 A_1' A_2 A_2'}$ that approximates the MEP with high fidelity with a tensor network representation
\begin{eqnarray}
	\Psi_{(\varepsilon)}^{A_1 A_1' A_2 A_2'}
		= & T^{A_1'}{}_{\gamma_1 f_1} U^{A_1}{}_{\bar{\gamma}_1 \bar{f}_1 \gamma_2 f_2} V^{A_2}{}_{\bar{\gamma}_2 \bar{f}_2 \gamma_3 f_3}
			W^{A_2'}{}_{\bar{\gamma}_3 \bar{f}_3} \times \nonumber \\
			& \quad
			\phi_{(1)}^{\gamma_1 \bar{\gamma}_1} \phi_{(2)}^{\gamma_2 \bar{\gamma}_2} \phi_{(3)}^{\gamma_3 \bar{\gamma}_3}
			\sigma_{(1)}^{f_1 \bar{f}_1} \sigma_{(2)}^{f_2 \bar{f}_2} \sigma_{(3)}^{f_3 \bar{f}_3} \label{eq:MEP_outer_product}
\end{eqnarray}
with bond dimensions given by:
\begin{eqnarray}
	\dim \mathcal{H}_{\gamma_1}
		& = & e^{S(A_1') - O(\sqrt{S(A_1')})}, \\
	\dim \mathcal{H}_{f_1}
		& = & e^{O(\sqrt{S(A_1')})}, \\
	\dim \mathcal{H}_{\gamma_2}
		& = & e^{S(A_1 A_1') - O(\sqrt{S(A_1 A_1')})}, \\
	\dim \mathcal{H}_{f_2}
		& = & e^{O(\sqrt{S(A_1 A_1')})}, \\
	\dim \mathcal{H}_{\gamma_3}
		& = & e^{S(A_2') - O(\sqrt{S(A_2')})}, \\
	\dim \mathcal{H}_{f_3}
		& = & e^{O(\sqrt{S(A_2')})}.
\end{eqnarray}

\begin{figure}[h]
\begin{center}
	\subfloat[\label{fig:MEP_tree}]{
		\begin{tikzpicture}[thick, >=latex, scale=1.2,
					cobwebs/.style={regular polygon, inner sep=0.5pt, fill=white, regular polygon sides=3, draw},
					EPR/.style={regular polygon, regular polygon sides=4, inner sep=0pt, fill=white, draw},
					tensor/.style={circle, draw, fill=white, minimum width=1.5em}
]
		\draw (-2.45, 0.5) to[out=78.5, in=180] (0, 2.5) to[out=0, in=101.5] (2.45, 0.5);

		\draw [blue] (2.45, 0.5) to[out=191.5, in=348.5] (-2.45, 0.5);

		\draw [purple] (0, 2.5) to (0, .2);

		\node at (-2.2, 2.5) (A1) {\large $A_1$};
		\node at (2.2, 2.5) (A2) {\large $A_2$};
		\node at (0, -3) (Ac) {\large};

		\node [EPR] at (-1.6, 0.4) (phi1) {$\phi$};
		\node [cobwebs] at (-0.9, 0.25) (sigma1) {$\sigma$};
		\node [EPR] at (0, 1) (phi2) {$\phi$};
		\node [cobwebs] at (0, 1.7) (sigma2) {$\sigma$};
		\node [EPR] at (1.6, 0.4) (phi3) {$\phi$};
		\node [cobwebs] at (0.9, 0.25) (sigma3) {$\sigma$};

		\node [tensor] at (-2, -0.35) (T) {$T$};
		\node [tensor] at (-1.2, 1.3) (U) {$U$};
		\node [tensor] at (1.2, 1.3) (V) {$V$};
		\node [tensor] at (2, -0.35) (W) {$W$};

		\node at (-1.5, -1.2) (A1p) {\large $A_1'$};
		\node at (1.5, -1.2) (A2p) {\large $A_2'$};

		\draw (T) edge[->] (A1p);
		\draw (U) edge[->] (A1);
		\draw (V) edge[->] (A2);
		\draw (W) edge[->] (A2p);

		\draw (phi1) edge[->] (T);
		\draw (phi1) edge[->] (U);
		\draw (sigma1) edge[->] (T);
		\draw (sigma1) edge[->] (U);
		\draw (phi2) edge[->] (U);
		\draw (sigma2) edge[->] (U);
		\draw (phi2) edge[->] (V);
		\draw (sigma2) edge[->] (V);
		\draw (phi3) edge[->] (V);
		\draw (sigma3) edge[->] (V);
		\draw (phi3) edge[->] (W);
		\draw (sigma3) edge[->] (W);
		\end{tikzpicture}
	}
	\hspace{1cm}
	\subfloat[\label{fig:MEP_loop}]{
		\begin{tikzpicture}[thick, >=latex, scale=1.2,
					cobwebs/.style={regular polygon, inner sep=0.5pt, fill=white, regular polygon sides=3, draw},
					EPR/.style={regular polygon, regular polygon sides=4, inner sep=0pt, fill=white, draw},
					tensor/.style={circle, draw, fill=white, minimum width=1.5em}]
		\draw (0, 0) circle (2.5cm);

		\draw [blue] (2.45, 0.5) to[out=191.5, in=348.5] (-2.45, 0.5);

		\draw [purple] (0, 2.5) to (0, .2);

		\node at (-2.2, 2.5) (A1) {\large $A_1$};
		\node at (2.2, 2.5) (A2) {\large $A_2$};
		\node at (0, -3) (Ac) {\large $A^c$};

		\node [EPR] at (-1.6, 0.4) (phi1) {$\phi$};
		\node [cobwebs] at (-0.9, 0.25) (sigma1) {$\sigma$};
		\node [EPR] at (0, 1) (phi2) {$\phi$};
		\node [cobwebs] at (0, 1.7) (sigma2) {$\sigma$};
		\node [EPR] at (1.6, 0.4) (phi3) {$\phi$};
		\node [cobwebs] at (0.9, 0.25) (sigma3) {$\sigma$};

		\node [tensor] at (-1.2, 1.3) (U) {$U$};
		\node [tensor] at (1.2, 1.3) (V) {$V$};
		\node [tensor] at (0, -1.1) (XTW) {$XTW$};

		\draw (U) edge[->] (A1);
		\draw (V) edge[->] (A2);
		\draw (XTW) edge[->] (Ac);

		\draw (phi1) edge[->] (XTW);
		\draw (phi1) edge[->] (U);
		\draw (sigma1) edge[->] (XTW);
		\draw (sigma1) edge[->] (U);
		\draw (phi2) edge[->] (U);
		\draw (sigma2) edge[->] (U);
		\draw (phi2) edge[->] (V);
		\draw (sigma2) edge[->] (V);
		\draw (phi3) edge[->] (V);
		\draw (sigma3) edge[->] (V);
		\draw (phi3) edge[->] (XTW);
		\draw (sigma3) edge[->] (XTW);
		\end{tikzpicture}
	}
	\caption{(a) A tensor network for the minimally entangled purification of two neighboring boundary regions in vacuum $AdS_3$, as given in equation 
			\eqref{eq:MEP_outer_product}. (b) The one-loop tensor network for the full boundary state given by equation \eqref{eq:MEP_isometry}, obtained by embedding the MEP
			isometrically into the global boundary. Note that in both figures, subscripts on the $\phi_{(i)}$ and $\sigma_{(i)}$ edge states have been suppressed. }
	\label{fig:MEP_networks}
\end{center}
\end{figure}
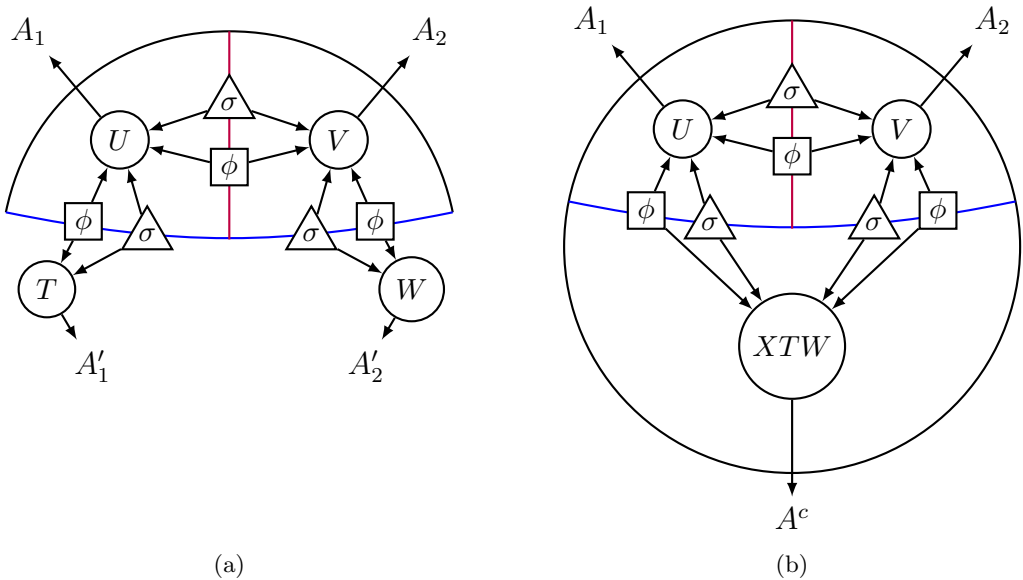

The outer product expression given in \eqref{eq:MEP_outer_product} is not particularly illuminating on its own, but has a natural interpretation in the geometric picture of tensor networks. In Figure \ref{fig:MEP_tree}, this tensor network is shown superposed over the geometric picture of the minimally entangled purification, with each network bond passing through its corresponding
bulk surface. Equations (\ref{surface_state_RT}), (\ref{purified_entropy_1}), and (\ref{purified_entropy_2})  imply that the bond dimensions of this tensor network match the areas of the surfaces $A_1', A_2',$ and $\Sigma_{A_1 : A_2}$ in Figure \ref{fig:EP_EW_connected}, justifying our interpretation of expression \eqref{eq:MEP_outer_product} as an approximate tensor network for the minimally entangled purification that matches the geometric properties of its holographic dual.

The tree network for the minimally entangled purification constitutes a tensor network for the ``top half'' of the bulk discretization shown in Figures \ref{fig:EP_EW_connected} and Figure \ref{fig:MEP_tree}. To find a geometric tensor network for the full boundary state, we need to find a tensor corresponding to the bulk region that lies between $A_1' \cup A_2'$ and the complementary boundary region $A^c$. Since the MEP $\ket{\Psi}_{A_1 A_1' A_2 A_2'}$ and the original global boundary state $\ket{\psi}_{A_1 A_2 A^c}$ are both purifications of the
reduced boundary state $\rho_{(A_1 A_2)}$, they are related by an isometry on the purifying space. That is, there exists an isometry\footnote{Formally, this isometry only exists if $\dim(\mathcal{H}_{A^c}) \geq \dim (\mathcal{H}_{A_1'} \otimes \mathcal{H}_{A_2'}).$ Since the dimension of the latter space is given to leading order by $e^{S(A^c)}$, and $S(A^c)$ is much smaller than $\log\dim(\mathcal{H}_{A^c})$, this bound is satisfied here and the isometry exists.}
\begin{equation}
	X: \mathcal{H}_{A_1'} \otimes \mathcal{H}_{A_2'} \hookrightarrow \mathcal{H}_{A^c}
\end{equation}
such that
\begin{equation}
	\ket{\psi}_{A_1 A_2 A^c}
		= (I_{A_1 A_2} \otimes X) \ket{\Psi}_{A_1 A_1' A_2 A_2'}.
\end{equation}
In tensor notation, this is
\begin{equation}
	\psi^{A_1 A_2 A^c}
		= X^{A^c}{}_{A_1' A_2'} \Psi^{A_1 A_1' A_2 A_2'}.
\end{equation}

Since $X$ is an isometry, the minimally entangled purification $\ket{\Psi}_{A_1 A_1' A_2 A_2'}$ can be replaced with the nearby state $\ket{\Psi_{(\varepsilon)}}_{A_1 A_1' A_2 A_2'}$ without any additional loss of precision. That is, the state
\begin{equation}
	\psi_{(\varepsilon)}^{A_1 A_2 A^c}
		= X^{A^c}{}_{A_1' A_2'} \Psi_{(\varepsilon)}^{A_1 A_1' A_2 A_2'}. \label{eq:MEP_isometry}
\end{equation}
approximates $\psi$ as well as $\Psi_{(\varepsilon)}$ approximates $\Psi.$ Plugging the tensor network expression for $\Psi_{(\varepsilon)}$ from equation \eqref{eq:MEP_outer_product} into equation \eqref{eq:MEP_isometry} (and contracting the tensors $X$, $T$, and $W$ into a single bulk tensor $XTW$) yields a tensor network description for the full (approximate) boundary state $\psi_{(\varepsilon)}$ with bond dimensions matching the areas of the surfaces $A_1'$, $A_2'$, and $\Sigma_{A_1 : A_2}$. This complete tensor network is sketched in Figure \ref{fig:MEP_loop}. By using the natural properties of the minimally entangled purification arising from the holographic entanglement of purification conjecture, we have managed to localize degrees of freedom within the Ryu-Takayanagi surface and hence to construct a non-tree tensor network for a generic holographic state with extremely high fidelity. 

With this tensor network now fully constructed, we must now ask an important question about its extension maps: in what sense can they be shown to be isometries? Are they exact isometries, as in Section \ref{sec:bipartite_networks}? Are they merely approximate isometries in the sense of Section \ref{sec:larger_tree_networks}? Or are they in fact neither of these things? Because the answer to this question will prove important both in constructing sub-AdS scale networks in Section \ref{sec:iteration} and in formulating the no-go theorem that we prove in Section \ref{sec:nogo_theorem}, we shall answer this question systematically for each extension map of the network shown in Figure \ref{fig:MEP_loop}.

Most of the extension maps in Figure \ref{fig:MEP_loop} (e.g. the maps outwards from $\Sigma_{A_1 : A_2}$ to $A_1 \cup A_1'$ and $A_2 \cup A_2'$ and the maps upwards from $A_1'$ to $A_1 \cup \Sigma_{A_1 : A_2}$ and from $A_2'$ to $A_1 \cup \Sigma_{A_1 : A_2}$) were also extension maps in the tree tensor network for the minimally entangled purification. As such, they are all at least approximate isometries in the sense of Section \ref{sec:larger_tree_networks}; depending on the order in which RT surfaces were added to the tree network for the MEP, several of them will be exact isometries. The extension map $XTW$ flowing downwards from the horizontal RT surface $A_1' \cup A_2'$ is not an extension map from the tree tensor network for the MEP; however, since it is a composition of an exact isometry $X$ with two MEP extension maps $T$ and $W$, and since $T$ and $W$ are exact isometries regardless of the order in which the edges in the tree tensor network for the MEP were distilled, $XTW$ is still an exact isometry in the final network.

There is one remaining extension map that one might also hope would be an isometry: the map upwards from $A_1' \cup A_2'$ to $A_1 \cup A_2$. In this case, we have no good argument that it should be an exact, or even an approximate, isometry. In particular, the $\ket{\phi}$ and $\ket{\sigma}$ edge states on $A_1'$ and $A_2'$ were constructed to approximate the reduced density matrices of the MEP on $\mathcal{H}_{A_1'}$ and $\mathcal{H}_{A_2'}$ \emph{individually}. The upwards extension map will be an approximate isometry if (and more importantly \emph{only} if) the product of these reduced density matrices approximates the reduced density matrix of the entire MEP on $\mathcal{H}_{A_1'} \otimes \mathcal{H}_{A_2'}$. Unfortunately, the assumptions we have made about the MEP thus far are only sufficient to show that the mutual information $I(A_1':A_2')$ is subleading in $G_N$; they do not imply that it is zero. We have no solid reason to believe that the reduced density matrix on $\mathcal{H}_{A_1'} \otimes \mathcal{H}_{A_2'}$ is actually close to a product state with respect to the trace norm. Furthermore, we will see in Section \ref{sec:nogo_theorem} that there is good reason to think that this \emph{cannot} be the case.\footnote{The fact that this map is not even an approximate isometry is somewhat problematic for interpreting the bottom-half state of the tensor network in Figure \ref{fig:MEP_loop} in terms of the surface-state correspondence. It implies that the bottom half state is \emph{not} an approximate purification of the reduced density matrix of the original holographic state on $\mathcal{H}_{A^c}$. However, we still hope that the relevant smooth entropies will behave correctly.}

Henceforth, we shall refer to an extension map of this kind as a ``moral'' isometry. In using this terminology, we mean that --- even though the extension map from $A_1' \cup A_2'$ to $A_1 \cup A_2$ in Figure \ref{fig:MEP_loop} is unlikely to be an exact or even approximate isometry --- we expect that small errors introduced in the ``bottom-half'' tensor $XTW$ will not blow up dramatically when mapped through the moral isometry to alter the ``top-half'' boundary state on $A_1 \cup A_2.$ Since the moral isometry preserves the normalization of the full-rank state $\ket{\phi_{(1)}}\ket{\sigma_{(1)}}\ket{\phi_{(2)}}\ket{\sigma_{(2)}}$, and also preserves the entanglement entropy of this state to leading order, it is tempting to think of the moral isometry as a combination of an exact isometry and some other, non-isometric operator that acts only on a subleading number of degrees of freedom. We will revisit the issue of moral isometries in Section \ref{sec:iteration}, where we invoke the moral isometry condition to argue that distilling the bottom-half tensor $XTW$ into a tree tensor network of its own will result in a global tensor network that still approximately reproduces the original boundary state of the CFT.


\subsection{Multiple Loops}

Of course, nothing in our construction thus far has limited us to considering the case of a bipartite boundary partition $A_1 : A_2.$ One might equally well wish to consider a more general multipartite partition, where a boundary region $A$ is partitioned into $n$ connected subregions as $A_1 : A_2 : \dots : A_n$. For simplicity, we assume that each subregion $A_i$ only shares a boundary with at most two neighbors, $A_{i-1}$ and $A_{i+1}.$ In $2+1$ spacetime dimensions, any connected partition can be ordered such that this is true. By analogy with the holographic entanglement of purification conjecture, one would expect that minimizing the quantity
\begin{equation}
	S(A_1 A_1') + S(A_1 A_1' A_2 A_2') + \dots + S(A_1 A_1' \dots A_{n-1} A_{n-1}') \label{multipartite_eop}
\end{equation}
over all possible purifications into spaces $A_1' \otimes \dots \otimes A_n'$ would correspond to minimizing the areas of surfaces that partition the entanglement wedge of $A = A_1 \cup \dots \cup A_n$ into $n$ distinct subregions.\footnote{In \cite{bao2018conditional}, a slightly different notion of multipartite entanglement of purification was shown to satisfy constraining inequalities involving linear combinations of entanglement entropies, which are also satisfied by the corresponding geometric quantity. Similar proof techniques, both holographic and information theoretic, should suffice to place analogous constraints on the quantity given in equation \eqref{multipartite_eop}. Since the holographic entanglement of purification conjecture was originally motivated in \cite{TU2017} by showing that the geometric bulk quantities and the information-theoretic boundary quantities satisfy the same constraints, our construction is equally well-motivated.} Each term in \eqref{multipartite_eop} should correspond to the area of one entanglement wedge cross-section, and since minimal surfaces cannot cross one another, the minimization of each area in the bulk can be performed independently up to subleading corrections.  If our extended conjecture is correct, this implies that each term in the entropy sum \eqref{multipartite_eop} can also be minimized independently, at leading order.

\begin{figure}[h]
\begin{center}
	\subfloat[\label{fig:tripartite_eop_partition}]{
		\begin{tikzpicture}[thick, scale=1.2, vertex/.style={draw, shape=circle, fill=black, scale=0.7}]
		\draw (0, 0) circle (2.5cm);

		\draw [blue] (2.45, 0.5) to[out=191.5, in=348.5] (-2.45, 0.5);

		\draw [purple] (-1.3, 2.14) to[out=301.3, in=86] (-0.84, .26);
		\draw [purple] (1.3, 2.14) to[out=238.7, in=94] (0.84, .26);

		\node at (-1.35, 1) (EW1) {$\Sigma_{A_1 : A_2}$};
		\node at (1.45, 1) (EW2) {$\Sigma_{A_2 : A_3}$};

		\node at (-2.3, 1.8) (A1) {\large $A_1$};
		\node at (0, 2.8) (A2) {\large $A_2$};
		\node at (2.3, 1.8) (A3) {\large $A_3$};
		\node at (0, -2.8) (Ac) {\large $A^c$};

		\node at (-1.6, 0) (A1p) {\large $A_1'$};
		\node at (0, -0.1) (A2p) {\large $A_2'$};
		\node at (1.6, 0) (A3p) {\large $A_3'$};

		\end{tikzpicture}
	}
	\hspace{1cm}
	\subfloat[\label{fig:tripartite_eop_tree}]{
		\begin{tikzpicture}[thick, >=latex, scale=1.2,
					cobwebs/.style={regular polygon, inner sep=0.5pt, fill=white, regular polygon sides=3, draw},
					EPR/.style={regular polygon, regular polygon sides=4, inner sep=0pt, fill=white, draw},
					tensor/.style={circle, draw, fill=white, minimum width=1.5em}]
		\draw (-2.45, 0.5) to[out=78.5, in=180] (0, 2.5) to[out=0, in=101.5] (2.45, 0.5);

		\draw [blue] (2.45, 0.5) to[out=191.5, in=348.5] (-2.45, 0.5);

		\draw [purple] (-1.3, 2.14) to[out=301.3, in=86] (-0.84, .26);
		\draw [purple] (1.3, 2.14) to[out=238.7, in=94] (0.84, .26);

		\node at (-2.3, 2) (A1) {\large $A_1$};
		\node at (0, 3) (A2) {\large $A_2$};
		\node at (2.3, 2) (A3) {\large $A_3$};
		\node at (0, -3) (Ac) {\large};

		\node [EPR] at (-1.3, 0.3) (phi1) {$\phi$};
		\node [cobwebs] at (-2, 0.4) (sigma1) {$\sigma$};
		\node [EPR] at (-0.85, 1) (phi2) {$\phi$};
		\node [cobwebs] at (-1.05, 1.6) (sigma2) {$\sigma$};
		\node [EPR] at (1.3, 0.3) (phi3) {$\phi$};
		\node [cobwebs] at (2, 0.4) (sigma3) {$\sigma$};
		\node [EPR] at (0.86, 1) (phi4) {$\phi$};
		\node [cobwebs] at (1.05, 1.6) (sigma4) {$\sigma$};
		\node [EPR] at (-0.4, 0.2) (phi5) {$\phi$};
		\node [cobwebs] at (0.4, 0.2) (sigma5) {$\sigma$};

		\node [tensor] at (-2, -0.35) (V1p) {$V_1'$};
		\node [tensor] at (-1.6, 1) (V1) {$V_1$};
		\node [tensor] at (1.6, 1) (V3) {$V_3$};
		\node [tensor] at (2, -0.35) (V3p) {$V_3'$};
		\node [tensor] at (0, 1.3) (V2) {$V_2$};
		\node [tensor] at (0, -0.5) (V2p) {$V_2'$};

		\node at (-1.5, -1.2) (A1p) {\large $A_1'$};
		\node at (0, -1.4) (A2p) {\large $A_2'$};
		\node at (1.5, -1.2) (A3p) {\large $A_3'$};

		\draw (V1p) edge[->] (A1p);
		\draw (V1) edge[->] (A1);
		\draw (V2) edge[->] (A2);
		\draw (V2p) edge[->] (A2p);
		\draw (V3) edge[->] (A3);
		\draw (V3p) edge[->] (A3p);

		\draw (phi1) edge[->] (V1p);
		\draw (phi1) edge[->] (V1);
		\draw (sigma1) edge[->] (V1p);
		\draw (sigma1) edge[->] (V1);
		\draw (phi2) edge[->] (V1);
		\draw (sigma2) edge[->] (V1);
		\draw (phi2) edge[->] (V2);
		\draw (sigma2) edge[->] (V2);
		\draw (phi3) edge[->] (V3);
		\draw (sigma3) edge[->] (V3);
		\draw (phi3) edge[->] (V3p);
		\draw (sigma3) edge[->] (V3p);
		\draw (phi4) edge[->] (V2);
		\draw (sigma4) edge[->] (V2);
		\draw (phi4) edge[->] (V3);
		\draw (sigma4) edge[->] (V3);
		\draw (phi5) edge[->] (V2);
		\draw (sigma5) edge[->] (V2);
		\draw (phi5) edge[->] (V2p);
		\draw (sigma5) edge[->] (V2p);
		\end{tikzpicture}
	}	\caption{A rough sketch of the procedure for constructing a ``3-to-one'' network for a holographic CFT state. The boundary subregion $A$ is given a connected partition $A_1 :
				A_2 : A_3$, and a minimally entangled purification for this partition is found by minimizing the sum $S(A_1 A_1') + S(A_1 A_1' A_2 A_2').$ In (a), this partition
				is shown along with the surfaces $\Sigma_{A_1 : A_2}$ and $\Sigma_{A_2 : A_3}$ whose minimal areas should correspond to the minimization of this sum. If the MEP
				has entropies corresponding to areas of surfaces in the entanglement wedge of $A_1 \cup A_2 \cup A_3$, then a ``top-half'' tree tensor network can be constructed
				as in (b), which can then be completed by an isometry on the purifying space to obtain a tensor network for the global CFT state.}
	\label{fig:tripartite_eop_network}
\end{center}
\end{figure}
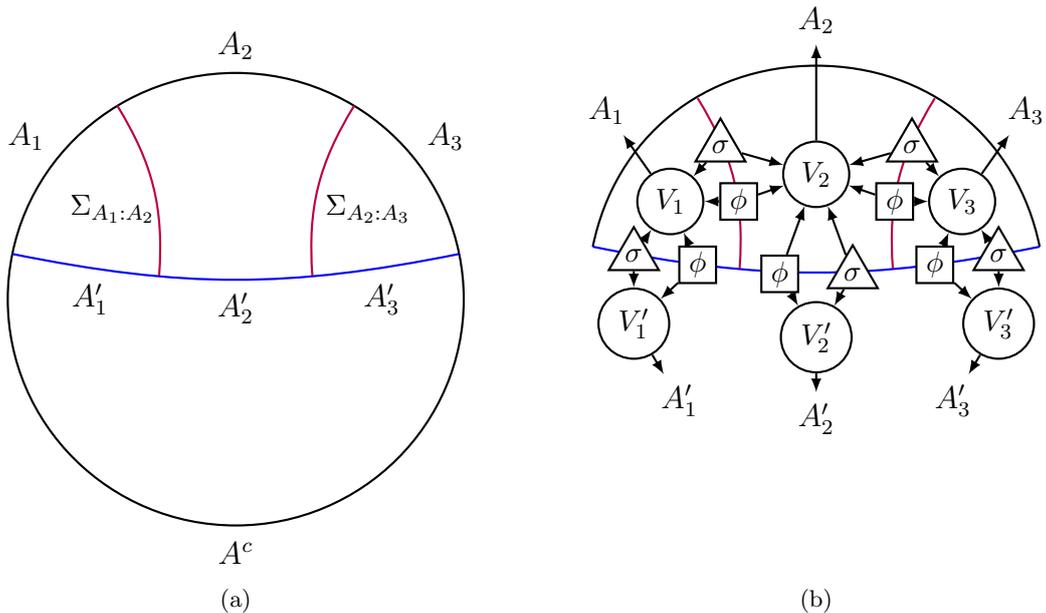

In the special case of two subregions (i.e., $n=2$), the quantity given in \eqref{multipartite_eop} reduces to the one minimized in defining the entanglement of purification \eqref{eop}. If the MEP of this partition has entropies given by the areas of extremal surfaces contained within the entanglement wedge, then one can readily construct an ``$n$-to-one'' network for the CFT state $\ket{\psi}$ by repeating the procedure detailed above. This is sketched for $n=3$ in Figure \ref{fig:tripartite_eop_network}.

Note that for $n \geq 3$, the $n$-to-one network contains bonds that correspond to extremal surface subregions with areas that remain finite even when the ultraviolet CFT regulator and corresponding bulk radius regulator are removed. If the holographic entanglement of purification conjecture and its extensions hold down to the string and Planck scales, as assumed, then the areas of these subregions can be chosen to be arbitrarily small relative to the AdS scale by choosing suitably small boundary subregions (so long as we remain above the string and Planck scales).\footnote{One important exception occurs when two neighboring entanglement wedge cross-sections $\Sigma_{A_1 : A_2}$ and $\Sigma_{A_2 : A_3}$ undergo a ``phase transition'' in the sense that they ``jump'' discontinuously across the RT surface even when the middle boundary region $A_2$ is made arbitrarily small. In this case, the region of the RT surface that is ``skipped over'' by this phase transition cannot be directly probed by a single application of the holographic entanglement of purification, and may in fact have an area well above the AdS scale. Nevertheless, techniques in Section \ref{sec:iteration} should still allow us to construct a sub-AdS tensor network within such a region by using multiple, iterated applications of the holographic entanglement of purification conjecture.  (This complication does not arise for bulk geometries that are close enough to a 2+1 AdS vacuum.)} The tensor network would therefore be capturing the bulk geometry at sub-AdS scales.

Thus far, we have implicitly assumed that we are working in $2+1$ spacetime dimensions. While the construction detailed above will certainly work in higher-dimensional spacetimes, it would no longer be completely accurate to claim that the $n$-to-one network captures the geometry at sub-AdS scales; while RT surface subregions can be chosen with finite, sub-AdS width, they also have at least one transverse direction that extends all the way to the boundary of the spacetime. Localizing the information on a single Ryu-Takayanagi surface to bounded, sub-AdS bulk regions in higher-dimensional spacetimes is a subtle procedure, and requires techniques from Section \ref{sec:iteration}. We will therefore comment on this generalization briefly in Section \ref{sec:arb_disc}.

\section{Iteration and Sub-AdS Locality} \label{sec:iteration}

In Section  \ref{sec:tree_networks}, we showed that the Ryu-Takayanagi formula, together with constraints on the smooth min- and max-entropies that follow from the extensive growth of the ordinary R\'{e}nyi entropies, is sufficient to construct geometrically appropriate tensor networks corresponding to an arbitrary discretization of the bulk by non-intersecting extremal surfaces. The resulting tensor networks are always tree tensor networks, where each bond of the network is associated to an entire extremal surface. In tree tensor networks, information is never localized within a single Ryu-Takayanagi surface.

In Section \ref{sec:MEP}, however, we were able to show that the holographic entanglement of purification conjecture can be used to associate network bonds to subregions of a single Ryu-Takayanagi surface. By assuming a natural extension of the holographic entanglement of purification conjecture to multipartite partitions of the boundary, these subregions could be made to have finite size even when the CFT and bulk regulators are removed. If the holographic entanglement of purification conjecture for multipartite boundary partitions holds up to stringy and quantum corrections, then these extremal surface subregions can be made arbitrarily small compared to the AdS scale in the semiclassical limit $G_N \to 0$ and $\lambda \to \infty$.\footnote{Note that since the bond dimension of the corresponding network edge goes like $e^{\operatorname{area}/4 G_N}$, the bond dimension will still diverge in the semiclassical limit.}

We would like to go further by achieving some form of sub-AdS locality not only in the sense of dividing bonds along a single Ryu-Takayanagi surface, but in the general granularity of the network. More precisely, we would like to construct tensor networks where each tensor is associated to a bulk subregion that occupies a volume well below $\ell_{AdS}^{d-1}$. We approach this problem by proposing a procedure to construct a tensor network for discretizations of the bulk whose discretization scale lies well below $\ell_{AdS}.$ 

\subsection{The Four-Tensor Network} \label{sec:four_tensor}

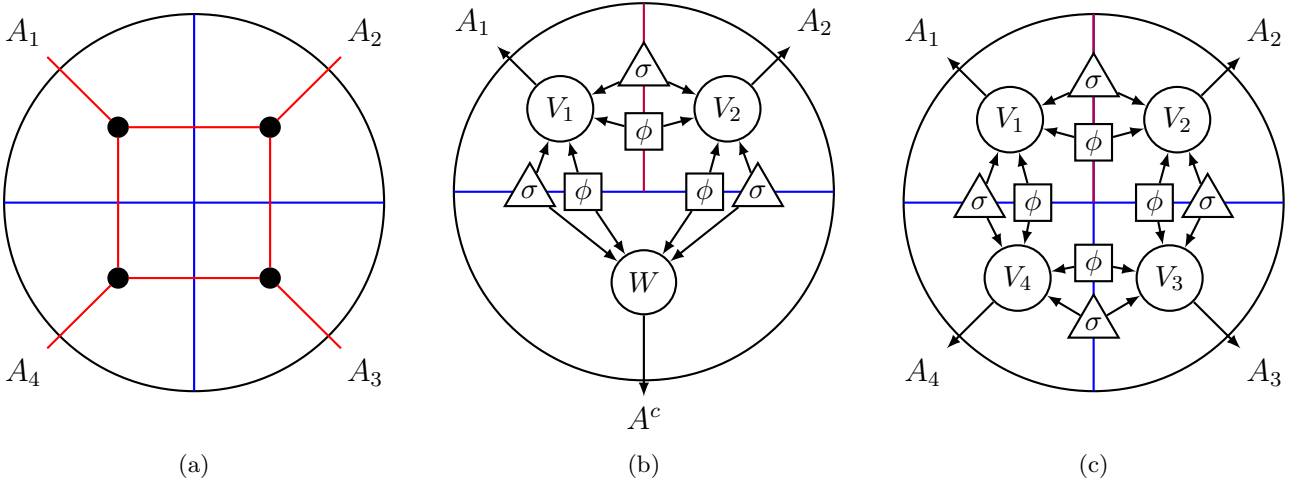
\begin{figure}[h]
\begin{center}
	\makebox[\textwidth][c]{
	\subfloat[\label{fig:plus_diagram}]{
		\begin{tikzpicture}[thick, vertex/.style={draw, shape=circle, fill=black, scale=0.7}]
		\draw (0, 0) circle (2.5cm);

		\draw [blue] (0, 2.5) to [out=270, in=90] (0, -2.5);
		\draw [blue] (2.5, 0) to [out=180, in=0] (-2.5, 0);

		\node [vertex] at (-1, 1) (1) {};
		\node [vertex] at (1, 1) (2) {};
		\node [vertex] at (1, -1) (3) {};
		\node [vertex] at (-1, -1) (4) {};

		\draw [red] (1) to (2);
		\draw [red] (2) to (3);
		\draw [red] (3) to (4);
		\draw [red] (4) to (1);

		\node at (-2.25, 2.25) (A1) {\large $A_1$};
		\node at (2.25, 2.25) (A2) {\large $A_2$};
		\node at (2.25, -2.25) (A3) {\large $A_3$};
		\node at (-2.25, -2.25) (A4) {\large $A_4$};
		\node at (0, -3) (Ac) {\large};

		\draw [red] (1) to (A1);
		\draw [red] (2) to (A2);
		\draw [red] (3) to (A3);
		\draw [red] (4) to (A4);
		\end{tikzpicture}
	}
	\hspace{0.3cm}
	\subfloat[\label{fig:plus_network_one_loop}]{
		\begin{tikzpicture}[thick, >=latex,
					cobwebs/.style={regular polygon, inner sep=0.5pt, fill=white, regular polygon sides=3, draw},
					EPR/.style={regular polygon, regular polygon sides=4, inner sep=0pt, fill=white, draw},
					tensor/.style={circle, draw, fill=white, minimum width=1.5em}]
		\draw (0, 0) circle (2.5cm);

		\draw [blue] (2.5, 0) to[out=180, in=0] (-2.5, 0);

		\draw [purple] (0, 2.5) to (0, 0);

		\node at (-2.25, 2.25) (A1) {\large $A_1$};
		\node at (2.25, 2.25) (A2) {\large $A_2$};
		\node at (0, -3) (Ac) {\large $A^c$};

		\node [EPR] at (-0.8, 0) (phi1) {$\phi$};
		\node [cobwebs] at (-1.5, 0.0) (sigma1) {$\sigma$};
		\node [EPR] at (0, 0.8) (phi2) {$\phi$};
		\node [cobwebs] at (0, 1.6) (sigma2) {$\sigma$};
		\node [EPR] at (0.8, 0) (phi3) {$\phi$};
		\node [cobwebs] at (1.5, 0) (sigma3) {$\sigma$};

		\node [tensor] at (-1.1, 1.1) (V1) {$V_1$};
		\node [tensor] at (1.1, 1.1) (V2) {$V_2$};
		\node [tensor] at (0, -1.2) (W) {$W$};

		\draw (V1) edge[->] (A1);
		\draw (V2) edge[->] (A2);
		\draw (W) edge[->] (Ac);

		\draw (phi1) edge[->] (W);
		\draw (phi1) edge[->] (V1);
		\draw (sigma1) edge[->] (W);
		\draw (sigma1) edge[->] (V1);
		\draw (phi2) edge[->] (V1);
		\draw (sigma2) edge[->] (V1);
		\draw (phi2) edge[->] (V2);
		\draw (sigma2) edge[->] (V2);
		\draw (phi3) edge[->] (V2);
		\draw (sigma3) edge[->] (V2);
		\draw (phi3) edge[->] (W);
		\draw (sigma3) edge[->] (W);
		\end{tikzpicture}
	}
	\hspace{0.3cm}
	\subfloat[\label{fig:plus_network_complete}]{
		\begin{tikzpicture}[thick, >=latex,
					cobwebs/.style={regular polygon, inner sep=0.5pt, fill=white, regular polygon sides=3, draw},
					EPR/.style={regular polygon, regular polygon sides=4, inner sep=0pt, fill=white, draw},
					tensor/.style={circle, draw, fill=white, minimum width=1.5em}]
		\draw (0, 0) circle (2.5cm);

		\draw [blue] (2.5, 0) to[out=180, in=0] (-2.5, 0);
		\draw [blue] (0, 2.5) to[out=270, in=90] (0, -2.5);

		\draw [purple] (0, 2.5) to (0, 0);

		\node at (-2.25, 2.25) (A1) {\large $A_1$};
		\node at (2.25, 2.25) (A2) {\large $A_2$};
		\node at (2.25, -2.25) (A3) {\large $A_3$};
		\node at (-2.25, -2.25) (A4) {\large $A_4$};
		\node at (0, -3) (Ac) {\large};

		\node [EPR] at (-0.8, 0) (phi1) {$\phi$};
		\node [cobwebs] at (-1.5, 0.0) (sigma1) {$\sigma$};
		\node [EPR] at (0, 0.8) (phi2) {$\phi$};
		\node [cobwebs] at (0, 1.6) (sigma2) {$\sigma$};
		\node [EPR] at (0.8, 0) (phi3) {$\phi$};
		\node [cobwebs] at (1.5, 0) (sigma3) {$\sigma$};
		\node [EPR] at (0, -0.8) (phi4) {$\phi$};
		\node [cobwebs] at (0, -1.6) (sigma4) {$\sigma$};

		\node [tensor] at (-1.1, 1.1) (V1) {$V_1$};
		\node [tensor] at (1.1, 1.1) (V2) {$V_2$};
		\node [tensor] at (1, -1) (V3) {$V_3$};
		\node [tensor] at (-1, -1) (V4) {$V_4$};

		\draw (V1) edge[->] (A1);
		\draw (V2) edge[->] (A2);
		\draw (V3) edge[->] (A3);
		\draw (V4) edge[->] (A4);

		\draw (phi1) edge[->] (V4);
		\draw (phi1) edge[->] (V1);
		\draw (sigma1) edge[->] (V4);
		\draw (sigma1) edge[->] (V1);
		\draw (phi2) edge[->] (V1);
		\draw (sigma2) edge[->] (V1);
		\draw (phi2) edge[->] (V2);
		\draw (sigma2) edge[->] (V2);
		\draw (phi3) edge[->] (V2);
		\draw (sigma3) edge[->] (V2);
		\draw (phi3) edge[->] (V3);
		\draw (sigma3) edge[->] (V3);
		\draw (phi4) edge[->] (V3);
		\draw (sigma4) edge[->] (V3);
		\draw (phi4) edge[->] (V4);
		\draw (sigma4) edge[->] (V4);
		\end{tikzpicture}
	}
	}
	\caption{Constructing a tensor network for the four-tensor discretization by an iterative procedure. (a) A discretization of the bulk into four regions by two intersecting Ryu-Takayanagi
			surfaces, along with the corresponding dual graph. (b) The one-loop network for the boundary partition $A_1 : A_2$ obtained from the minimally entangled purification of
			this partition. (c) The full four-tensor network, obtained from (b) by distilling entanglement out of the bottom tensor $W$.}
\end{center}
\end{figure}

We begin by considering the simplest tensor network that our prior techniques were unable to address, namely the four-tensor square network shown in Figure \ref{fig:plus_diagram} that corresponds to a discretization of the bulk by two complete, intersecting extremal surfaces. One way to construct such a network involves a process of iteration: begin by constructing a one-loop network like the one shown in Figure \ref{fig:MEP_loop}, where the ``bottom half'' of the discretization in \ref{fig:plus_diagram} is represented by a single tensor, then divide this tensor into two tensors that represent the bulk subregions in the discretization. We reproduce the one-loop network in Figure \ref{fig:plus_network_one_loop} with a relabeling of the tensors that is slightly more convenient for our current purposes. This process of iteration is functionally almost identical to the inductive procedure for constructing tree networks detailed in Section \ref{sec:larger_tree_networks}.

The bulk state assigned to the ``bottom half'' of the one-loop network in Figure \ref{fig:plus_network_one_loop}, as defined in Section \ref{sec:tree_networks}, is the state comprised of the bulk tensor $W$ along with the edge states $\phi$ and $\sigma$ that correspond to neighboring extremal surfaces. As in \ref{sec:tree_networks}, this state can be approximated by a tree tensor network in which the tensor $W$ is replaced by an expression of the form
\begin{equation}
	W^{A_3 A_4}{}_{A_1' A_2'} \approx V_3^{A_3}{}_{A_2' \gamma f} V_4^{A_4}{}_{A_1' \bar{\gamma} \bar{f}} \phi^{\gamma \bar{\gamma}} \sigma^{f \bar{f}},
\end{equation}
where $\phi$ is a maximally entangled state on a space of dimension $e^{S^\varepsilon_{\text{max}}(A_1' A_4)},$ and $S^{\varepsilon}_{\text{max}}(A_1' A_4)$ is the smooth max-entropy of the bottom-half bulk state in the subregion $A_1' A_4.$ The operators $V_3$ and $V_4$ are not themselves isometries. However, when combined with the edge state $\phi$ and $\sigma$ operators on the horizontal edges of the network, they become exact isometries flowing outwards from the newly created vertical edge.

Such a network can always be constructed for any state by entanglement distillation; however, the resulting network is only geometrically appropriate for the discretization given in Figure \ref{fig:plus_diagram} if the smooth min- and max-entropies satisfy
\begin{equation} \label{eq:bottom_half_min_max}
	S^{\varepsilon}_{\text{max}}(A_1' A_4)
		= \frac{\operatorname{Area}(\Sigma_{A_3 :A_4})}{4 G_N} + O\left(\frac{1}{\sqrt{G_N}}\right)
	\quad
	\text{and}
	\quad
	S^{\varepsilon}_{\text{min}}(A_1' A_4)
		= \frac{\operatorname{Area}(\Sigma_{A_3 :A_4})}{4 G_N} + O\left(\frac{1}{\sqrt{G_N}}\right).
\end{equation}

Arguments given in Section \ref{sec:smoothminmax} would imply equations \eqref{eq:bottom_half_min_max} if the tensor network state for the bulk subregion represented by $W$ has entropies given by the areas of extremal surfaces in the entanglement wedge of $A_3 \cup A_4$. However, unlike the minimally entangled purification, this state has not previously been conjectured to have this property. Nevertheless, such a conjecture is closely analogous to the conjectures used in Section \ref{sec:MEP}, and follows intuitively from the surface-state conjecture \cite{MT2015}. The state defined by $W$ and its neighboring edge states is naturally associated to the ``bottom half'' bulk subregion, and so the area of the extremal surface dividing $A_3$ and $A_4$ is unquestionably the ``natural'' geometric quantity that would be associated to the smooth min- and max-entropies of this state.

If this is indeed the case, and the bulk tensor $W$ can be distilled into a tree network for the bottom half of the four-tensor discretization from Figure \ref{fig:plus_diagram}, then the resulting expression is a tensor network for the four-tensor discretization whose bond dimensions match the areas of extremal surfaces in the bulk. This completed network is sketched in Figure \ref{fig:plus_network_complete}. The only remaining question to ask is whether the boundary state of this network accurately reproduces the original ``target'' CFT state.

Arguments given in Section \ref{sec:MEP} imply that the one-loop network of Figure \ref{fig:plus_network_one_loop} well approximates the original CFT state. Since the tree network for the bottom half state well approximates the original bottom half state in the one-loop network of Figure \ref{fig:plus_network_one_loop}, the final network should still well approximate the boundary CFT state, so long as the error induced in the bottom half state from its approximation as a tree network doesn't dramatically increase in size when the rest of the network is added. 

Of course, here we run into something of a problem. The extension map upwards from the bottom half state is not an exact isometry. Indeed, as we discussed in Section \ref{sec:TN_MEP}, it is not even an approximate isometry in the sense of Section \ref{sec:larger_tree_networks}; instead we proposed that it should be called a ``moral'' isometry. We therefore will not have good control over the total accumulated error between the state produced by the four-tensor network and the target state. This is in contrast to the networks in Section \ref{sec:tree_networks} (and Section \ref{sec:MEP}), where we could precisely control the total error that could be accumulated, so long as the RT surfaces were added in an appropriate order.

On the other hand, the edge states on the horizontal RT surface and the top half state are both normalized quantum states. This means that the upwards flowing map preserves the norm of the edge states on the horizontal RT surface, which are fully entangled. Hence we can be relatively hopeful that a generic perturbation to the bottom half state will not be dramatically blown up in size by the upwards flowing map, and the final state produced by the network in Figure \ref{fig:plus_network_complete} should approximately reproduce the original CFT state. 

Another way of seeing that the four-tensor network should approximately reproduce the correct boundary state is to compare the exact isometry $W$ in Figure \ref{fig:plus_network_one_loop} to the final downwards-flowing extension map made up of tensors $V_3$ and $V_4$ and bottom-half edge states $\ket{\phi}$ and $\ket{\sigma}$ in Figure \ref{fig:plus_network_complete}. For the sake of this argument, let the full downwards-flowing extension map be denoted by $B$. Since the network in Figure \ref{fig:plus_network_complete} was obtained from the network in Figure \ref{fig:plus_network_one_loop} by tree network distillation, $B$ and $W$ must have approximately the same action on the reduced density matrices of the $\ket{\phi}$ and $\ket{\sigma}$ states on the horizontal RT surface, i.e.,
\begin{equation} \label{eq:XW_marginal}
	\lVert (B - W) \phi_1 \sigma_1 \phi_2 \sigma_2 \rVert_2 \ll 1.
\end{equation}
The condition for the four-tensor network to approximately reproduce the original holographic state is for $B$ and $W$ to approximately agree on the \emph{full} state for the top half of the four-tensor network, i.e.,
\begin{equation} \label{eq:XW_joint}
	\lVert (B - W) \rho_T^{1/2} \rVert_2 \ll 1,
\end{equation}
where $\rho_T$ is the truncated tensor network state for the top half of Figures \ref{fig:plus_network_one_loop} and \ref{fig:plus_network_complete}. These conditions are inequivalent, as we do not expect $\rho_T^{1/2}$ to be a product state across the two halves of the horizontal RT surface, because the upwards-flowing extension map is only a moral isometry. However, our version of the holographic entanglement of purification conjecture implies that the correlations between these two halves are subleading in $G_N$ in the sense of the mutual information. Equation \eqref{eq:XW_marginal} can be interpreted as taking a weighted average of $(B-W)$ over the product of the marginal distributions on each half of $\rho_T$, while equation \eqref{eq:XW_joint} can be interpreted as taking a weighted average over the joint distribution. We expect that, barring unlikely disasters, equation \eqref{eq:XW_marginal}, which follows from conjectures given in Section \ref{sec:MEP}, should  imply equation \eqref{eq:XW_joint}.

We will revisit the question of exact, approximate, and moral isometries in Section \ref{sec:nogo_theorem}, where it is shown that any geometrically appropriate tensor network of the form shown in Figure \ref{fig:plus_network_complete} \emph{must} have some moral isometries, as constructing such a tensor network with stronger bulk-to-boundary isometry conditions is inconsistent with the dynamics of the original CFT state on the boundary.

\subsection{Arbitrarily Fine Discretizations} \label{sec:arb_disc}

\begin{figure}[h]
\begin{center}
	\subfloat[\label{fig:generic_discretization_grid}]{
	\begin{tikzpicture}[thick, vertex/.style={draw, shape=circle, scale=0.4, fill=black}]
	\draw (0, 0) circle (2.5cm);

	\draw [blue] (-1.5, 2) arc (180+36.9:360-36.9:1.875);
	\draw [blue] (-2.3, 0.98) arc (180+66.9:360-66.9:5.87);
	\draw [blue] (-2.3, -0.98) arc (180-66.9:66.9:5.87);
	\draw [blue] (-1.5, -2) arc (180-36.9:36.9:1.875);

	\node at (-1.5, 2.6) (A) {};
	\node at (-0.5, 2.9) (B) {};
	\node at (0.5, 2.9) (C) {};
	\node at (1.5, 2.6) (D) {};
	\node at (-2.3, 1.8) (E) {};
	\node at (2.3, 1.8) (F) {};
	\node at (-3, 0) (G) {};
	\node at (3, 0) (H) {};
	\node at (-2.3, -1.8) (I) {};
	\node at (2.3, -1.8) (J) {};
	\node at (-1.5, -2.6) (K) {};
	\node at (-0.5, -2.9) (L) {};
	\node at (0.5, -2.9) (M) {};
	\node at (1.5, -2.6) (N) {};

	\draw [purple] (-1, 2.29) arc (23.59:13.92:5.72);
	\draw[purple] (-0.69, 1.39) arc (6.02:-6.78:3.63);
	\draw [purple] (1, 2.29) arc (180-23.59:180-13.92:5.72);
	\draw[purple] (0.69, 1.39) arc (180-6.02:180+6.78:3.63);

	\draw [purple] (-0.692, 0.55) arc (7.14:-7.14:4.423);
	\draw [purple] (0.692, 0.55) arc (180-7.14:180+7.14:4.423);

	\draw [purple] (-1, -2.29) arc (-23.59:-13.92:5.72);
	\draw[purple] (-0.69, -1.39) arc (-6.02:6.78:3.63);
	\draw [purple] (1, -2.29) arc (180+23.59:180+13.92:5.72);
	\draw[purple] (0.69, -1.39) arc (180+6.02:180-6.78:3.63);

	\draw [purple] (0, 2.5) to (0, -2.5);

	\end{tikzpicture}
	}
	\hspace{1cm}
	\subfloat[\label{fig:generic_discretization_graph}] {
	\begin{tikzpicture}[thick, vertex/.style={draw, shape=circle, scale=0.4, fill=black}]
	\draw (0, 0) circle (2.5cm);

	\node [vertex] at (-1.1, 1.9) (1) {};
	\node [vertex] at (-0.4, 1.85) (2) {};
	\node [vertex] at (-1.4, 1.2) (3) {};
	\node [vertex] at (-0.35, 0.9) (4) {};
	\node [vertex] at (-1.5, 0) (5) {};
	\node [vertex] at (-0.35, 0) (6) {};
	\node [vertex] at (-1.4, -1.2) (7) {};
	\node [vertex] at (-0.35, -0.9) (8) {};
	\node [vertex] at (-1.1, -1.9) (9) {};
	\node [vertex] at (-0.4, -1.85) (10) {};
	\node [vertex] at (1.1, 1.9) (11) {};
	\node [vertex] at (0.4, 1.85) (12) {};
	\node [vertex] at (1.4, 1.2) (13) {};
	\node [vertex] at (0.35, 0.9) (14) {};
	\node [vertex] at (1.5, 0) (15) {};
	\node [vertex] at (0.35, 0) (16) {};
	\node [vertex] at (1.4, -1.2) (17) {};
	\node [vertex] at (0.35, -0.9) (18) {};
	\node [vertex] at (1.1, -1.9) (19) {};
	\node [vertex] at (0.4, -1.85) (20) {};

	\node at (-1.5, 2.6) (A) {};
	\node at (-0.5, 2.9) (B) {};
	\node at (0.5, 2.9) (C) {};
	\node at (1.5, 2.6) (D) {};
	\node at (-2.3, 1.8) (E) {};
	\node at (2.3, 1.8) (F) {};
	\node at (-3, 0) (G) {};
	\node at (3, 0) (H) {};
	\node at (-2.3, -1.8) (I) {};
	\node at (2.3, -1.8) (J) {};
	\node at (-1.5, -2.6) (K) {};
	\node at (-0.5, -2.9) (L) {};
	\node at (0.5, -2.9) (M) {};
	\node at (1.5, -2.6) (N) {};

	\draw [red] (1) to (2) to (4) to (3) to (5) to (6) to (8) to (7) to (9) to (10) to (20) to (19) to (17) to (18) to (16) to (15) to (13) to (14) to (12) to (11);
	\draw [red] (1) to (3);
	\draw [red] (4) to (6);
	\draw [red] (5) to (7);
	\draw [red] (2) to (12);
	\draw [red] (4) to (14);
	\draw [red] (6) to (16);
	\draw [red] (8) to (18);
	\draw [red] (8) to (10);
	\draw [red] (18) to (20);
	\draw [red] (11) to (13);
	\draw [red] (14) to (16);
	\draw [red] (15) to (17);

	\draw [red] (1) to (A);
	\draw [red] (2) to (B);
	\draw [red] (3) to (E);
	\draw [red] (5) to (G);
	\draw [red] (7) to (I);
	\draw [red] (9) to (K);
	\draw [red] (10) to (L);
	\draw [red] (12) to (C);
	\draw [red] (11) to (D);
	\draw [red] (13) to (F);
	\draw [red] (15) to (H);
	\draw [red] (17) to (J);
	\draw [red] (20) to (M);
	\draw [red] (19) to (N);

	\end{tikzpicture}
	}
	\caption{(a) A grid discretization for vacuum $AdS_3$, for which one can find a tensor network by iterating an ``$n$-to-one'' loop network from the top to the bottom of the grid. As 
			explained in the text, not all surfaces in this discretization will meet at right angles. If the network is constructed ``top-to-bottom'', then the bottom of each vertical
			segment will meet the corresponding horizontal RT surface at a right angle.
			(b) The dual graph of this discretization, which forms the underlying geometry for a holographic tensor network.}
	\label{fig:generic_discretization}
\end{center}
\end{figure}
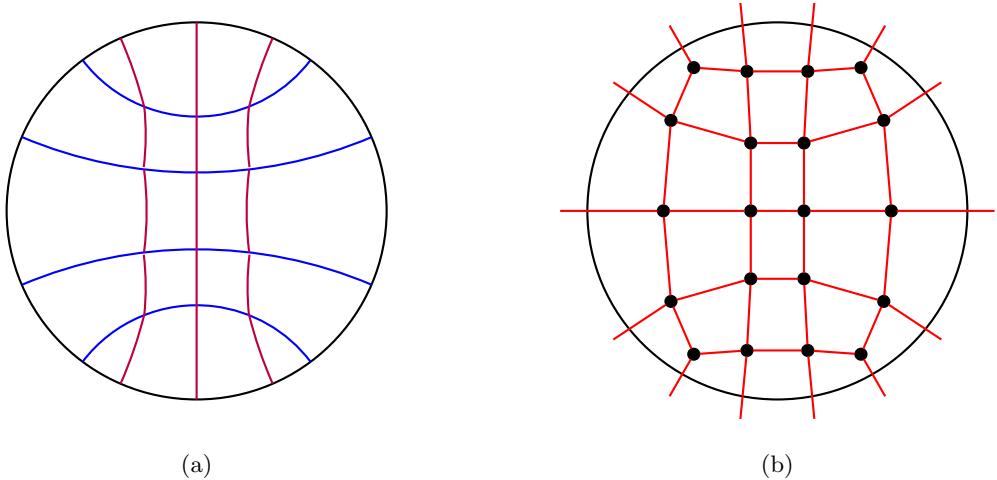

It is fairly easy to extend this construction to tensor networks on arbitrarily fine grid discretizations of the bulk. To iterate the one-loop (now $n$-loop) network, a grid is chosen like the one in Figure \ref{fig:generic_discretization}, where the horizontal surfaces are extremal surfaces, and each vertical segment is an extremal surface linking the horizontal RT surfaces on either of its endpoints. The vertical surfaces are chosen via a ``top-to-bottom'' inductive procedure, where the top point of each segment is fixed at the bottom of the previous segment, while the bottom point is chosen to minimize the total area of the surface. In other words, each vertical segment is the minimal surface connecting neighboring horizontal RT surfaces subject to the constraint that it must continue the vertical segment above it.

To extend the one-loop iteration procedure to this discretization, we begin by constructing an $n$-to-one network for the top ``row'' of the grid, the bottom tensor of which is then distilled into a grid network for the remainder of the bulk by induction. By assuming that the bulk state represented by the bottom tensor satisfies the surface-state correspondence, and hence the holographic entanglement of purification conjecture, this top-row distillation procedure can be repeated until a network is produced for the entire grid. Note that because the bulk spacetime is curved, the grid on which we discretize cannot have right angles everywhere: in a negatively-curved spacetime, there cannot exist a quadrilateral, bounded by geodesics, with each corner having an angle of $\pi/2$. The iterative procedure for constructing a tensor network on this grid will generally produce vertical surfaces that have kinks as they pass through each horizontal RT surface.

It is worth noting that this procedure can be used to generalize the $n$-to-one networks of Section \ref{sec:MEP} to higher-dimensional spacetimes. In Section \ref{sec:MEP}, the holographic entanglement of purification conjecture was used to construct a tree tensor network for the entanglement wedge of a boundary region that was partitioned as $A_1 : \dots : A_n.$ This tree tensor network was then extended to the global spacetime by an isometry on the purifying space. Generalizing this procedure to higher dimensions requires constructing a tensor network for a boundary partition that is a \emph{grid}, not simply a one-dimensional chain. Such a partition cannot be represented by a tree tensor network, and will generally require a grid network that looks more like the one sketched in Figure \ref{fig:generic_discretization}. Since we now know how to construct grid networks in $2+1$ dimensions by iteration, however, we are able to construct $n$-to-one networks in higher-dimensional spacetimes; one simply constructs the minimally entangled purification for a ``chain'' partition $A_1 : \dots : A_n,$ then distills the other grid directions using the iterative techniques explained above. By repeating this procedure inductively, it is possible to construct a tensor network for a grid discretization in spacetimes of arbitrary dimension.

All our constructions in this section, as well as those in Section \ref{sec:MEP}, are built on the holographic entanglement of purification conjecture, which itself has yet to be proven and may not be exactly true as stated. Moreover, even if the holographic entanglement of purification conjecture is itself valid, the various generalizations of it that we used in this section could be one step too far. However, we believe that the second possibility is considerably less likely than the first. All of our constructions follow from the same basic guiding principle as the holographic entanglement of purification conjecture itself: that there should exist a state associated to any convex bulk surface (the ``surface-state correspondence'' \cite{MT2015}) and that by minimizing an entropy or a sum of entropies over all possible purifications of a reduced density matrix, we can obtain something close to the state associated to the surface that minimizes the corresponding area or sum of areas.

Having completed our description of the details of our general construction, we offer two final motivations for believing that it is the most natural way to construct a geometrically appropriate tensor network, assuming one exists. The first is that radial flow in AdS/CFT has long been understood to be a form of renormalization group flow for the boundary CFT state. Since renormalization of a state is best understood as a process of disentangling and removing redundant degrees of freedom \cite{Vidal2007,Swingle2012-1,Swingle2012-2}, the radial flow of a boundary subregion to a Ryu-Takayanagi surface should correspond to performing some position-dependent RG flow on the boundary. Our procedure for constructing holographic tensor networks consists of disentangling and discarding as many degrees of freedom as possible on a boundary subregion without changing the reduced density matrix on the complementary subregion of the boundary. If renormalization group flow is a philosophically correct approach to describing the bulk in AdS/CFT, then our constructions should also be valid, with the additional benefit that they have the potential to work at sub-AdS scales.

A second supporting motivation for our construction is that when we construct tensor networks through entanglement minimization and distillation, the resulting networks have bulk legs that are ``as small as possible'' while still approximately preserving the boundary state. The usual Ryu-Takayanagi inequalities for tensor networks \cite{Swingle2012-1} imply that the bond dimensions of a tensor network have lower bounds that are determined by the bulk geometry. If \emph{any} geometrically accurate tensor network exists for a given CFT boundary state, then it should be found by a maximally efficient minimization procedure; we believe that our minimization procedure is the most obvious one to consider.

\section{Quantum Geometry} \label{sec:quantum_geo}

In this section we consider the effects of quantum fluctuations of the spacetime geometry, which we mentioned briefly in Section \ref{sec:bipartite_networks} when discussing subleading entanglement in our tensor networks.  We first argue that these fluctuations are best understood as quantum superpositions of tensor networks.  We then point out the existence of a quantum uncertainty relation between the areas of intersecting holographic entropy surfaces, whereby a very precise measurement of the area of one such surface causes the area of the other surface to grow dramatically in size.  This poses some issues for the interpretation of tensor networks whose underlying bulk discretizations contain intersecting Ryu-Takayanagi surfaces, such as those constructed in Section \ref{sec:iteration}. We close the section by formalizing these issues in the form of a no-go theorem that limits the isometry conditions that can be imposed on the bulk-to-boundary maps of such networks.

\subsection{Superpositions of Tensor Networks: Sweeping Away the ``Cobwebs''} \label{sec:cobwebs}

In this paper, we have generally aimed to describe a holographic boundary state with a \emph{single} tensor network that was expected to capture the bulk geometry. In a full theory of quantum gravity, however, the bulk geometry itself is expected to be quantum mechanical, and therefore subject to quantum fluctuations around some semiclassical background. In Section \ref{sec:bipartite_networks}, we proposed that the necessity of including in our networks some subleading edge states $\ket{\sigma}$, which are not maximally entangled, is intimately related to the existence of these fluctuations. Specifically, we argued that the $O(1/\sqrt{G_N})$ log rank of these states suggests that they correspond to fluctuations in the areas of extremal surfaces in AdS/CFT.

Another approach would be to have a single tensor network encode a single, non-fluctuating bulk geometry. In this interpretation, a holographic boundary state should be described not as a single tensor network that encodes the quantum fluctuations in the geometry, but as a weighted quantum superposition of networks, each of which describes a (very slightly different) non-fluctuating bulk geometry. The idea that fluctuations over different geometries correspond to taking a quantum superposition of different tensor networks has been previously discussed in \cite{QYY2017, HH2017, AR, DHM}.

We can replace our single network with a superposition of tensor networks by reinterpreting its subleading tensors.  In Section \ref{sec:bipartite_networks}, we suggested that because of their relatively small dimension, the $\sigma$-legs could be thought of as thin ``cobwebs'' adhering to the thick ``girders'' of the main network of  $\phi$-legs.\footnote{In such an approach it would be natural to represent long range entanglement of geometry fluctuations using cobwebs that have legs extending nonlocally to many different girders.} Instead of thinking of the cobwebs as part of the tensor network, however, we can choose to interpret them as determining the weights with which the many different ``girder-only'' networks are superposed against one another.  In doing so, one would eliminate the cobwebs associated with geometric fluctuations in any single network, instead using them to weight a superposition of ``fixed'' geometries. One major advantage of this approach, as we will see, is that the holographic state can be described, with high accuracy, by a superposition of only $O(1/\sqrt{G_N})$ fixed-geometry tensor networks; this is a huge improvement over the $e^{O(1/\sqrt{G_N})}$-rank Hilbert space of fluctuations that was necessary in Section \ref{sec:bipartite_networks}.

To be more precise, consider an edge state $\ket{\sigma}$ in a holographic tensor network. By measuring this state in its Schmidt basis, we obtain a tensor network with no cobwebs on the corresponding edge. If we measure all cobwebs in the network according to this procedure, then the resulting network for any measurement outcome has no cobwebs on any edge. To write the original holographic state in terms of these measured networks, we need to use a superposition of the networks associated with every possible measurement outcome, with each network weighted by the corresponding Schmidt coefficient in each measured cobweb state $\ket{\sigma}.$ Given the absence of the cobwebs that we associated with fluctuations in geometry, one might reasonably suggest that each of these tensor network corresponds to a fixed, non-fluctuating geometry. The original, holographic state can then be written as a superposition of appropriately weighted geometries.

So far, this ``superpositions'' framework is merely a different way of interpreting the tensor networks we already constructed in Section \ref{sec:tree_networks}. However, interpreting the full, semiclassical tensor network as an ensemble of ``girder-only'' networks with slightly different geometries suggests that we should allow the dimensions of the girders to vary for different networks in the superposition. This can be accomplished with a slight adjustment to the block-averaging procedure from Section \ref{sec:bipartite_networks}, in which we allow different blocks to have different widths. In fact, we will now show that for any fixed error $\varepsilon$, the ``target'' holographic state can be approximated to within tolerance $\varepsilon$ by choosing the block widths to be proportional to $\varepsilon e^E O(\sqrt{G_N})$, where $E$ is the block-averaged eigenvalue of the modular Hamiltonian $K = -\log \rho$. This makes the total number of blocks, and hence the total number of girder-only networks in the superposition, order $O(1/\sqrt{G_N}).$

To define a general block-averaging prescription, let $n$ be an index that labels the eigenvalues of $K$ as in Section \ref{sec:bipartite_networks}, and let $w_n$ be the width of the block containing the $n^{\mathrm{th}}$ eigenvalue. If $p_n = e^{-E}$ are the eigenvalues of the density matrix $\rho$, then the one-norm error induced by replacing each $p_n$ with the average eigenvalue within its block is given by
\begin{equation} \label{eq:gen-averaging-error}
	\varepsilon = \int |p_n - p_n^{\varepsilon}|\, \text{d}n \approx - \frac{1}{4} \int \frac{dp_n}{dn} w_n\, \text{d}n,
\end{equation}
where we have approximated $p_n$ as being roughly linear within each block and approximated the index $n$ by a smooth function. To find the optimal block-averaging procedure for representing $\rho$, we want to minimize the number of blocks---and hence the total number of tensor networks that must be superposed to describe $\rho$---subject to the constraint of a fixed error $\varepsilon$. The total number of blocks is given by
\begin{equation} \label{eq:nblocks}
	N_{\mathrm{blocks}}
		= \int \frac{1}{w_n}\, \text{d}n,
\end{equation}
and so the function $w_n$ that minimizes the total number of blocks subject to the constraint given by \eqref{eq:gen-averaging-error} satisfies
\begin{equation} \label{eq:lagrange}
	\frac{1}{w_n^2} = \frac{\lambda}{4} \frac{dp_n}{dn},
\end{equation}
where $\lambda$ is a Lagrange multiplier that does not depend on $n$. We may solve for the value of $\lambda$ by plugging this expression back into \eqref{eq:gen-averaging-error}, yielding
\begin{equation} \label{eq:lambda-prelim}
	\sqrt{-\lambda} = \frac{1}{2 \varepsilon} \int \sqrt{- \frac{dp_n}{dn}}\, \text{d}n.
\end{equation}
With respect to the smooth index $n$, the density of states is given by
\begin{equation}
	D(E) = \frac{dn}{dE},
\end{equation}
and so the eigenvalues $p_n = - e^{-E}$ of the density matrix satisfy
\begin{equation}
	\frac{dp_n}{dn} = - \frac{e^{-E}}{D(E)}.
\end{equation}
It follows from \eqref{eq:lambda-prelim}, then, that the Lagrange multiplier $\lambda$ is given by
\begin{equation} \label{eq:lambda-final}
	\sqrt{-\lambda} = \frac{1}{2 \varepsilon} \int \sqrt{D(E) e^{-E}}\, \text{d}E.
\end{equation}
Using expression \eqref{eq:lagrange} for the optimal block widths $w_n$, we find that the optimal number of blocks is given by
\begin{equation} \label{eq:opt-blocks}
	N_{\mathrm{blocks}}
		= \int \frac{1}{w_n}\, \text{d}n
		= \frac{1}{4 \varepsilon} \left[ \int \sqrt{D(E) e^{-E}}\, \text{d}E \right]^2.
\end{equation}

In Section \ref{sec:smoothminmax}, we argued that the spectrum of $K$ is tightly constrained around the leading saddle point in $D(E) e^{-E}$. Near this saddle point, the function can be approximated by a normalized Gaussian of width $O(1/\sqrt{G_N})$, i.e.,
\begin{equation} \label{eq:density-of-states}
	D(E) e^{-E} = O(\sqrt{G_N}) e^{-O(G_N) (E - S)^2}.
\end{equation}
The optimal number of blocks \eqref{eq:opt-blocks} may therefore be computed as
\begin{equation}
	N_{\mathrm{blocks}}
		= \frac{1}{\varepsilon} \frac{1}{O(\sqrt{G_N})},
\end{equation}
as we claimed above. The optimal block widths, $w_n$, satisfy
\begin{equation}
	w_n = \frac{2}{\sqrt{-\lambda}} \sqrt{- \frac{dn}{dp_n}} = \varepsilon O(G_N^{1/4}) \sqrt{D(E) e^{E}}.
\end{equation}
Near the saddle point, the variation in the right hand side of \eqref{eq:density-of-states} is subleading and so we have
\begin{align}
D(E) \approx O(\sqrt{G_N}) e^{E}.
\end{align}
It follows that the block widths in the optimal block-averaging procedure are proportional to $e^E$. Assuming that the modular energy $E$ has a holographic interpretation as $A/4G_N$, where $A$ is the area of the minimal surface, this is the right size for the tensor network geometry to match the semiclassical geometry of the holographic state. 

Unlike our construction in Section \ref{sec:tree_networks}, however, the optimal block-averaging procedure represents the semiclassical holographic geometry as a superposition of $O(1/\sqrt{G_N})$ rather than $e^{O(1/\sqrt{G_N})}$ ``fixed-geometry'' networks. We take this as a suggestion that the superposition framework, in which one allows the different tensor networks in the ensemble to have slightly different ``fixed'' geometries, is a more efficient and informative description of the holographic state.

In this superposition framework, it is natural to interpret a single tensor network in the superposition as corresponding to an (approximate) eigenstate of the bulk area operator \cite{AR, DHM, ADS2016, BO2017}.  If the usual bulk state generated by a path integral corresponds to a canonical ensemble of the area operator, then the state of a single tensor network corresponds to a microcanonical ensemble which takes values over a tiny range of areas.\footnote{We do not expect that the area spectrum will contain large exact degeneracies, but because of the $\epsilon$-smoothing, we can approximate an area spectrum with exponentially small gaps with a degenerate spectrum.}  This interpretation is compelling because it makes the flat entanglement spectrum of many tensor network models (in which all R\'{e}nyi entropies are equal) into a feature rather than a bug.  In a single tensor network, the R\'{e}nyi entropy \emph{is} flat; to get a state with a non-flat spectrum, one must take superpositions of different geometries.

\subsection{Uncertainty Relations for Intersecting Ryu-Takayanagi Surfaces}

Having said all this, we will now identify a serious issue with the approach of \cite{AR, DHM}.  Namely, area-eigenvalue states with a flat or nearly-flat entanglement spectrum do not correspond to a bulk geometry that is similar to the original state.  As a result, it is not possible to construct geometrically appropriate tensor networks that are in a microcanonical area ensemble for multiple directions simultaneously.

The problem is that in general relativity, the area and boost angle are canonically conjugate quantities obeying a Heisenberg uncertainty relation \cite{BTZ1994, CT1995}:
\begin{align} \label{eq:boost_uncertainty}
\Delta E \Delta t \ge \frac{1}{2},
\end{align}
where the modular Hamiltonian is $E = \mathrm{area}/4G_N + O(1)$ \cite{JLMS} and the conjugate time $t$ is the boost angle (in hyperbolic radians). Obtaining a flat entanglement spectrum in a holographic state requires constraining $\Delta E$ to a small, $O(1)$ number, and hence measuring the area of the corresponding surface to within a tolerance of $O(G_N).$

\begin{figure}[h!]
	\begin{center}
	\includegraphics[scale=0.5]{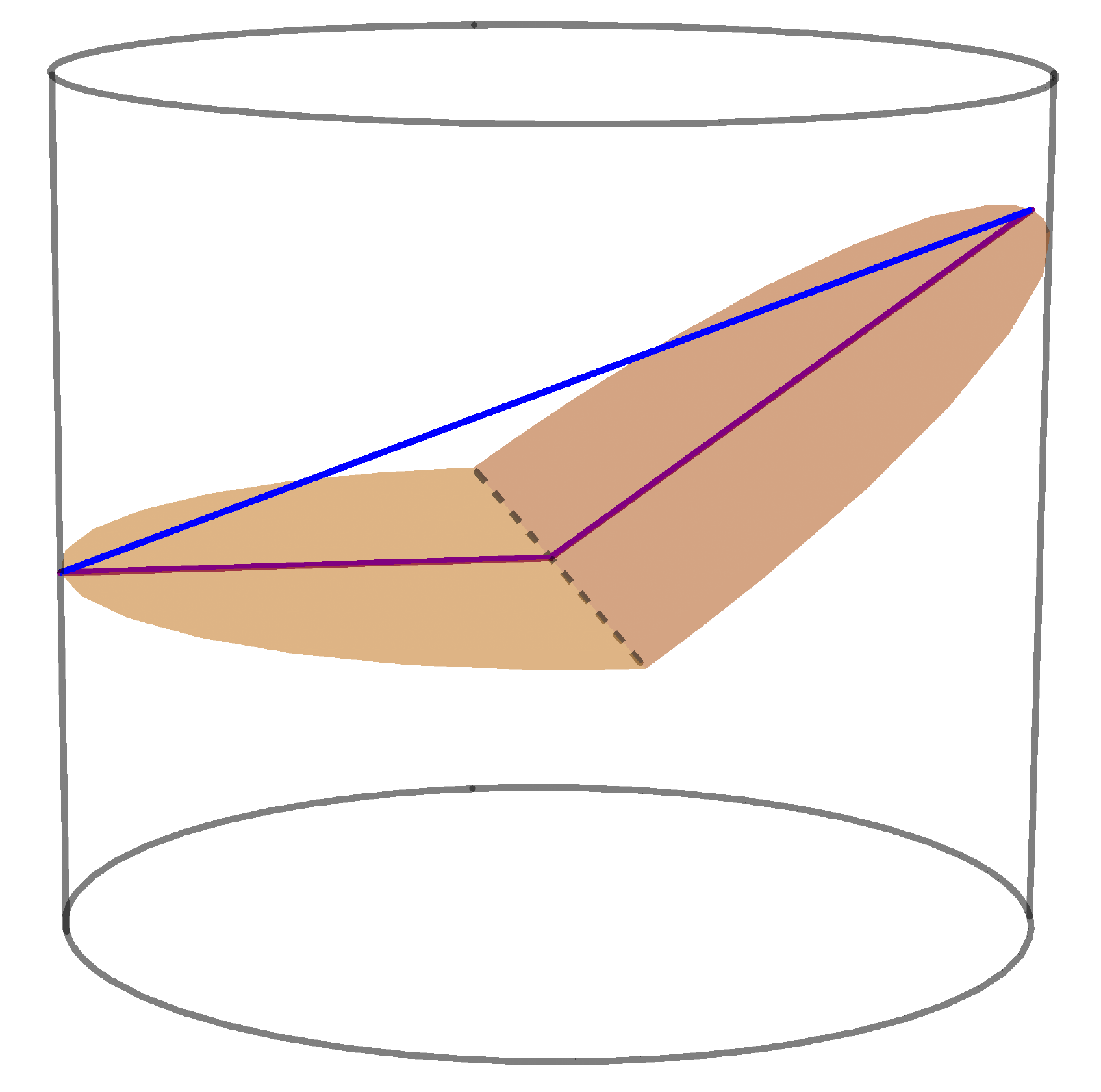}
	\caption{A spacelike slice of vacuum $AdS_3$ formed by boosting one half of the $t=0$ static slice. This boost introduces a nontrivial extrinsic curvature at the dashed line in the surface.
			The original RT surface for a particular boundary region on the $t=0$ slice, sketched by a purple line, is no longer the entangling surface for the corresponding boundary
			region in the half-boosted state. Instead, one must consider the HRT surface, sketched in blue, which has strictly larger area.}
	\label{fig:boost}
	\end{center}
\end{figure}

Since a single bin in our construction measures $E$ to an accuracy of $O(\epsilon)$, \eqref{eq:boost_uncertainty} implies that the uncertainty in the corresponding boost angle is $O(1/\epsilon)$.  This is quite large, and in fact it is large enough to take us out of the validity of the static slice regime.  In particular, introducing a large crease of extrinsic curvature at the horizontal Ryu-Takayanagi surface of Figure \ref{fig:plus_network_complete} will make it so that the area of the vertical HRT surface (which follows a \emph{spacetime} geodesic and therefore no longer lies on the creased slice) will have a significantly greater area than the minimal  surface on the original static slice, as shown in Figure \ref{fig:boost}.\footnote{In evaluating the entropy of the HRT surface, it is helpful to use a boost-invariant UV cutoff surface, so that the area of surfaces on the slice is independent of the boost angle.}  Since this is true for either sign of the boost angle, the expectation value of the area given an uncertain boost is also larger.

This uncertainty principle implies that if we start by distilling information on the horizontal surface, consider \emph{just one term} of the resulting superposition, and then attempt a ``vertical'' distillation, then the leading $O(1/G_N)$ part of the vertical entropies will be larger than on the original static slice.  In other words, it will not be possible to construct a geometrically appropriate tensor network for a single term of the superposition.\footnote{In the exact AdS/CFT bulk state, this spuriously large entropy must disappear when we take all terms in the superposition, due to destructive interference.  However, it can be difficult to keep approximations under control when there is destructive interference among a large number of terms, since small errors can accumulate and leave a substantial remainder.}  This problem is related to the issues for dynamical tensor networks that we discuss in section \ref{dyn}.

\subsection{A No-Go Theorem for the Four-Tensor Network} \label{sec:nogo_theorem}

The uncertainty relationship discussed in the previous section shows that we cannot simultaneously measure the areas of horizontal and vertical Ryu-Takayanagi surfaces with high accuracy.  This might make one wonder whether it is really possible, in a single tensor network, to localize information on both the vertical and horizontal Ryu-Takayanagi surfaces simultaneously, as required for our iterative constructions in \ref{sec:iteration} to be geometrically accurate.  

In this section, we will prove that there is indeed such an obstruction preventing the construction of certain kinds of geometrically accurate tensor networks with crossing Ryu-Takayanagi surfaces, even in the simplest case of a four-tensor network (section \ref{sec:four_tensor}).  However, our no-go theorem is only valid for tensor networks having approximate isometry properties that are stronger than those of our actual network.  Thus, we can optimistically hope that our tensor network constructions are still valid.\footnote{If not, then we believe the construction in \ref{sec:MEP}, where we localize information on a single Ryu-Takayanagi surface, will still be valid.}

When constructing sub-AdS scale tensor networks in Sections \ref{sec:MEP} and \ref{sec:iteration}, we found that the bulk-to-boundary ``extension maps'' associated with a particular network were not generally exact isometries. In the $n$-to-one loop networks of Section \ref{sec:MEP}, at least the ``downward-flowing'' map could be shown to be an exact isometry --- or, depending on the order in which the ``top-half'' RT surfaces were distilled in constructing a tree network for the MEP, at least an approximate isometry in the sense of equation \eqref{eq:2ndapproxisom}. In the full sub-AdS network of Section \ref{sec:iteration}, however, we found that the bulk-to-boundary maps in our network could \emph{not} generally be proven to be exact or approximate isometries, and were in fact only ``morally'' isometric in the sense that they preserved the normalization of the state and preserved the entanglement entropy to leading order. The non-exactness of these isometries stands in contrast to the holographic tensor network toy models of AdS/CFT introduced in \cite{HaPPY}.

In this section, we prove that geometrically appropriate tensor networks for generic bulk discretizations of static states in AdS/CFT \emph{cannot}, in fact, have bulk-to-boundary extension maps which are all exact or even approximate isometries. We formalize this for the four-tensor network for vacuum $AdS_3$ of section \ref{sec:iteration} in the following theorem:

\begin{thm} \label{thm:nogo}
	No four-tensor network of the form shown in Figure \ref{fig:plus_network_complete} can simultaneously satisfy the following four properties:

	\begin{enumerate}[(i)]
		\item Each leading edge state $\ket{\phi}$ is maximally entangled on a Hilbert space of dimension $e^{S\pm o(S)}$, where $S$ is proportional to the area of the
			corresponding bulk surface, satisfying
			\begin{equation}
				S = \frac{\operatorname{area}}{4 G_N} + o\left(\frac{1}{G_N}\right).
			\end{equation}
		\item Each subleading edge state $\ket{\sigma}$ is submaximally entangled on a Hilbert space of dimension $e^{o(S)}.$
		\item The four extension maps that map either side of either RT surface to the boundary are all approximate isometries in the sense that
			\begin{equation}
				V^{\dagger} V \approx \mathds{1}
			\end{equation}
			for any such map $V$, where this approximation means that $V^{\dagger} V$ is close to the identity in the operator norm.
		\item The boundary state of the tensor network approximately reproduces the boundary state of the $AdS_3$ vacuum with high fidelity.
	\end{enumerate}
\end{thm}

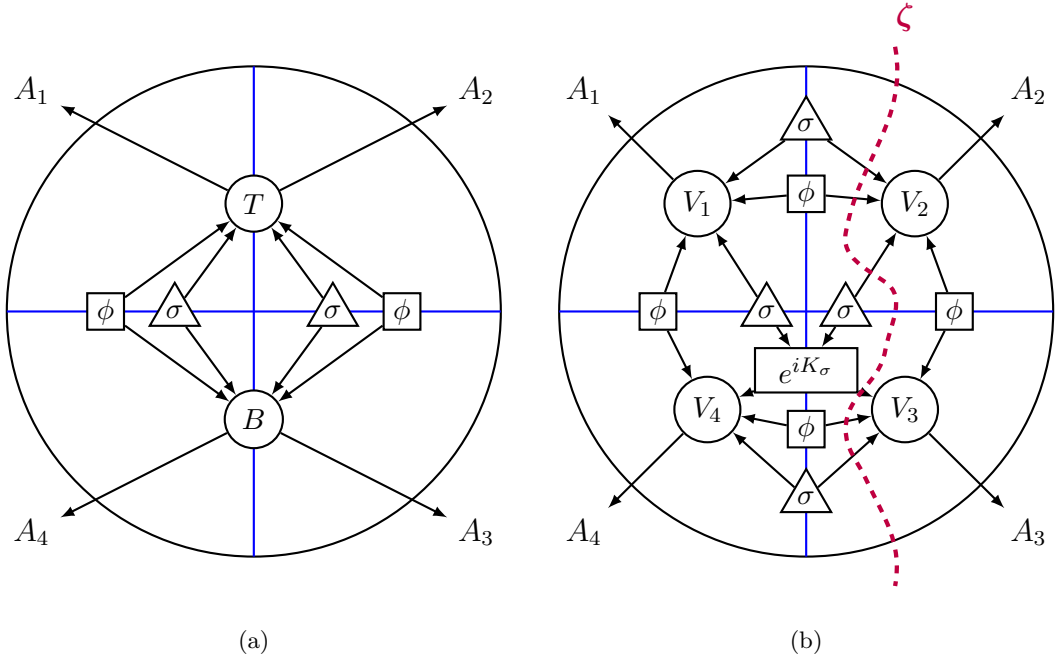
\begin{figure}[h]
\begin{center}
	\subfloat[\label{fig:nogo_a}]{
		\begin{tikzpicture}[scale=1.3, thick, >=latex,
					cobwebs/.style={regular polygon, inner sep=0.5pt, fill=white, regular polygon sides=3, draw},
					EPR/.style={regular polygon, regular polygon sides=4, inner sep=0pt, fill=white, draw},
					tensor/.style={circle, draw, fill=white, minimum width=1.5em}]
		\draw (0, 0) circle (2.5cm);

		\draw [blue] (2.5, 0) to[out=180, in=0] (-2.5, 0);
		\draw [blue] (0, 2.5) to (0, -2.5);

		\node at (-2.25, 2.25) (A1) {\large $A_1$};
		\node at (2.25, 2.25) (A2) {\large $A_2$};
		\node at (2.25, -2.25) (A3) {\large $A_3$};
		\node at (-2.25, -2.25) (A4) {\large $A_4$};
		\node at (0, -3) (Ac) {\large};

		\node [EPR] at (-1.5, 0.0) (phi1) {$\phi$};
		\node [cobwebs] at (-0.8, 0) (sigma1) {$\sigma$};
		\node [EPR] at (1.5, 0) (phi3) {$\phi$};
		\node [cobwebs] at (0.8, 0) (sigma3) {$\sigma$};

		\node [tensor] at (0, 1.1) (T) {$T$};
		\node [tensor] at (0, -1.1) (B) {$B$};

		\draw (T) edge[->] (A1);
		\draw (T) edge[->] (A2);
		\draw (B) edge[->] (A3);
		\draw (B) edge[->] (A4);

		\draw (phi1) edge[->] (B);
		\draw (phi1) edge[->] (T);
		\draw (sigma1) edge[->] (B);
		\draw (sigma1) edge[->] (T);
		\draw (phi3) edge[->] (T);
		\draw (sigma3) edge[->] (T);
		\draw (phi3) edge[->] (B);
		\draw (sigma3) edge[->] (B);
		\end{tikzpicture}
	}
	\hspace{0.3cm}
	\subfloat[\label{fig:nogo_b}]{
		\begin{tikzpicture}[scale=1.3, thick, >=latex,
					cobwebs/.style={regular polygon, inner sep=0.5pt, fill=white, regular polygon sides=3, draw},
					EPR/.style={regular polygon, regular polygon sides=4, inner sep=0pt, fill=white, draw},
					tensor/.style={circle, draw, fill=white, minimum width=1.5em},
					boost/.style={rectangle, fill=white, draw, minimum width=3.5em},]
		\draw (0, 0) circle (2.5cm);

		\draw [blue] (2.5, 0) to[out=180, in=0] (-2.5, 0);
		\draw [blue] (0, 2.5) to[out=270, in=90] (0, -2.5);

		\node at (-2.25, 2.25) (A1) {\large $A_1$};
		\node at (2.25, 2.25) (A2) {\large $A_2$};
		\node at (2.25, -2.25) (A3) {\large $A_3$};
		\node at (-2.25, -2.25) (A4) {\large $A_4$};
		\node at (0, -3) (Ac) {\large};

		\node [EPR] at (-1.5, 0.0) (phi1) {$\phi$};
		\node [cobwebs] at (-0.4, 0) (sigma1) {$\sigma$};
		\node [EPR] at  (0, 1.2) (phi2) {$\phi$};
		\node [cobwebs] at (0, 1.9) (sigma2) {$\sigma$};
		\node [EPR] at (1.5, 0) (phi3) {$\phi$};
		\node [cobwebs] at (0.4, 0) (sigma3) {$\sigma$};
		\node [EPR] at (0, -1.2) (phi4) {$\phi$};
		\node [cobwebs] at (0, -1.9) (sigma4) {$\sigma$};

		\node [tensor] at (-1.1, 1.1) (V1) {$V_1$};
		\node [tensor] at (1.1, 1.1) (V2) {$V_2$};
		\node [tensor] at (1, -1) (V3) {$V_3$};
		\node [tensor] at (-1, -1) (V4) {$V_4$};

		\node [boost] at (0, -0.6) (boost) {$e^{i K_{\sigma}}$};

		\draw (V1) edge[->] (A1);
		\draw (V2) edge[->] (A2);
		\draw (V3) edge[->] (A3);
		\draw (V4) edge[->] (A4);

		\draw (phi1) edge[->] (V4);
		\draw (phi1) edge[->] (V1);
		\draw (sigma1) edge[->] (boost);
		\draw (boost) edge[->] (V4);
		\draw (sigma1) edge[->] (V1);
		\draw (phi2) edge[->] (V1);
		\draw (sigma2) edge[->] (V1);
		\draw (phi2) edge[->] (V2);
		\draw (sigma2) edge[->] (V2);
		\draw (phi3) edge[->] (V2);
		\draw (sigma3) edge[->] (V2);
		\draw (phi3) edge[->] (V3);
		\draw (sigma3) edge[->] (boost);
		\draw (boost) edge[->] (V3);
		\draw (phi4) edge[->] (V3);
		\draw (sigma4) edge[->] (V3);
		\draw (phi4) edge[->] (V4);
		\draw (sigma4) edge[->] (V4);

		\draw [purple, ultra thick, dashed] (0.9, 2.7) to[out=280, in=80] (0.4, 0.8) to[out=260, in=145] (0.8, 0.3)
							to[out=325, in=80] (0.8, -0.4) to[out=260, in=90] (0.4, -1.2)
							to[out=270, in=115] (0.6, -1.7)
							to[out=295, in=80] (0.9, -2.8);
		\node [purple] at (1, 3) (zeta) {\large$\bm{\zeta}$};
		\end{tikzpicture}
	}
	\caption{Figures to accompany the no-go theorem for the four-tensor network. (a) The extension maps for the top and bottom halves of the horizontal RT surfaces are drawn explicitly
			as tensors $T$ and $B$. (b) The boost operator on the ``bottom half'' of the boundary, $A_3 A_4$, can be represented on the tensor network by a unitary operator
			$e^{i K_{\sigma}}$ that acts only on the subleading edge states $\ket{\sigma}.$ The graph cut $\zeta$, sketched here, has dimension given to leading order by
			the combined size of the $\ket{\phi}$ Hilbert spaces on the vertical RT surface.}
	\label{fig:nogo}
\end{center}
\end{figure}

Suppose that the four-tensor network shown in Figure \ref{fig:plus_network_complete} \emph{does} satisfy all conditions given in Theorem \ref{thm:nogo}. We denote the ``upwards-pointing'' extension map from the horizontal RT surface as $T$, and the ``downwards-pointing'' map as $B$. In other words, if we were to collapse all the tensors in the top and bottom halves of Figure \ref{fig:plus_network_complete}, excluding the tensors on the horizontal RT surface itself, then the resulting tensor network would have only $T$ as its top-half tensor and $B$ as its bottom-half tensor (sketched in Figure \ref{fig:nogo_a}). We denote the actual CFT state as $\ket{\psi^{\text{CFT}}}$, and the tensor network state as $\ket{\psi^{\text{TN}}}$. Assumption (iv) of Theorem \ref{thm:nogo} ensures
\begin{equation}
	\ket{\psi^{\text{CFT}}} \approx \ket{\psi^{\text{TN}}}.
\end{equation}
From the form of the tensor network shown in Figure \ref{fig:nogo_a}, we see that $\ket{\psi^{\text{TN}}}$ has the form
\begin{equation} \label{eq:TB_TN_state}
	\ket{\psi^{\text{TN}}} = (T \otimes B) \ket{\phi \sigma},
\end{equation}
where $\ket{\phi \sigma}$ represents the combined pure state of all edge states $\ket{\phi}$ and $\ket{\sigma}$ on the horizontal RT surface.

Let $K^{\text{CFT}}_B$ be the modular Hamiltonian of the bottom-half boundary state $\psi^{\text{CFT}}_{A_3 A_4}$, and $e^{i K^{\text{CFT}}_B}$ the corresponding boost operator. Since this operator is unitary, we have
\begin{equation}
	e^{i K^{\text{CFT}}_B} \ket{\psi^{\text{CFT}}} \approx e^{i K^{\text{CFT}}_B} \ket{\psi^{\text{TN}}}.
\end{equation}
Since the reduced state of the tensor network on $A_3 A_4$ approximately reproduces the reduced state of the CFT on the same region, their modular Hamiltonians approximately agree.\footnote{The modular Hamiltonians may not actually be close in the sense of the operator norm; however, their action on the global states $\ket{\psi^{\text{TN}}}$ and $\ket{\psi^{\text{CFT}}}$ are approximately the same.} It follows that the action of the modular Hamiltonian on the CFT state can be represented in the tensor network as
\begin{equation} \label{eq:TN_mod_flow}
	e^{i K^{\text{CFT}}_B} \ket{\psi^{\text{CFT}}} \approx e^{i K^{\text{TN}}_B} \ket{\psi^{\text{TN}}},
\end{equation}
where $K^{\text{TN}}_B$ is the modular Hamiltonian of the bottom-half boundary state $\psi^{\text{CFT}}_{A_3 A_4}$ in the tensor network.

We see immediately from the form of the tensor network state given in equation \eqref{eq:TB_TN_state} that this modular Hamiltonian takes the explicit form
\begin{equation}
	K^{\text{TN}}_B = - \log \psi^{\text{TN}}_{A_3 A_4} = - \log \tr_{A_1 A_2} (T B \phi \sigma B^{\dagger} T^{\dagger}).
\end{equation}
Condition (iii) of Theorem \ref{thm:nogo} ensures that $T^{\dagger} T$ is close to the identity, ensuring that the partial trace over $A_1 A_2$ in the above expression can be replaced by a partial trace over the domain of $T$, which we call $\mathcal{H}_{f} \otimes \mathcal{H}_{\gamma}$. Here, as in Section \ref{sec:tree_networks}, $\mathcal{H}_{f}$ corresponds to the subleading states $\ket{\sigma}$ while $\mathcal{H}_{\gamma}$ corresponds to the maximally entangled states $\ket{\phi}.$ The modular Hamiltonian of the reduced tensor network state on $A_3 A_4$ therefore satisfies
\begin{equation}
	K^{\text{TN}}_B \approx - B (\log \tr_{f \gamma} \phi \sigma) B^{\dagger}.
\end{equation}
In other words, the modular Hamiltonian of the tensor network on the bottom-half boundary region $A_3 A_4$ can be approximately represented by
\begin{equation}
	K^{\text{TN}}_B \approx B K_{\phi \sigma} B^{\dagger},
\end{equation}
where $K_{\phi \sigma}$ is the modular Hamiltonian of the reduced RT surface-state $\tr_{f \gamma} \phi \sigma.$

Since the edge states $\ket{\phi}$ are maximally entangled, their modular Hamiltonian when restricted to either side of the RT surface is simply a multiple of the identity. This contributes only a normalization factor to the overall network. We therefore write the modular Hamiltonian of the tensor network on $A_3 A_4$ as
\begin{equation}
	K^{\text{TN}}_B \approx B K_{\sigma} B^{\dagger},
\end{equation}
up to an additive constant coming from the normalization, which we ignore. Here $K_{\sigma}$ is an operator that acts only on the subleading boundary states $\ket{\sigma},$ and only acts on one side of the RT surface (in this case, the bottom half).

Returning to the tensor network expression for the modular flow of the CFT given in equation \eqref{eq:TN_mod_flow}, we see that the boost operator on the bottom half of the boundary CFT state can be represented on the tensor network as
\begin{equation}
	e^{i K^{\text{CFT}}_B} \ket{\psi^{\text{CFT}}} \approx B e^{i K_{\sigma}} B^{\dagger} \ket{\psi^{\text{TN}}}.
\end{equation}
From the expression for the tensor network state given in \eqref{eq:TB_TN_state}, we may rewrite this expression as
\begin{equation} \label{eq:sigma_flow}
	e^{i K^{\text{CFT}}_B} \ket{\psi^{\text{CFT}}} \approx (T \otimes B) e^{i K_{\sigma}} \ket{\phi \sigma},
\end{equation}
where we have used condition (iii) of Theorem \ref{thm:nogo} to ensure that $B^{\dagger} B$ is close to the identity. This final tensor network representation for the boosted CFT state is sketched in Figure \ref{fig:nogo_b}.

Equation \ref{eq:sigma_flow} essentially tells us that the modular flow of the CFT on the boundary region $A_3 A_4$ can be represented by a unitary operator that acts only on the subleading states $\ket{\sigma}$ on the horizontal RT surface. This conclusion, however, contradicts assumptions (i) and (ii) of Theorem \ref{thm:nogo}, which restrict the bond dimensions of the network. To see this contradiction, consider the entanglement entropy of the boundary region $A_2 A_3$ in the CFT state $e^{i K^{\text{CFT}}_B} \ket{\psi^{\text{CFT}}}$. In the bulk, this operator acts as a boost on the entanglement wedge of $A_2 A_3.$ In vacuum $AdS_3$, it is easy to show that the entangling surface of $A_2 A_3$ in the boosted state has greater area in the boosted state than in the unboosted state. By the HRT formula \cite{HRT2007}, it follows that the entanglement entropy of $A_2 A_3$ in the boosted state must be greater than the corresponding entropy in the unboosted state at leading order, i.e., $S(A_2 A_3) > \operatorname{area}(\operatorname{vert})/4 G_N$, where $\operatorname{area}(\operatorname{vert})$ represents the area of the vertical RT surface (see Fig. \ref{fig:boost}).

\emph{However}, this is in direct contradiction a bound derived by Swingle in \cite{Swingle2012-1}! There it was shown that the entanglement entropy of a boundary subregion $A$ in any tensor network is bounded above by $\log\dim\zeta$ for any graph cut $\zeta$ that partitions $A$ from its complement. From the tensor network representation of the boosted state given in equation \eqref{eq:sigma_flow} and the graph cut sketched in Figure \ref{fig:nogo_b}, conditions (i) and (ii) of Theorem \ref{thm:nogo} imply that the entanglement entropy of $A_2 A_3$ in the boosted state satisfies
\begin{equation} \label{eq:boosted_swingle_bound}
	S(A_2 A_3) \leq \frac{\operatorname{area}(\operatorname{vert})}{4 G_N}
\end{equation}
to leading order in $G_N.$ The bound given in \eqref{eq:boosted_swingle_bound} clearly contradicts the HRT formula for the boosted state.

We conclude that no geometrically appropriate four-tensor network can be constructed for the $AdS_3$ vacuum with approximate isometries from each RT surface to the boundary. This provides partial justification for our relatively weak isometry conditions: both the state-dependent approximate isometry condition of Section \ref{sec:tree_networks} and, in particular, the ``moral'' isometries of Sections \ref{sec:MEP} and \ref{sec:iteration}. While one might initially think that stronger isometry conditions should be possible in a tensor network construction of a holographic state, it turns out that such conditions are incompatible with the dynamics of AdS/CFT. The geometrically appropriate four-tensor network constructed in Section \ref{sec:iteration} avoids our no-go theorem precisely \emph{because} some of its extension maps are only moral isometries.\footnote{A more pessimistic interpretation would be that either the entropy of purification conjecture, or one of the additional conjectures that we made in Section \ref{sec:iteration}, fails to hold in the form necessary for the constructions in Section \ref{sec:iteration} to go through.}

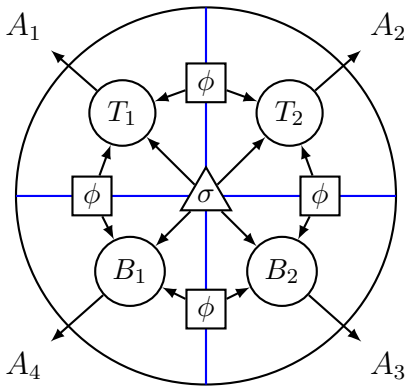
\begin{figure}[h]
\begin{center}
		\begin{tikzpicture}[thick, >=latex,
					cobwebs/.style={regular polygon, inner sep=0.5pt, fill=white, regular polygon sides=3, draw},
					EPR/.style={regular polygon, regular polygon sides=4, inner sep=0.3pt, fill=white, draw},
					tensor/.style={circle, draw, fill=white, minimum width=1.5em}]
		\draw (0, 0) circle (2.5cm);

		\draw [blue] (2.5, 0) to [out=180, in=0] (-2.5, 0);
		\draw [blue] (0, 2.5) to (0, -2.5);

		\node at (-2.4, 2.25) (A1) {\large $A_1$};
		\node at (2.4, 2.25) (A2) {\large $A_2$};
		\node at (2.4, -2.25) (A3) {\large $A_3$};
		\node at (-2.4, -2.25) (A4) {\large $A_4$};

		\node [EPR] at (-1.5, 0.0) (phi1) {$\phi$};
		\node [cobwebs] at (0, 0) (sigma) {$\sigma$};
		\node [EPR] at (0, 1.5) (phi2) {$\phi$};
		\node [EPR] at (1.5, 0) (phi3) {$\phi$};
		\node [EPR] at (0, -1.5) (phi4) {$\phi$};

		\node [tensor] at (-1.1, 1.1) (V1) {$T_1$};
		\node [tensor] at (1.1, 1.1) (V2) {$T_2$};
		\node [tensor] at (1, -1) (V3) {$B_2$};
		\node [tensor] at (-1, -1) (V4) {$B_1$};

		\draw (V1) edge[->] (A1);
		\draw (V2) edge[->] (A2);
		\draw (V3) edge[->] (A3);
		\draw (V4) edge[->] (A4);

		\draw (phi1) edge[->] (V4);
		\draw (phi1) edge[->] (V1);
		\draw (sigma) edge[->] (V4);
		\draw (sigma) edge[->] (V1);
		\draw (sigma) edge[->] (V3);
		\draw (sigma) edge[->] (V2);
		\draw (phi2) edge[->] (V1);
		\draw (phi2) edge[->] (V2);
		\draw (phi3) edge[->] (V2);
		\draw (phi3) edge[->] (V3);
		\draw (phi4) edge[->] (V3);
		\draw (phi4) edge[->] (V4);
		\end{tikzpicture}
	\caption{A four-tensor network with explicit entanglement between the right and left sides of the horizontal RT surface, where all four tensors share a common cobweb state.
			Adding subleading entanglement like this cannot be used to subvert our no-go theorem.}
	\label{fig:entangled_cobwebs}
\end{center}
\end{figure}

Note that while our proof of the no-go theorem required the existence of a subleading state on the horizontal RT surface to absorb the action of the modular Hamiltonian, it did not necessarily require that the subleading states take the exact form shown in Figure \ref{fig:plus_network_complete}. In fact, the same proof would apply for many different kinds of cobweb entanglement (using the terminology of Section \ref{sec:cobwebs}, where we referred to maximally entangled states as ``girders'' and subleading states as ``cobwebs''), so long as the cobweb states are able to absorb the modular Hamiltonian on the horizontal RT surface and do not interfere with the size of the girders. In particular, one could consider a four-tensor network with a generic cobweb state that is entangled among all four quadrants of the network, sketched in Figure \ref{fig:entangled_cobwebs}. Such a tensor network still satisfies all the necessary conditions of our no-go theorem, and so cannot have approximate bulk-to-boundary isometries while accurately reproducing the dynamics of the boundary CFT state.

\section{Discussion} \label{sec:discussion}

In this article, we have given a constructive procedure for distilling the spatial geometry of a static spacetime from the quantum entanglement of a boundary CFT, including structure at sub-AdS scales.  We described a general procedure for constructing large-scale ``tree tensor networks'' (Sec. \ref{sec:tree_networks}), explained how the holographic entanglement of purification conjecture could be used to partition the information on a single Ryu-Takayanagi surface below the AdS scale (Sec \ref{sec:MEP}), and proposed an iterative approach for constructing a sub-AdS tensor network by distilling the geometry incrementally (Sec. \ref{sec:iteration}). We showed that tree tensor networks are always quantum error correcting in the appropriate sense (Sec. \ref{sec:qec}), and suggested that our more general, sub-AdS constructions should have similar properties. We also proved an important no-go theorem (Sec. \ref{sec:quantum_geo}) showing that the bulk-to-boundary maps of any geometrically appropriate tensor network for AdS/CFT cannot be exact or approximate isometries in the usual sense.

While our constructive procedures can always be performed so long as the smooth min- and max-entropies of various surface subregions agree at leading order, our holographic conjectures ensure that the resulting tensor network geometry also matches the geometry of the corresponding AdS/CFT state in the sense that the Hilbert space dimensions of its legs match the areas of corresponding spacetime surfaces. This constitutes significant data about the spacetime geometry, in the sense that a sufficiently large set of bulk area observables can be used to reconstruct the metric, as has been shown explicitly in four spacetime dimensions \cite{BCFK} and likely holds more generally. Furthermore, our holographic conjectures ensure that any two choices of iterative construction for a single bulk discretization should produce tensor networks that differ only at subleading orders in $1/G_N$; in the limit as the discretization scale is taken arbitrarily small, any two tensor networks corresponding to \emph{different} discretization schemes should converge up to subleading corrections.  So if our conjectures hold, then for the first time we have used tensor networks to obtain the ``it'' of continuous spatial geometry from the ``qubit'' of quantum entanglement for states in full AdS/CFT.

In fact, our dictionary for constructing a tensor network uses \emph{only} quantum information properties of the boundary state.  This implies that, if we regulate the CFT on a lattice, acting on any lattice points with local unitaries does not change the resulting bulk geometry.  This raises some philosophical puzzles given that entanglement is not a linear quantum observable and therefore cannot be measured by a normal quantum observation \cite{MW2012}.  These puzzles may be related to the AMPS firewalls paradox \cite{amps1, amps2, mathurfirewall, braunsteincurtain} and claims that the construction of geometry is necessarily state-dependent \cite{PR2014, MS2013}, but in this article we make no claims about which geometrical features are truly measurable.

\subsection{Entanglement Shadows}

In describing our constructions, we usually had in mind either the vacuum state or small perturbations around it. As a result, we have mostly ignored some subtleties than can appear in more complicated states.  However, we anticipate that our results can be extended to essentially arbitrary spacetimes.  In particular, it should be possible to use our results to probe the geometry of the so-called ``entanglement shadow,'' i.e., a bulk region that cannot be probed by normal Ryu-Takayanagi surfaces \cite{CKNV2012, H2012, EW2014-2, BCCD2015}.  Such entanglement shadow regions appear, for example, in the region surrounding a massive star or black hole. We can probe such regions using our iterative construction in Section \ref{sec:iteration}, beginning by dividing RT surfaces that pass just outside the entanglement shadow into pieces using the holographic entanglement of purification.  By iteratively constructing new surfaces anchored to these RT surfaces, it is clear that we can get farther into the bulk than we could get by starting on the boundary.\footnote{For a similar discussion of how entanglement shadows can be probed by extremal surfaces anchored to points in the bulk, see \cite{NRS2018}.}

This raises the question of how deep into a general bulk we can probe.  It is particularly interesting to ask this question in the context of wormholes extended between multiple asymptotic CFT regions \cite{BHMMR2014, SS2014, PR2017, BCDR2018}.  Because our construction relies on taking consecutive bipartite divisions of the system, our construction always requires starting with the full entangled state of all the asymptotic CFT regions.  We can then easily construct a Hilbert space representing the compact RT surface of an entire connected boundary component.  It is not clear, however, that there will always be nontrivial entanglement wedge cross-sections extending to the global RT surface, because it may be the case that the minimal cross-section of the entanglement wedge will close off on the boundary rather than traveling down the wormhole throat.

In light of the above reflections, we believe that the only possible obstruction to continuing deeper into the bulk using our iterative procedure would be a locally minimal area surface $\Sigma$ for an entire connected component of the boundary with the property that for any bipartite division of $\Sigma = \Sigma_1 \cup \Sigma_2$, the minimal area surface separating $\Sigma_1$ and $\Sigma_2$ is always whichever of $\Sigma_1$ or $\Sigma_2$ has the least area.  Such a surface $\Sigma$ would be analogous to a Haar-random pure state, for which the entanglement entropy of any bipartite division scales as the volume of the smaller subsystem.  Even in these cases, however, one might sometimes be able to go deeper if there exists a nontrivial entanglement wedge cross-section from $\Sigma$ to the true global minimum.  It would be interesting to confirm this intuition using geometrical proofs similar to \cite{EW2014-2}.

\subsection{Bit Threads}

Another potential avenue for future work is the apparent connection between our construction and the bit-thread formalism for holographic entanglement \cite{FH2017}. Bit threads describe boundary entanglement in terms of smooth flows on the bulk geometry, with intuition inherited from the machinery of maximal flows on discrete graphs. In some sense, then, the bit thread formalism is inspired by the notion of discretizing the bulk geometry of a holographic state; discretizing the bulk geometry of a holographic state while preserving its entanglement structure is exactly what we have done in this paper by constructing tensor networks for AdS/CFT. Since our construction is largely based on the Ryu-Takayanagi formula and its generalizations, and bit threads interpretations of holographic entanglement are formally equivalent to RT, our work could be reframed in the language of bit threads.

The bit threads formalism suffers from the fact that  known holographic entropy inequalities \cite{hayden2013holographic, bao2015holographic} seem generally more difficult to prove in the language of bit threads than in the usual Ryu-Takayanagi picture \cite{cui2018bit, hubeny2018bulk}. It is possible, however, that the bit thread formalism is more useful than the Ryu-Takayanagi formula for describing localization of boundary entanglement to small ``cells'' of the bulk. Many times in this work, especially in Section \ref{sec:larger_tree_networks}, we have appealed to notions of ``information flow'' toward or away from bulk subregions to describe our construction intuitively. This notion is best expressed in the language of bit threads, and one might expect that reframing our work in the bit thread formalism could yield new insight into the information content of bulk subregions in tensor networks with sub-AdS locality. In particular, one might ask the following: what happens when one tries to define a spacetime flow that maximizes the flow through a single bulk subregion in our tensor network construction, or a combination of bulk subregions? Does this maximal bulk flow have a natural interpretation in terms of boundary information? Flows of this type were recently discussed in \cite{ADP2018}. We leave further analysis of this question for future work.

\subsection{Fiber Directions}

An important aspect of the holographic bulk that cannot be decomposed in our construction is the geometry of the Kaluza-Klein fiber directions.  For example, the ABJM model vacuum is dual to $AdS_4 \times S_7$ \cite{ABJM}, while the $\mathcal{N} = 4$ Super Yang-Mills vacuum is dual to $AdS_5 \times S_5$ \cite{Maldacena1999}.  In such states, we cannot possibly use entropy of purification to subdivide the $S_n$ factor into smaller pieces, because spherical symmetry implies that all entanglement cross-sections will be symmetrical.\footnote{Interestingly, it might be possible to do better in excited geometries which are not spherically symmetric.}  In our construction, these fiber directions simply go along for the ride, without being subdivided, even though their radius of curvature is of the same order as that of the AdS factor, which we do subdivide.\footnote{See \cite{MST2014, KU2015} for a possible idea for how to understand the fiber directions in terms of entanglement between different field degrees of freedom.  If this idea is correct it would be interesting to combine it with our construction to obtain a discretization of the full space.}

Note that, as in all theories with large extra dimensions, the 10 or 11 dimensional Planck length will be parametrically larger than the effective Planck scale of the Kaluza-Klein reduced theory on AdS.  This raises the question of whether ignoring the fiber directions might actually allow us to subdivide the AdS factor at a finer scale than we could if we were also subdividing the sphere.  After all, our construction depends on the Hilbert space dimension being large, and including all the modes of the sphere makes the Hilbert space larger than it would be otherwise.  It seems likely that higher-curvature/stringy corrections to the Ryu-Takayanagi formula would prevent this from working, but this subject bears further investigation.

\subsection{Dynamics}\label{dyn}

Another important unanswered question is how to extend our construction to dynamical settings.  The main inhibition to extending the construction to holographic states with dynamical bulk spacetimes is that while boundary entropies are still given by the areas of extremal surfaces in the bulk, those extremal surfaces can generally not all be chosen to lie in a single spacelike bulk slice \cite{HRT2007}. In a static spacetime with Killing time parameter $t$, by contrast, all Ryu-Takayanagi surfaces of the $t=0$ boundary slice can be made to lie in the $t=0$ slice of the bulk.  Such a proposal would presumably require either (i) discretizing time to produce a $d$-dimensional tensor network for a $d$-dimensional spacetime \cite{QY2018, HH2017}, or else (ii) constructing different tensor networks for different Cauchy slices of the bulk \cite{May2017}, which nevertheless give rise to the same  boundary state.  In the latter case, it is tempting to identify the gauge equivalence of tensor networks with the Hamiltonian constraint of the continuum bulk general relativity, since both involve an equivalence of states at different times.

Unfortunately, it is impossible for tensor networks on dynamical Cauchy slices to be geometrically appropriate in the same sense as on static slices.  Let us consider any boundary region $R$, and attempt to construct a tensor network on a Cauchy slice $\Sigma$ which does \emph{not} include the HRT surface $X_R$.  By the maximin construction \cite{maximin}, the minimal area cut $\gamma$ always has less area than the HRT surface: $\operatorname{area}(\gamma) < \operatorname{area}(X_R)$ (for an example, see Figure \ref{fig:boost}).  On the other hand, the Swingle bound \cite{Swingle2012-1} requires that $\operatorname{area}(\gamma_R) \ge S(R) = \operatorname{area}(X_R)$.  This is a contradiction, unless we allow the log of the bond dimension to exceed the area even at leading order.

Perhaps, then, dynamical tensor networks are described by tensor networks that include long, nonlocal links connecting different parts of the network, which carry $O(1/G_N)$ amounts of information.  Presumably we could still construct a tensor network by our minimization procedures.  But for a general boundary slice, only a single family of nonintersecting HRT surfaces could be simultaneously placed on the same Cauchy slice and represented by edges with geometrically appropriate bond dimensions.  The other directions would have to have information flow exceeding their area.  

It may still be possible to construct such tensor networks in a compelling way by using the modular flow techniques of \cite{CDLQ2018, FLW2018}.  In this picture, when two flat slices meet at an HRT surface with a nonzero boost angle, there is a nonlocal exchange of information along the RT surface due to the modular flow associated with the boost.  (Otherwise, the maximin principle would be violated.)  This clarifies that, although our construction localizes the information of the RT surface at sub-AdS scales, this localization must be understood as only being valid in a particular Lorentz frame of reference.

On the other hand, since the modular flow relates the dynamical and static cases, the dynamical problems posed by the Swingle bound are to some degree already present in the static case. This observation led to our no-go theorem in Section \ref{sec:nogo_theorem}, which shows that geometrically appropriate tensor networks for the AdS/CFT correspondence cannot generally be expected to have approximate bulk-to-boundary isometries when Ryu-Takayanagi surfaces intersect in the bulk.

\subsection{Geometry from Entanglement}

The fact that our distillation procedure depends only on the entanglement structure of the state suggests that it may be more broadly applicable to other kinds of entangled states, perhaps e.g. those that are hypothesized to live on so-called holographic screens \cite{Bousso1999, SW2016} in cosmology.  Given a quantum state on a lattice, one need only check that its smooth min- and max-entropies of purification have favorable properties; if so, it will project a holographic state onto its interior.  The surface-state conjecture implies that the quadrilaterals of our tensor network grid (see e.g. Figure \ref{fig:generic_discretization_grid}) have suitable holographic states living on their boundaries.  In principle, we could use our construction to determine these boundary states explicitly.

We do not expect our quadrilaterals to be associated with ``perfect tensors'' \cite{HaPPY}, since cutting a quadrilateral along the diagonal results in a surface with less area than the two other sides of the triangle, allowing a nontrivial distillation to be performed along the diagonal.  But we do expect that there will be an approximate isometry (in the sense of Section \ref{sec:larger_tree_networks}) mapping any \emph{one} of the edges to the other three edges. In general, it will be important to prove as many isometry-like relations as possible (subject to the constraints of our no-go theorem), both for the purposes of quantum error correction and to determine how sensitive the tensor network is to the precise order in which distillations are performed.

This article goes in the direction of starting with a boundary state and analyzing what the holographic tensor network must be.  A complementary approach would be to start with the tensors associated with different kinds of geometries, and then synthesize them back together into an arbitrary geometry.  In doing so it would be important to check that all expected isometries continue to hold.  It would also be critical to show that, to high accuracy, the tensor network associated with a geometrical region does not significantly depend on either its external spatial context, or the methodology used to construct it.

If this can be done, then the holographic principle would finally be freed from the straitjacket of asymptotically AdS boundary conditions.  It could be applied equally well to universes with other asymptotic structures, or even to closed cosmologies!  In the latter case, the tensor network could be evaluated to give some complex number for each possible choice of spatial geometry.  Such a ``tensor network partition function'' would in effect define a special cosmological state over the space of 3-metrics.  It would be interesting to determine what relationship this special state might have to other proposals for special initial conditions, e.g. the Hartle-Hawking state.

\section*{Acknowledgments}
{\small We would like to acknowledge useful conversations with Ahmed Almheiri, Chris Akers, Raphael Bousso, Xi Dong, William Donnelly, Dan Harlow, Patrick Hayden, Ted Jacobson, Isaac Kim, Aitor Lewkowycz, Juan Maldacena, Don Marolf, Masamichi Miyaji, Ali Mollabashi, Xiao-Liang Qi, Dan Ranard, Eva Silverstein, Steve Shenker, Brian Swingle, Tadashi Takayanagi, and Guifre Vidal.  We especially thank Patrick Hayden, Brian Swingle, and Michael Walter for sharing with us their unpublished work on smooth min- and max-entropies in quantum field theory.  GP, JS, and AW were supported by the Simons Foundation (``It from Qubit''), AFOSR grant number FA9550-16-1-0082, and the Stanford Institute for Theoretical Physics.  AW was also supported in part by the John Templeton Foundation, Grant ID\# 60933, and would like to thank the Centro Atómico Bariloche, the California Institute of Technology, and the Kavli Institute for Theoretical Physics (supported in part by NSF grant PHY-1748958) for hospitality at key stages of this project.  NB is supported by the National Science Foundation under grant number 82248-13067-44-PHPXH and by the Department of Energy under grant number DE-SC0019380. He would also like to thank the SITP for hospitality while part of this work was completed.  The opinions expressed are those of the authors and do not necessarily reflect the views of funding agencies.}

\bibliographystyle{JHEP}
\bibliography{bibliography}

\end{document}